\documentclass[journal]{IEEEtran}
\usepackage{amsmath,amssymb,amsfonts}
\usepackage[ruled,linesnumbered,lined]{algorithm2e}
\usepackage{algorithmic}
\usepackage{array}
\usepackage[caption=false,font=normalsize,labelfont=sf,textfont=sf]{subfig}
\usepackage{textcomp}
\usepackage{stfloats}
\usepackage{url}
\usepackage{verbatim}
\usepackage{graphicx}
\usepackage{cite}
\usepackage{bm}
\usepackage{color, multirow}
\usepackage[table]{xcolor}
\usepackage{colortbl}
\usepackage[colorlinks, linkcolor=blue, citecolor=blue]{hyperref}
\usepackage{pifont, utfsym}
\usepackage{enumitem}
\setlist[itemize]{leftmargin=*}
\begin{document}

\title{Interference Management for Integrated Sensing and Communications: A Multiple Access Perspective}

\author{Kexin Chen, Yijie Mao,~\IEEEmembership{Member,~IEEE,} Wonjae Shin,~\IEEEmembership{Senior Member,~IEEE,}\\ Bruno Clerckx,~\IEEEmembership{Fellow,~IEEE,} and Christos Masouros,~\IEEEmembership{Fellow,~IEEE}
\thanks{This work has been supported in part by the National Nature Science Foundation of China under Grant 62571331. (\textit{Corresponding author: Yijie Mao})}
\thanks{K. Chen and Y. Mao are with the School of Information Science and Technology, ShanghaiTech University, Shanghai 201210, China (email: \{chenkx2023, maoyj\}@shanghaitech.edu.cn).}
\thanks{W. Shin is with the School of Electrical Engineering, Korea University, Seoul 02841, South Korea (email: wjshin@korea.ac.kr).}
\thanks{B. Clerckx is with the Department of Electrical and Electronic Engineering, Imperial College London, SW7 2AZ London, U.K. (email: b.clerckx@imperial.ac.uk).}
\thanks{C. Masouros is with the Department of Electronic and Electrical Engineering, University College London, WC1E 7JE London, U.K. (email: c.masouros@ucl.ac.uk).}
%\vspace{-5mm}
}

%\markboth{Journal of \LaTeX\ Class Files,~Vol.~0, No.~0, September~2024}%
\maketitle
\begin{abstract}
%% ISAC
The integrated sensing and communication (ISAC) technique has been considered a key enabler for 6G radio access networks. 
ISAC fulfills a brand new paradigm shift in wireless networks via the seamless interplay between communication and sensing within a unified network. 
However, the tight integration of these functionalities inevitably gives rise to various types of interference, posing significant challenges to existing ISAC waveform designs and rendering interference management a critical concern. 
Inspired by the development trajectory of wireless communications, different multiple access (MA) techniques, such as orthogonal multiple access (OMA), space-division multiple access (SDMA), and more recently, non-orthogonal multiple access (NOMA) and rate-splitting multiple access (RSMA), have been demonstrated to play a pivotal role in efficiently utilizing limited spectrum resources, designing ISAC waveforms, as well as managing inter-user interference and inter-functionality interference in ISAC.
Notably, the interplay between MA and ISAC presents mutually beneficial integration. 
On the one hand, ISAC helps MA techniques better exploit their interference management
capability beyond the communication-only networks. On the other hand, different MA techniques serve as promising solutions for inter-functionality and inter-user interference management in ISAC.
In this paper, we deliver the first comprehensive tutorial of MA techniques in ISAC networks.  
Specifically, we illustrate the fundamental principles of ISAC, classify the diverse types of interference in different ISAC systems, and compare MA-assisted ISAC designs, highlighting their respective advantages and limitations.
Moreover, we provide an outlook on the emerging applications and future research directions of different MA-assisted ISAC.
\end{abstract}

\begin{IEEEkeywords}
Integrated sensing and communication (ISAC), multiple access (MA) techniques, interference management, orthogonal multiple access (OMA), space-division multiple access (SDMA), non-orthogonal multiple access (NOMA), rate-splitting multiple access (RSMA).
\end{IEEEkeywords}
%%%%%%%%%%%%%%%%%%%%%%%%%%%%%%%%%%
\section{Introduction}\label{section1} 
%% introduction of ISAC
% overview
Integrated sensing and communication (ISAC) has attracted substantial attention as a promising enabler for the sixth generation (6G) wireless networks and beyond. 
% IMT-2030
Recently, the Radio Communication Division of the International Telecommunication Union (ITU-R) has incorporated ISAC as one key usage scenario for International Mobile Telecommunication 2030 (IMT-2030/6G) \cite{Kaushik24_ISACstandard,itu21}.
Meanwhile, leading standard development organizations (SDOs) in wireless communications, such as the 3rd Generation Partnership Project (3GPP), have prioritized efforts to establish a unified framework for ISAC \cite{Kaushik23_ISACstandard,singh24_ISACstandard,3gpp22}.
%%%%%%%%%%%%%%%%%
The development of ISAC is indeed noteworthy, since it enables the seamless integration of communication and sensing functionalities into a unified system. By optimizing the shared utilization of hardware platforms, spectrum and signal processing pipelines, ISAC offers multiple benefits \cite{lf21_CRBoptimization,lf23_seventyISAC,zd25_ISAC}.
% pros: new applications
Specifically, the unified system opens up innovative applications to future networks, including but not limited to augmented reality, smart home, and autonomous driving. 
This attributes to its accommodation of the unprecedented demands for high-resolution sensing and high-quality wireless connectivity. 
% pros: efficiency
Additionally, the dual usage of resources for communication and sensing significantly enhances energy efficiency (EE) and spectral efficiency (SE).
This thereby alleviates the resource rivalry between the two functionalities \cite{lf20_jointCandS}.
% pros: gain
Moreover, as communication and sensing integration deepens, ISAC surpasses traditional single-functionality systems, driving coordination and integration gains that significantly boost overall system performance while ensuring seamless interplay between the two \cite{lf22_ISACsurvey}.
%%%%%%%%%%%%%%%%%
\par
% cons
Although substantial efforts have been made to facilitate ISAC systems, there still remain a host of challenges to be solved \cite{zandrew21_ISACsp, lf22_ISACsurvey,lf23_seventyISAC,lsh24_ISACtenchallenges}.
% cons: waveform design
One critical challenge lies in designing appropriate waveforms, since communication focuses more on efficient and reliable data transmission, whereas radar requires high-resolution sensing with favourable auto-correlation properties \cite{lsh24_ISACtenchallenges}.
% cons: interference
Another major challenge in ISAC systems is the management of various types of interference, which arises from the overlap of time-frequency resources between communication and sensing functionalities.
As the integration of communication and sensing progresses from coexistence to shared resources and joint signal processing, ISAC systems become increasingly vulnerable to severe interference, making efficient solutions for managing interference a high priority.
%%%%%%%%%%%%%%%%%
\par
% MA
Fortunately, recent studies have shown that different multiple access (MA) techniques, 
originally developed in communication-only networks to enable multi-user wireless access \cite{Tse05_MA}, are capable to address interference among users or between communication and sensing in ISAC systems \cite{xcc21_RSMAISAC, ylf22_RSMApart2, mxd23_NOMAISACmagazine}. 
Thanks to these interference management capabilities, the incorporation of MA techniques in ISAC opens the door to enhancing both communication and sensing performance, thereby motivating the exploration of various MA-assisted ISAC approaches.
% following subsections
In the following subsections, we will first introduce different types of interference in ISAC systems, followed by a discussion on interference management strategies tailored to each type of interference, especially the need for MA-assisted ISAC. Finally, we summarize the related works and the major contributions of this work.
%%%%%%%%%%%%%%%%%%%%%%%%%%%%%%%%%%
\subsection{Diverse Interference Types in ISAC} \label{section1A}
As mentioned, the core mission of ISAC is to transcend isolated functionalities and enable seamless synergy between communication and sensing, thereby optimizing performance across the entire framework.
%% NO-ISAC/O-ISAC
In fact, such synergy spans a hierarchy of integration levels—from a low level of integration, i.e., orthogonal ISAC (O-ISAC), where communication and sensing operate independently via orthogonal radio resources, to a high level, i.e., non-orthogonal ISAC (NO-ISAC), where the two functionalities share radio resources for unified operation \cite{mxd23_NOMAISACmagazine, zc23_SemiISAC}.
Compared to conventional sensing-only and communication-only systems, there exists increasingly severe interference in ISAC, which can be categorized into the following five types:
%% diverse interference types
%%%%%%%%%%%%%%%%%
%% clutter interference
\subsubsection{Clutter interference}
This interference arises from signals reflected by unrelated objects in the surrounding environment, which distorts both communication and sensing performance.
%%%%%%%%%%%%%%%%%
%% self interference
\subsubsection{Self-interference}
This occurs at the base station (BS) with co-located transmit and receive antennas, where sensing echoes arriving at the receive antennas interfere with ongoing transmit signals \cite{la22_ISAClimitsurvey,wzq23_ISACsurvey_SI}.
%%%%%%%%%%%%%%%%%
%% inter-target interference
\subsubsection{Inter-target interference}
Inter-target interference has emerged as one challenging obstacle encountered in ISAC involving multiple targets. 
The echo signals reflected from various targets may overlap on the path back to the receive antennas, making it difficult to distinguish among targets. This interference becomes particularly severe when the targets are closely spaced. Moreover, in cases where there is a large disparity in echo strength, the sidelobes of strong targets may mask the echoes of weaker ones, significantly degrading the performance of target discrimination \cite{Ruggiano12_intertarget,zxj25_intertarget}.
%%%%%%%%%%%%%%%%%
%% inter-user interference
\subsubsection{Inter-user interference}
This refers to the interference among co-channel communication users. It has become another inevitable challenge in existing ISAC systems supporting multiple communication users. And such interference is likely to become increasingly severe in future ISAC networks as the number of communication users surges, posing a risk to system performance degradation.
%%%%%%%%%%%%%%%%%
%% inter-functionality interference
\subsubsection{Inter-functionality interference}
%% definition
Such interference arises from the overlapping radio resource utilization between communication and sensing functionalities, leading to two distinct subtypes: \textit{communication-to-sensing interference} and \textit{sensing-to-communication interference}. 
In fact, the presence of inter-functionality interference is influenced by the integration level between the two functionalities.
Specifically, inter-functionality interference is marginal when the two functionalities are allotted orthogonal radio resources, i.e., in O-ISAC systems. 
In contrast, NO-ISAC systems, which feature a high level of integration between communication and sensing, inevitably suffer from severe inter-functionality interference \cite{mxd23_NOMAISACmagazine}. 
\par 
Collectively, these types of interference hinder the performance, reliability, and scalability of ISAC systems, posing challenges to their design and deployment. Therefore, it is necessary to design efficient and effective interference management strategies to address and mitigate these interference issues.
\subsection{Interference Management Strategies in ISAC} \label{section1B}
Effective interference management is essential for ensuring robust performance in ISAC systems. 
Various strategies have been developed to address different interference types, each requiring tailored solutions to maintain system efficiency and reliability. This subsection details the interference management strategies tailored to each type of interference in ISAC systems. 
% clutter-interference
\subsubsection{Management for clutter interference}
As for clutter interference, it can be directly managed by adjusting antenna arrays to boost direct-path signals or installing clutter-rejecting structures \cite{Lazaro14_clutter1}.
Advanced filter design techniques, such as space-time adaptive processing (STAP) \cite{hjf20_clutterSTAP}, further improve the system performance by accurately estimating the clutter covariance matrix (CCM), which characterizes the clutter environment. 
Additionally, beamforming methods are employed to minimize the energy transmitted toward high-clutter areas \cite{lr22_clutterwaveform}. 
It is also worth noting that clutter can be exploited as a valuable source for environment monitoring and reconstruction \cite{Fung20_clutterreconstruction}.
% SI
\subsubsection{Management for self-interference}
For managing self-interference in ISAC, the strategies can be generally classified into propagation, analog, and digital domains.
The self-interference avoidance in the propagation domain aims to reduce the coupling between the transmit and receive paths by leveraging spatial isolation techniques based on antenna separation \cite{Barneto19_SIpropagation1}, cross-polarization \cite{Everett14_propagation2}, and beamforming design \cite{lz23_SIpropagation3}.
Additionally, one typical analog-domain method, implemented prior to the analog-to-digital converter (ADC), is to generate a negative copy of the self-interference signal and subtract it from the received signal \cite{zys19_SIcancellation}.
The final defense against self-interference, i.e., digital-domain mitigation techniques, involves linear or nonlinear adaptive filters that generate cancellation signals to subtract residual self-interference after the ADC \cite{wzl21_SIdigital}.
For detailed discussions on clutter interference and self-interference management, readers can refer to \cite{wzq23_ISACsurvey_SI,tam24_FDISACinterference,nyy24_ISACinterference}. 
% inter-target interference
\subsubsection{Management for inter-target interference}
When confronting inter-target interference caused by closely spaced targets, one common method is to leverage super-resolution parameter estimation algorithms including but not limited to subspace methods such as multiple signal classification (MUSIC) \cite{Schmidt86_MUSIC}, and compressed sensing (CS)-based methods such as Newtonized orthogonal matching pursuit (NOMP) \cite{xmh23_NOMP}. 
This contrasts with direct interference mitigation in communication networks, since radar targets are typically passive reflectors incapable of controlling their interaction with transmit signals. 
Moreover, multiple-input multiple-output (MIMO) radar waveform designs with favorable auto- and cross-correlation properties play a crucial role for alleviating inter-target interference \cite{sjx16_MIMOradarwaveform,Alaee19_MIMOradarwaveform}. 
This is achieved by designing orthogonal emitted waveforms via time, frequency, Doppler, or code-division schemes \cite{Alaee19_MIMOradarwaveform,shb14_TDFDCDMAradar,Nguyen20_DDRDMAradar}, while maintaining low auto-correlation levels, thereby preventing the sidelobes of strong targets from masking the weak targets. 
Another effective approach involves iteratively estimating and cancelling strong signals, which facilitates the discrimination of weaker targets \cite{fb13_CLEAN,zxj25_intertarget}.
% inter-user/inter-functionality interference
\subsubsection{Management for inter-user and inter-functionality interference}
To address inter-user and inter-functionality interference, symbol-level precoding (SLP)-based interference exploitation \cite{la20_SLPtutorial} has recently emerged as an effective solution. 
To be specific, by exploiting the knowledge of channel state information (CSI) and communication data symbols, SLP—one non-linear precoding approach—provides additional degrees of freedom (DoF) for ISAC waveform design in both the time and spatial domains \cite{lr22_SLPMIMO}. 
This enables the transformation of originally detrimental inter-user and inter-functionality (especially sensing-to-communication) interference into constructive interference (CI), which pushes the received signals away from the decision boundaries of the modulation constellation and thus enhances the symbol detection performance \cite{cyw24_SLPISAC,ww25_SLPnomaISAC,wyr25_SLPISAC,snc25_SLPsecureISAC}.
It is also worth noting that although primarily intended for inter-user and inter-functionality interference exploitation, the temporal design flexibility of SLP enables the suppression of range-Doppler sidelobes in the ambiguity function, potential to alleviate the masking effect caused by sidelobes of strong targets, i.e., aforementioned inter-target interference \cite{lps25_SLPISAC}.
\par
Another promising solution to alleviate inter-user and inter-functionality interference, and to meet other critical requirements like massive connectivity, high data throughput, low latency, and enhanced transmission reliability, is the integration of MA techniques based on the classical block-level precoding (BLP) \cite{yangp19_6Gcommunication, bruno24_MA}.
Contrary to interference exploitation in SLP, BLP alleviates interference with reduced computational complexity thanks to its symbol-independent nature, where a constant precoding matrix is utilized within a channel coherence interval\footnote{Note that a symbol-dependent variant of BLP has recently been explored in ISAC systems, where a constant precoding matrix is applied over a block of symbols shorter than the channel coherence interval \cite{wyr25_SLPISAC}. By partially exploiting CI, this approach achieves a more favorable trade-off between system performance and computational cost. For clarity, BLP in this work refers to the conventional and widely adopted symbol-independent implementation.}.
Therefore, this work focuses on MA at the block level unless otherwise specified. Readers are referred to \cite{ww25_SLPnomaISAC} for a study on SLP-based MA in ISAC, which, however, is still at an early stage.
BLP-based MA techniques are not only essential but also pivotal in advancing ISAC for 6G networks, ensuring optimized performance and scalability in next-generation wireless systems.
%%%%%%%%%%%%%%%%%
In the past several years, aligned with the development trajectory of wireless communications, MA techniques have evolved significantly, from orthogonal MA (OMA) to space-division MA (SDMA) in multi-antenna networks, and more recently, the investigation of non-orthogonal MA (NOMA) and rate-splitting MA (RSMA). 
They are essential for enabling efficient multi-user connectivity, facilitating massive access, and mitigating inter-user interference. Different MA schemes enable different interference management capabilities. Typically, we categorize these interference management approaches as follows \cite{m22_RSMAtutorial}:
%%%%%%%%%%%%%%%%%
\begin{itemize}
    \item \textit{Avoiding interference via scheduling users in orthogonal radio resources:} As in OMA, users are allocated with orthogonal radio resources, which thereby avoids the inter-user interference. The primary advantage of this approach is its adaptability to all levels of interference, whether strong or weak. However, the main disadvantage is that the limited time/frequency resources cannot accommodate the growing number of users.
    %%%%%%%%%%%%%%%%%
    \item \textit{Fully treating interference as noise:} As in SDMA\footnote{Here, SDMA refers specifically to SDMA based on multi-user linear precoding (MU-LP), unless otherwise specified.}, multiple users share the same time-frequency resource blocks, where each user is allowed to decode its intended communication streams directly by fully treating the residual interference as pure noise. This approach is well-suited for low interference levels, but becomes less effective when the interference is moderate or strong.
    %%%%%%%%%%%%%%%%%
    \item \textit{Fully decoding interference:} For instance, NOMA\footnote{Various strategies have been proposed for NOMA, e.g., with users being superposed via the power domain (i.e., power-domain NOMA) or code domain (i.e., code-domain NOMA) \cite{makki20_NOMAtutorial}. Note that this work focuses on power-domain NOMA, which is simplified as NOMA in the following. To provide a holistic view, other unprevailing MA schemes in the code domain or other domains are briefly discussed in Section \ref{section7}. Besides, readers can refer to \cite{bruno24_MA} for a more comprehensive discussion on various types of MA schemes.} tackles inter-user interference with at least one communication user mandated to decode and remove interference from other users. In contrast to fully treating interference as noise, this approach is only suited for strong interference levels. When the interference is moderate or weak, it is less desirable due to the loss of spatial multiplexing gain \cite{bruno21_MA}.
    %%%%%%%%%%%%%%%%%
    \item \textit{Partially decoding interference and partially treating interference as noise:} This is a brand new routine as taken in RSMA, where user messages are split into common and private parts, with the common parts decoded by multiple users and the private parts decoded by the corresponding individual users. This is an ideal and flexible strategy since it dynamically adjusts the amount of interference to be decoded and treated as pure noise based on varying interference levels. In contrast to the interference management approaches used in SDMA and NOMA, such approach employed by RSMA is well-suited for all levels of interference. Specifically, when interference is weak or strong, RSMA automatically reduces to SDMA or NOMA by turning off either the common or private streams, effectively bridging SDMA and NOMA. In cases of medium interference—neither strong nor weak—RSMA can still significantly enhance performance, making it a promising MA scheme for future networks \cite{m17_RSMAdownlink, m22_RSMAtutorial}.
\end{itemize}
%%%%%%%%%%%%%%%%%
\par
The aforementioned approaches to alleviate inter-user interference is the fundamental difference among OMA, SDMA, NOMA, and RSMA. 
The unique interference management capabilities of different MA schemes offer significant potential for ISAC applications, where communication and sensing functionalities coexist and often compete for resources. Integrating these MA schemes into ISAC systems holds significant potential to effectively manage inter-functionality and inter-user interference, and enhance overall performance\footnote{It is important to note that MA techniques were originally developed to support multiple communication users and have primarily been employed in ISAC systems to mitigate communication-related interference. 
While existing studies have investigated advanced MA techniques tailored for inter-target interference management, e.g., time/frequency/code-division MA proposed in \cite{shb14_TDFDCDMAradar} and Doppler/range-division MA proposed in \cite{Nguyen20_DDRDMAradar}, these approaches were specifically designed for conventional radar-only networks and remain unexplored in ISAC contexts. Since our work focuses on the dual-functional nature of ISAC systems, such radar-centric MA methods fall outside the scope of this study.}.
%% brief summary: win-win integration
In fact, the synergies between ISAC and MA bring a win-win integration. On the one hand, ISAC helps MA schemes better exploit their interference management capability beyond the conventional communication-only networks. On the other hand, different MA techniques serve as promising solutions for inter-functionality and inter-user interference management in ISAC. 
%%%%%%%%%%%%%%%%%%
%% other questions: how/what
However, integrating these MA schemes into ISAC systems remains challenging, raising three key questions that require further in-depth exploration:
\begin{itemize}
    \item[-] \textit{How can various MA techniques be effectively leveraged to tackle different types of interference in ISAC?}
    \item[-] \textit{What are the primary advantages and disadvantages of integrating various MA techniques into ISAC?}
    \item[-] \textit{What is the performance comparison among different MA-assisted ISAC systems?}
\end{itemize}
In the remainder of this work, we will delve deeply into these questions.
%%%%%%%%%%%%%%%%%%%%%%%%%%%%%%%%%%
\setlength{\textfloatsep}{8pt} 
\begin{table*}[t]
\centering
\caption{A Summary of Existing Surveys\&Tutorials\&Magazines on MA and/or ISAC}
\label{Table:existing works}
\begin{tabular}{|p{0.3cm}cp{0.9cm}p{2.1cm}p{1.6cm}p{10.2cm}|}\hline
%\multirow{2}{*}{\begin{tabular}[c]{@{}c@{}}\textbf{Ref.}\end{tabular}} & \multirow{2}{*}{\textbf{Type}} & \multicolumn{2}{c|}{\textbf{Related Topic}} & \multirow{2}{*}{\begin{tabular}[c]{@{}c@{}}\textbf{Main Discovery}\end{tabular}} \\ \cline{3-4}
\textbf{Year} & \textbf{Ref.} & \textbf{Type} & \textbf{MA Scheme} & \textbf{ISAC System} & \textbf{Main Contribution} \\ \hline
2011 & \cite{sturm11_ISACtutorial} & Tutorial & \usym{1F5F4} & NO-ISAC & Presented different waveform designs for ISAC with thorough operability and performance evaluation.\\ \hline
2016 & \cite{hassanien16_ISACtutorial} & Magazine & \usym{1F5F4} & NO-ISAC & Presented various information-embedding strategies for dual-functional waveform design. \\ \hline
2017 & \cite{islam17_NOMAtutorial} & Survey & NOMA & \ding{55} & Surveyed the progress of power-domain NOMA, and discussed potential NOMA designs and open challenges.\\ \hline
2019 & \cite{mishra19_ISACtutorial} & Magazine & \usym{1F5F4} & O-ISAC,\newline NO-ISAC & Presented various aspects of implementing mmWave ISAC, with a focus on the waveform design.\\ \hline
2020 & \cite{makki20_NOMAtutorial} & Survey & NOMA & \ding{55} & Surveyed the current status of NOMA, and summarized its 3GPP discussions.\\ \hline
2020 & \cite{mdy20_ISACtutorial} & Magazine & \usym{1F5F4} & NO-ISAC & Summarized ISAC studies in the context of autonomous vehicles, and discussed its challenges and research directions. \\ \hline
2021 & \cite{bruno21_MA} & Tutorial & Various MA,\newline especially NOMA\newline and RSMA & \ding{55} & Presented the multiplexing gains of multiple-input single-output (MISO) NOMA, MIMO NOMA, and RSMA, and showed the superiority of RSMA.\\ \hline
2021 & \cite{zandrew21_ISACsp} & Tutorial & \usym{1F5F4} & NO-ISAC & Presented the implementation of ISAC systems from the perspective of signal processing. \\ \hline
2021 & \cite{cyh21_ISACsurvey} & Magazine & \usym{1F5F4} & O-ISAC,\newline NO-ISAC & Presented the concept of signaling layer in internet of things (IoT) architectures, and discussed existing ISAC solutions for IoT across various layers. \\ \hline
2022 & \cite{zandrew22_ISACsurvey} & Survey & \usym{1F5F4} & NO-ISAC & Surveyed ISAC studies in the mobile network context, and discussed its motivation, methodologies, challenges and future directions. \\ \hline
2022 & \cite{la22_ISAClimitsurvey} & Survey & \usym{1F5F4} & NO-ISAC & Surveyed existing progress on fundamental limits in ISAC, and discussed performance metrics and bounds, open challenges and research directions. \\ \hline
2022 & \cite{lf22_ISACsurvey} & Survey & \usym{1F5F4} & O-ISAC,\newline NO-ISAC & Surveyed pertinent studies on ISAC involving its historical evolution, key applications, performance trade-offs, signal processing techniques, and explored its synergy with other wireless technologies. \\ \hline
2022 & \cite{deSena22_RSMARIS} & Magazine & RSMA & \ding{55} & Presented the synergy of RSMA and reconfigurable intelligent surface (RIS), and discussed three potential improvements and its use cases.\\ \hline
2022 & \cite{m22_RSMAtutorial} & Survey & Various MA,\newline especially RSMA & Briefly\newline mentioned & Surveyed the pertinent studies on RSMA, and discussed its detailed basics, superiority over other MA, applications and future directions.\\ \hline
2022 & \cite{Mishra22_RSMApart1} & Tutorial & RSMA & Briefly\newline mentioned & Presented the basics of RSMA and its applications in 6G.\\ \hline
2022 & \cite{ylf22_RSMApart2} & Tutorial & RSMA & NO-ISAC & Presented the interplay between RSMA and ISAC, with RSMA-aided downlink communication and mono-static sensing as an example.\\ \hline
2022 & \cite{lhy22_RSMApart3} & Tutorial & RSMA & \ding{55} & Presented the interplay between RSMA and RIS with a general model, and discussed its advantages and challenges.\\ \hline
2022 & \cite{lyw22_NOMA} & Magazine & NOMA & Briefly\newline mentioned & Presented the use of NOMA in 6G, with a focus on its basic principle, new requirements and applications.\\ \hline
2023 & \cite{wzq23_ISACsurvey_SI} & Survey & \usym{1F5F4} & NO-ISAC & Surveyed ISAC studies from aspects including signal design, processing and optimization.\\ \hline 
2023 & \cite{mxd23_NOMAISACmagazine} & Magazine & NOMA & NO-ISAC & Summarized various downlink/uplink NOMA-aided NO-ISAC systems.\\ \hline
2023 & \cite{lf23_seventyISAC} & Tutorial & \usym{1F5F4} & NO-ISAC & Presented basic principles and signal processing theories in ISAC, and discussed its development trends in spectrum and antenna aspects, as well as future directions. \\ \hline
2024 & \cite{tam24_FDISACinterference} & Magazine & \usym{1F5F4} & NO-ISAC & Presented interference issue in full-duplex ISAC, and outlined potential solutions such as radio architecture design.\\ \hline
2024 & \cite{Kaushik24_ISACstandard} & Magazine & \usym{1F5F4} & O-ISAC,\newline NO-ISAC & Presented several innovative aspects in ISAC from the perspective of 6G standardization. \\ \hline
2024 & \cite{park24_RSMAtutorial} & Magazine & Various MA,\newline especially RSMA & Briefly\newline mentioned & Presented the theoretical foundation of RSMA, and discussed its four key benefits and ten promising applications and scenarios.\\ \hline
2024 & \cite{bruno24_MA} & Tutorial, Survey & Various MA & Briefly\newline mentioned & Surveyed past, emerging and future MA for multi-functional 6G, and discussed artificial intelligence (AI) for MA, MA for AI, applications and standardization of MA.\\ \hline
2024 & \cite{lyx24_MAISAC} & Survey & Various MA & NO-ISAC & Surveyed various MA from orthogonal/non-orthogonal perspective for managing interference in NO-ISAC with mono-static sensing, and discussed its future directions. \\ \hline
2024 & \cite{nyy24_ISACinterference} & Survey & Briefly mentioned & NO-ISAC & Surveyed various interference management techniques in ISAC, including interference avoidance, suppression and exploitation.\\ 
\hline
2025 & \cite{lxw25_ISAC} & Survey & \usym{1F5F4} & NO-ISAC & Surveyed existing ISAC studies with a focus on layered architectures and enabling technologies, and discussed standard efforts, prototypes and open issues.\\ 
\hline 
\end{tabular}
\end{table*}
\subsection{Related Works}
\par
%% overview
A summary of existing related surveys, tutorials and magazines on MA and/or ISAC is provided in Table \ref{Table:existing works}. 
% works: MA
Tutorials, surveys or magazines \cite{islam17_NOMAtutorial, makki20_NOMAtutorial,bruno21_MA,m22_RSMAtutorial,Mishra22_RSMApart1, lhy22_RSMApart3,deSena22_RSMARIS,lyw22_NOMA,park24_RSMAtutorial, bruno24_MA} mainly focus on MA techniques, highlighting their distinctive features and diverse applications in advancing 6G technologies. 
Only \cite{m22_RSMAtutorial,Mishra22_RSMApart1,lyw22_NOMA, park24_RSMAtutorial, bruno24_MA} briefly discussed the use of MA schemes in ISAC, but they failed to provide an in-depth discussion on how different MA techniques address interference in ISAC.
% works: ISAC
While tutorials, surveys or magazines \cite{sturm11_ISACtutorial,hassanien16_ISACtutorial,mishra19_ISACtutorial,mdy20_ISACtutorial,cyh21_ISACsurvey,zandrew21_ISACsp,zandrew22_ISACsurvey,la22_ISAClimitsurvey,lf22_ISACsurvey,lf23_seventyISAC,Kaushik24_ISACstandard,lxw25_ISAC} focus on ISAC techniques, they overlook the critical issue of interference management within ISAC systems. 
% works: interference management in ISAC
Although \cite{wzq23_ISACsurvey_SI, tam24_FDISACinterference, nyy24_ISACinterference} focus on interference management issues in ISAC, they have their own limitations. 
For instance, the authors of \cite{wzq23_ISACsurvey_SI} primarily addressed the issue from a signal optimization perspective, focusing exclusively on alleviating inter-functionality interference and self-interference through interference cancellation (such as selective interference
cancellation), and interference avoidance (such as duplex) techniques. 
While \cite{tam24_FDISACinterference} analyzed the challenging issues of self-interference and inter-functionality interference, the study only briefly outlined the potential solutions such as radio architecture design (i.e., antenna cancellation and analog cancellation), beamforming, and mode selection (i.e., the target and user pairing), without providing detailed illustrations or performance evaluations.
The authors of \cite{nyy24_ISACinterference} briefly summarized interference management techniques in ISAC systems, involving interference avoidance (such as waveform design methods), interference suppression (such as filter design), and interference exploitation (such as power exploitation). However, the study lacks an in-depth discussion of these interference management approaches in ISAC, nor does it summarize their respective advantages and disadvantages.
Moreover, none of these works thoroughly explore the integration of various MA schemes to address the interference management issue.
% works: ISAC&ISAC
Although some surveys, tutorials or magazines investigate the integration of MA and ISAC \cite{mxd23_NOMAISACmagazine, ylf22_RSMApart2, lyx24_MAISAC}, the authors of \cite{ylf22_RSMApart2} only considered downlink RSMA for certain NO-ISAC systems, and \cite{mxd23_NOMAISACmagazine} focused solely on NOMA for ISAC. 
Recently, the survey paper \cite{lyx24_MAISAC} indeed explored various MA schemes in NO-ISAC systems for upcoming 6G networks. 
However, it lacks a detailed overview of different interference types in ISAC and is limited by a narrow focus on NO-ISAC. 
Moreover, there is a lack of discussion on interference management, particularly regarding how different MA schemes handle interference in ISAC.
To the best of our knowledge, no existing survey, tutorial or magazine paper thoroughly explains why and how to integrate different MA techniques with various ISAC scenarios, which calls for in-depth exploration.

%%%%%%%%%%%%%%%%%%%%%%%%%%%%%%%%%%%%
%% purpose
\subsection{Contributions} 
The main goal of this tutorial is to provide, for the first time, a holistic overview of existing MA techniques for managing interference in ISAC. This includes an in-depth discussion of ISAC principles, diverse interference types in ISAC, a comprehensive review of various MA-assisted ISAC systems with performance comparison, as well as emerging applications and future research directions for different MA-assisted ISAC.
%% contributions
Specifically, our major contributions are summarized as
follows:
% principles of ISAC (integration level); and MA (interference management approaches).
%%%%%%%%%%%%%%%%%%
% motivation. why
\par
\textit{First}, diverse interference types in ISAC systems are thoroughly summarized for the first time in this work, which specifically include clutter interference, self-interference, inter-target, inter-user and inter-functionality interference.
In contrast to existing studies \cite{ylf22_RSMApart2,mxd23_NOMAISACmagazine,wzq23_ISACsurvey_SI,tam24_FDISACinterference,nyy24_ISACinterference, lyx24_MAISAC}, our work highlights the unique interference characteristics across various ISAC scenarios by considering different integration levels, i.e., the O-ISAC, NO-ISAC, and semi-ISAC (S-ISAC), both uplink and downlink communications as well as mono-static and bi-static sensing modes.
The need for effective interference management in ISAC naturally drives the incorporation of MA schemes in ISAC systems.
%%%%%%%%%%%%%%%%%%
% compare the various MA-assisted ISAC systems. how/what
\par
\textit{Second}, we identify how different MA schemes, i.e., OMA, SDMA, NOMA, and RSMA, can be leveraged to alleviate the interference encountered in various ISAC systems. Unlike existing surveys or tutorials such as \cite{mxd23_NOMAISACmagazine, ylf22_RSMApart2} which focus on the superiority of specific MA scheme NOMA or RSMA in certain ISAC scenarios, this is the first tutorial paper delivering a detailed review of the interplay between diverse ISAC systems and various MA schemes. 
Specifically, we first illustrate the basic O-ISAC systems assisted by OMA, with a focus on time-division MA (TDMA), and orthogonal-frequency-division MA (OFDMA)-assisted O-ISAC.
Additionally, NO-ISAC systems assisted by more efficient MA techniques such as downlink SDMA, NOMA and RSMA are delineated and compared with respective advantages and disadvantages. 
To provide a holistic view, we further summarize uplink MA techniques such as uplink NOMA and RSMA for managing interference in NO-ISAC systems. Moreover, this is the first work to provide a comprehensive performance comparison of various MA-assisted ISAC. 
%%%%%%%%%%%%%%%%%%
% application
\par
\textit{Third}, we summarize some open challenges in MA-assisted ISAC such as precoding toward random ISAC signals, followed by its emerging applications and research directions such as vehicle-to-everything (V2X) networks, millimeter-wave (mmWave) and terahertz (THz), edge intelligence, etc. This differs from other tutorials which concentrate solely on MA schemes (e.g., \cite{Mishra22_RSMApart1,m22_RSMAtutorial}) or ISAC systems (e.g., \cite{lf22_ISACsurvey}) for some particular applications.
\par
We hope this tutorial will clarify the fundamental principles of MA-assisted ISAC, strengthen the foundation of existing research on the interplay between MA and ISAC, serve as a valuable reference on the use of MA in ISAC, and ultimately expedite advance research in this field.
%%%%%%%%%%%%%%%%%%%%%%%%%%%%%%%%%%%%
\subsection{Organization and Notation}
The remainder of this work is organized as follows. 
In Section \ref{section2}, we introduce the principles of different ISAC systems, along with the interference encountered in each system.
Section \ref{section3} reviews OMA-assisted O-ISAC systems, with a focus on TDMA/OFDMA-assisted O-ISAC.
Section \ref{section4} delineates different NO-ISAC systems assisted by more efficient downlink MA techniques such as SDMA, NOMA and RSMA in the downlink, highlighting their respective advantages and disadvantages. 
In Section \ref{section5}, we concentrate on uplink MA techniques for interference management in NO-ISAC systems.
To provide a full picture, Section \ref{section7} explores the incorporation of other advanced MA techniques for managing interference in future ISAC systems.
Furthermore, Section \ref{section6} provides a performance comparison among different MA-assisted ISAC. 
In Section \ref{section8}, we discuss the open challenges of MA-assisted ISAC such as the precoding toward random ISAC signals, as well as its emerging applications and future research directions in various promising directions such as V2X networks and edge intelligence, etc. 
Section \ref{section9} concludes this paper.
%The outline of this paper is presented in Table \ref{Table:paper outline}.
%%
\par
\textit{Notations:} Matrices and vectors are denoted as the boldface letters in upper case and lower case, respectively. The subscript $\Re$ and $\Im$ represent the real and imaginary part of a complex. $\mathrm{diag}(\mathbf{a})$ constructs a square matrix where the entries of vector $\mathbf{a}$ are on the diagonal, and all off-diagonal elements are set to zero. $\mathrm{tr}(\cdot)$, $(\cdot)^{-1}$, $(\cdot)^{T}$, $(\cdot)^{\ast}$ and $(\cdot)^{H}$ represent trace, inverse, transpose, conjugate and conjugate-transpose, respectively. The operator $\left \| \cdot  \right \|_{2}$ means the Euclidean norm.
%%%%%%%%%%%%%%%%%%%%%%%%%%%%%%%%%%%%
\section{Interference in Various ISAC Systems}\label{section2}
This section provides an overview of various ISAC systems, including NO-ISAC, O-ISAC, and S-ISAC, highlighting different types of interference encountered in each system. 
%%%%%%%%%%%%%%%%%%
\par
%% overview: O-ISAC,NO-ISAC,Semi-ISAC
% tendency: higher intergration level
As mentioned previously, beyond the mere coexistence of communication and sensing, the future tendency for ISAC lies in an increasingly higher level of integration between the two functionalities. 
% category
In general, ISAC systems can be conceived based on different integration levels, ranging from a low level of integration, i.e., O-ISAC, and a medium level, i.e., S-ISAC, and finally to a high level, i.e., NO-ISAC \cite{mxd23_NOMAISACmagazine,zc23_SemiISAC}.
% O-ISAC
Specifically, O-ISAC systems allocate orthogonal radio resources to communication and sensing, such that there exists no inter-functionality interference. 
% NO-ISAC
In contrast, NO-ISAC schedules the communication and sensing functionalities with non-orthogonal radio resources, which fulfills promising dual-functional trade-offs with effective interference management.
% Semi-ISAC
Additionally, S-ISAC allows a portion of the wireless resources being exclusively utilized for either communication or sensing, and the rest being shared by the two functionalities \cite{zc23_SemiISAC}. This thereby bridges the aforementioned O-ISAC and NO-ISAC in terms of both interference levels and SE, as illustrated in Table \ref{Table:ISAC overview}.
\begin{table}[t]
\centering
\caption{Comparison of Different ISAC systems in terms of interference levels and SE.}
\label{Table:ISAC overview}
\begin{tabular}{|cccc|}
\hline
\rowcolor{gray!20}
\textbf{ISAC System} & \begin{tabular}[c]{@{}c@{}}\textbf{Resource}\\\textbf{Allocation}\end{tabular} & \begin{tabular}[c]{@{}c@{}}\textbf{Interference}\\\textbf{Level}\end{tabular}  & \begin{tabular}[c]{@{}c@{}}\textbf{Spectral}\\\textbf{Efficiency}\end{tabular} \\ \hline
\begin{tabular}[c]{@{}c@{}}Orthogonal ISAC\\(O-ISAC)\end{tabular} & \begin{tabular}[c]{@{}c@{}}Dedicated\\resources\end{tabular} & Low & Low \\ \hline
\begin{tabular}[c]{@{}c@{}}Non-orthogonal ISAC\\(NO-ISAC)\end{tabular} & \begin{tabular}[c]{@{}c@{}}Fully shared\\resources\end{tabular} & High  & High \\ \hline
\begin{tabular}[c]{@{}c@{}}Semi-ISAC\\(S-ISAC)\end{tabular} & \begin{tabular}[c]{@{}c@{}}Partially shared\\resources\end{tabular} & Medium & Medium \\ \hline  
\end{tabular}
\end{table}
%%%%%%%%%%%%%%%%%%
\par
For each integration level, ISAC systems can be further categorized based on their communication and sensing modes. 
Specifically, wireless communication operates in either uplink or downlink transmission modes, while sensing—referring to radar sensing in this work—can be classified into mono-static and bi-static modes\footnote{Note that the multi-static sensing mode, as an extension of the bi-static mode, is discussed in Section \ref{section8A}.} \cite{la22_ISAClimitsurvey,ywf24_sensingmode}. 
In the mono-static sensing mode, the transmit and receive antennas are equipped at the same BS, while the transmit and receive antennas are separated at different sites in the bi-static mode \cite{la22_ISAClimitsurvey,ywf24_sensingmode}.
% downlink and uplink sensing
Moreover, in alignment with downlink and uplink communication, downlink and uplink sensing have recently been introduced in ISAC for clarity \cite{zandrew21_ISACsurvey,zandrew22_ISACsurvey}. Specifically, in downlink sensing, the sensing signal (e.g., the dual-functional or dedicated sensing signals) is transmitted from the BS. In uplink sensing, the sensing signal originates from the users, meaning that the sensing functionality is achieved by leveraging the uplink communication signals.
Based on different communication and sensing modes, the ISAC architectures can be further classified into the following five different types\footnote{It is important to note that we exclude the uplink mono-static sensing scenario (i.e., where the uplink signal from a communication user is used for sensing, with the user also acting as a radar receiver) from our classification. This exclusion is based on the fact that users with small sizes typically exhibit limited sensing capabilities, as discussed in \cite{lf22_ISACsurvey}. Such an ISAC model is rarely considered in practice due to practical challenges.}: \textit{1) uplink communication and downlink mono-static sensing} \cite{mxd23_NOMAISACmagazine, oycj22_uplinkISAC, oycj22_uplinkdownlinkISAC, qq24_uplinkISAC, wxy22_FDsensing}, \textit{2) uplink communication and downlink bi-static sensing } \cite{Kim22_upc_bis,lxw25_ISAC}, \textit{3) downlink communication and downlink mono-static sensing} \cite{lf21_CRBoptimization, mxd23_NOMAISACmagazine,hhc23_ISAC_radarsequence,lx20_ISAC_radarsequence, ylf22_RSMApart2,lhc24_downc_downlinkmono}, \textit{4) downlink communication and downlink bi-static sensing }\cite{Pucci22_downlinkbistatic, Zabini23_downlinkbistatic, zqm23_downlinkbistatic,mth25_downlinkbistatic}, \textit{5) uplink communication and uplink bi-static sensing} \cite{pjh23_uplinkc_uplinkbistatics,Tapio24_uplinkbistaticISAC,hjm25_uplinkbistaticISAC}.
These ISAC architectures at varying integration levels encounter distinct types of interference, necessitating tailored interference management strategies. The subsequent subsections will elaborate these architectures for each integration level in detail.
% other three
% uplink communication and uplink mono-static sensing
% (downlink communication and uplink mono-static sensing)
% (downlink communication and uplink bi-static sensing)
%%
%%%%%%%%%%%%%%%%%%
%% NO-ISAC
\subsection{Non-orthogonal ISAC}\label{section2A}
\begin{figure}[tb]
\centering
\includegraphics[width=0.9\linewidth]{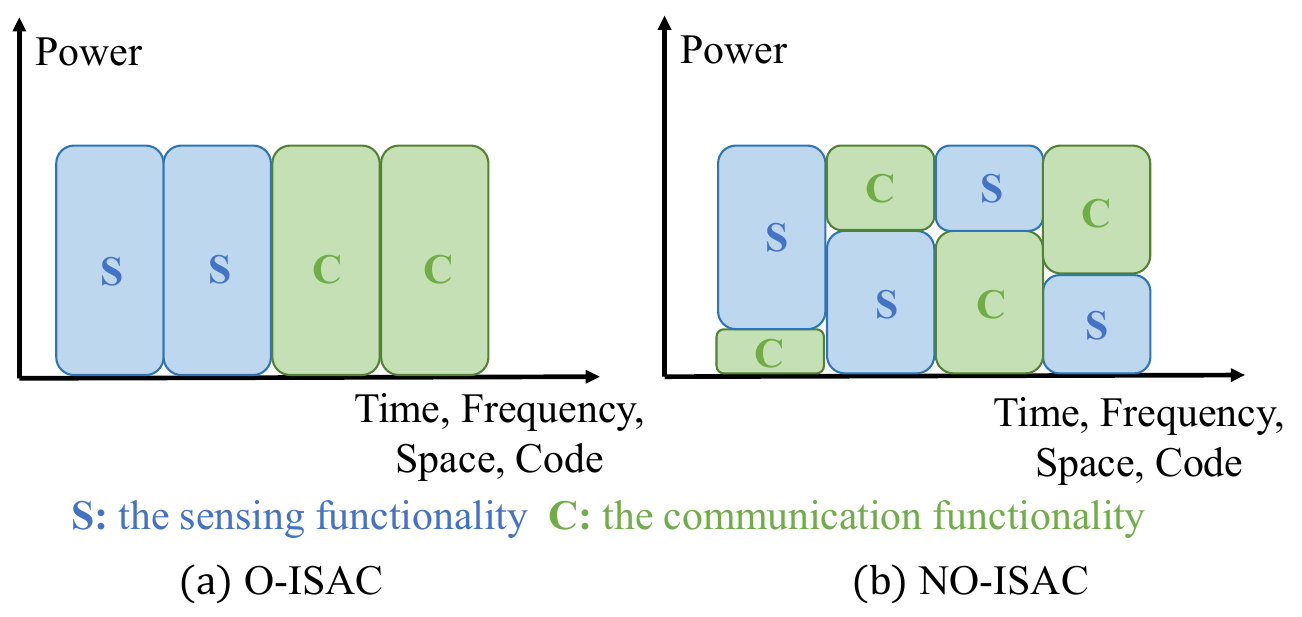}
\caption{The typical O-ISAC and NO-ISAC models.}
\label{fig:NO-ISAC}
\end{figure}

% definition
We begin by introducing the NO-ISAC system, which experiences the highest level of interference. 
As shown in Fig. \ref{fig:NO-ISAC}, contrary to the orthogonal radio resource allocation in O-ISAC, NO-ISAC schedules communication and sensing via non-orthogonal radio resources.
Such integration leads to mutual interference between the two functionalities, but also enables ISAC to exploit its full potential, where enhanced communication and sensing trade-offs can be achieved under appropriate waveform design.
%% dual-functional waveform design/ five architectures
As illustrated in Fig. \ref{fig:ISACcases}, there are five distinct ISAC architectures at this integration level, which can be classified into two categories: architectures with and without dual-functional waveform design. 
Specifically, in architectures without dual-functional waveform design, i.e., Fig. \ref{fig:ISACcases}(a) and (b), communication and sensing functionalities are achieved leveraging separate signals. While in Fig. \ref{fig:ISACcases}(c)-(e), the design of a dual-functional transmit signal is essential to support both communication and sensing functionalities, which thereby enables maximum integration by exploiting the same signal.
It is worth noting that different dual-functional waveform designs result in varying types of interference in certain NO-ISAC architectures, which will be discussed in detail later. 
In the following, we outline the five distinct NO-ISAC architectures and the specific interference challenges associated with each system. 
%%%%%%%%%%%%%%%%%%%%%%%
%% different ISAC architectures: 5 types
\subsubsection{ISAC architectures without dual-functional waveform design} 
\begin{figure*}
    \centering
    \includegraphics[width=0.85\linewidth]{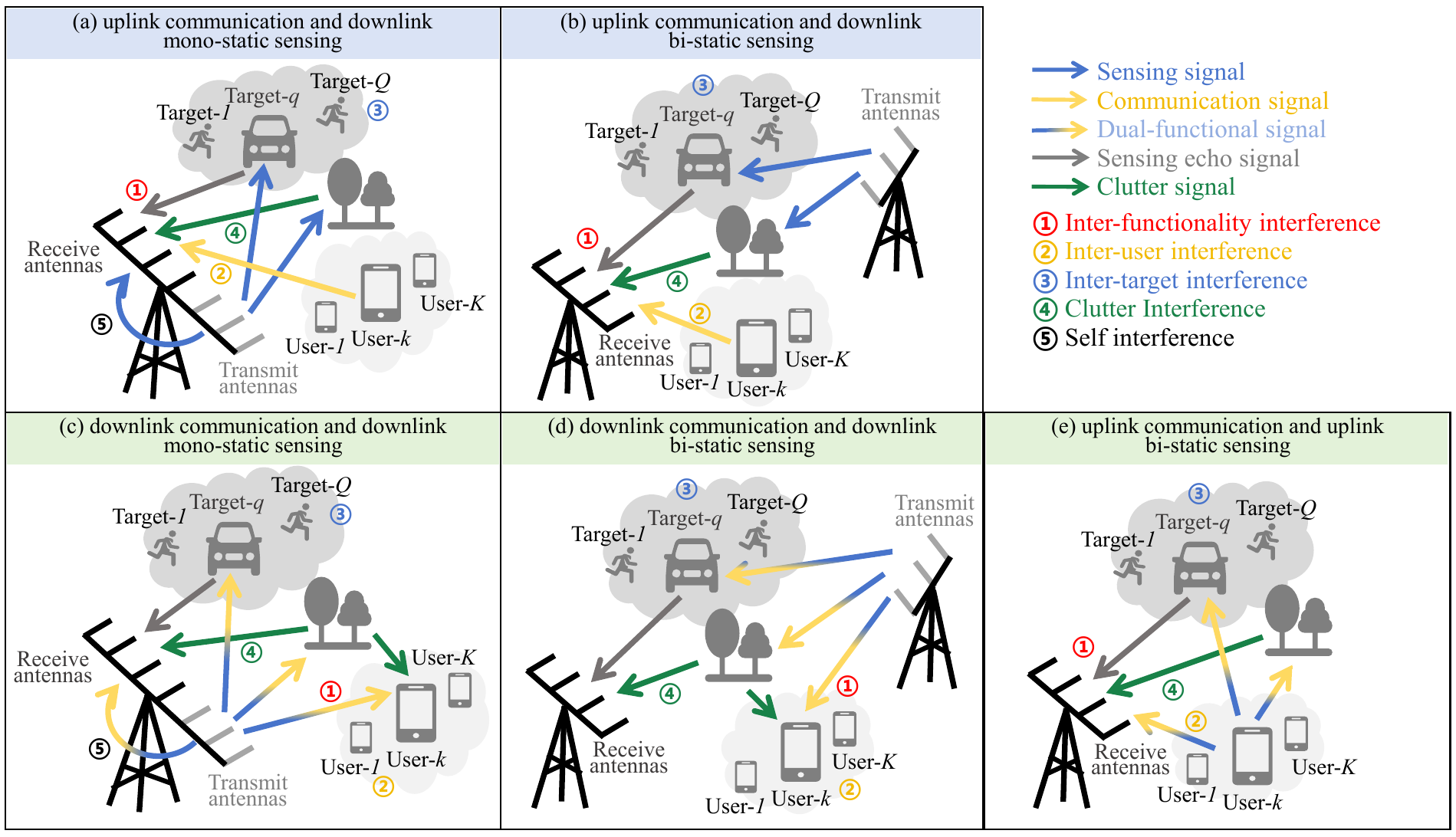}
    \caption{Various types of interference observed in five distinct NO-ISAC architectures.}
    \label{fig:ISACcases}
\end{figure*}

\begin{itemize}
    %%%%%%%%%%%%%%%%%%%%%%%
    \item \textit{Uplink communication and downlink mono-static sensing \cite{mxd23_NOMAISACmagazine, oycj22_uplinkISAC, oycj22_uplinkdownlinkISAC, qq24_uplinkISAC, wxy22_FDsensing}:}
    As illustrated in Fig. \ref{fig:ISACcases}(a), in this architecture, the BS only transmits sensing waveforms to targets. 
    In the meanwhile, it receives both information signals from communication users and sensing echo signals from targets.
    In this architecture, there exist inter-functionality, inter-user, inter-target, clutter interference as well as self-interference.
    Specifically, this architecture typically encounters both \textit{communication-to-sensing} and \textit{sensing-to-communication interference}, since radar sensing and uplink information decoding are simultaneously carried out at the same BS.
    This architecture also experiences \textit{self-interference} mentioned in Section \ref{section1A} due to mono-static sensing, where radar echo signals are prone to be interfered by transmit signals.
    %%%%%%%%%%%%%%%%%%%%%%%
    \item \textit{Uplink communication and downlink bi-static sensing \cite{Kim22_upc_bis,lxw25_ISAC}:} 
    As illustrated in Fig. \ref{fig:ISACcases}(b), there is one BS sending sensing signals to targets. In the meanwhile, there is another BS that simultaneously receives communication signals from users and radar echo signals from targets. 
    Such architecture encounters inter-functionality, inter-user, inter-target and clutter interference.
    Similar to Fig. \ref{fig:ISACcases}(a), there occurs \textit{communication-to-sensing interference} as well as \textit{sensing-to-communication interference}, which arise from the simultaneous use of BS for communication and sensing signal processing.
    Moreover, in such bi-static sensing architecture, \textit{self-interference} is largely alleviated at the cost of extra hardware cost and more parameters to be estimated \cite{la22_ISAClimitsurvey}.
\end{itemize}
% joint design category
\begin{figure}[tb]
    \centering
    \includegraphics[width=0.9\linewidth]{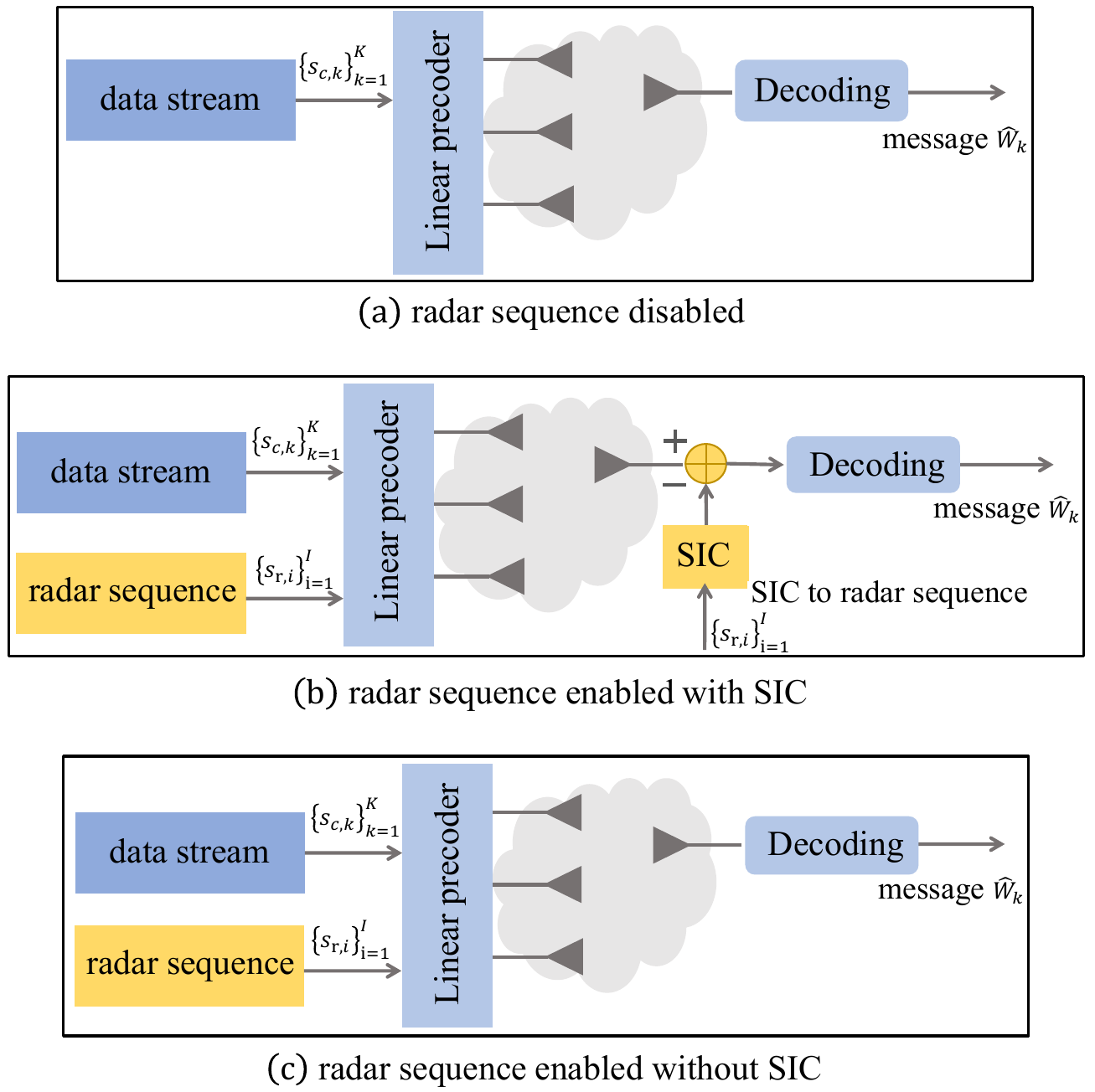}
    \caption{Typical three schemes of joint dual-functional waveform design in NO-ISAC \cite{hhc23_ISAC_radarsequence}.}
    \label{fig:NOISACcategory}
\end{figure}
%%%%%%%%%%%%%%%%%%
\subsubsection{ISAC architectures with dual-functional waveform design} The architectures shown in Fig. \ref{fig:ISACcases}(c)-(e) demand the design of a dual-functional transmit signal, so as to support both communication and sensing functionalities.
Considering that different dual-functional waveform designs lead to varying interference types in these architectures, we first provide an overview of these design schemes.
\par
In general, dual-functional waveform design can be typically classified into three categories: communication-centric, sensing-centric, and joint design \cite{lf22_ISACsurvey, mishra19_ISACtutorial, mdy20_ISACtutorial, lf23_seventyISAC, zandrew21_ISACsp, cyh21_ISACsurvey}. 
In contrast to the other two schemes that prioritize a single functionality, the joint design scheme enables flexible trade-offs between the two functionalities and has emerged as an important research direction in ISAC waveform design \cite{lf21_CRBoptimization, mxd23_NOMAISACmagazine, lx20_ISAC_radarsequence, hhc23_ISAC_radarsequence, ylf22_RSMApart2, hzy23_FDISAC}.
%It is important to note that, in this tutorial, we mainly focus on the joint design scheme, since it has become increasingly popular in recent ISAC studies \cite{lf21_CRBoptimization, mxd23_NOMAISACmagazine, lx20_ISAC_radarsequence, hhc23_ISAC_radarsequence, ylf22_RSMApart2, hzy23_FDISAC}. Readers can refer to \cite{lf22_ISACsurvey, mishra19_ISACtutorial, mdy20_ISACtutorial, lf23_seventyISAC, zandrew21_ISACsp, cyh21_ISACsurvey} for a more in-depth discussion on various design approaches.
% definition
% general
In particular, constrained by both communication and sensing performance, the joint dual-functional waveform can be conceived through reusing information signals or combining dedicated signals for the two functionalities, i.e., exclude or include additional radar sequences besides the information signals \cite{lx20_ISAC_radarsequence, hhc23_ISAC_radarsequence}.
% no radar sequence
Specifically, when extra radar sequences are excluded, information signals are utilized to implement both communication and sensing functionalities.
% with radar sequence
Conversely, with extra radar sequences, the dedicated communication and sensing signals are enabled to better fulfill the two functionalities. 
% about SIC
We should note that there contains no signaling information in radar sequences, which implies that they can be previously known at communication users, and consequently be eliminated via successive interference cancellation (SIC). 
% category
As illustrated in Fig. \ref{fig:NOISACcategory}, depending on the presence of additional radar sequences and the use of SIC at the receiver, the joint dual-functional waveform design can be further classified as follows:
\par
\textit{i)} \textit{Radar-sequence-disabled waveform design:}
% definition
Without radar sequences, both communication and sensing tasks are realized by reusing the information signals.
% further explain
In other word, for a multi-antenna ISAC serving $K$ single-antenna users, the $K$-rank data matrix containing data streams for $K$ communication users is utilized for the two functionalities after linearly precoded by the $K$-rank precoding matrix.  
% DoF: rank(definition); deficient; 
When considering an underloaded setting, i.e., the number of transmit antennas $N_T$ is larger than the number of communication users $K$, the available DoF for such MIMO radar waveform is equal to $K$, i.e., the rank of the transmit signal. 
% cons
However, this approach is likely to lead to a performance loss in sensing functionality, as the DoF available for sensing is limited by the number of users $K$ \cite{lx20_ISAC_radarsequence}. 
%%%%%%%%%%%%%%%%%%%%%%%%%%%%%%%%%%%%%%%%
\par
\textit{ii)} \textit{Radar-sequence-enabled waveform design with SIC:}
% definition, radar sequence
With the aim of achieving both satisfactory communication and sensing performance, this approach enables joint transmission of radar sequence and information signals. However, it faces inter-functionality interference caused by the dedicated radar sequences. Fortunately, communication users can mitigate this interference using SIC before decoding their intended information messages.
This attributes to the fact that the radar sequences involving no useful information can be pre-designed and shared between the transmitter and communication users before transmission \cite{hhc23_ISAC_radarsequence}.
% pros/cons
Overall, this scheme potentially extends the available DoFs of the radar waveform to its maximum value, i.e., $N_T$, leading to an enhanced radar performance \cite{Attiah24_sensingsequence}.
And it effectively eliminates the interference caused by additional radar sequences, albeit at the expense of high receiver complexity.
%%%%%%%%%%%%%%%%%%
%%%%%%%%%%%%%%%%%%%%%%%%%%%%%%%%%%%%%%%% 
\par
\textit{iii)} \textit{Radar-sequence-enabled waveform design without SIC:}
% definition
Similar to the previous radar-sequence-enabled scheme with SIC, this approach also facilitates the simultaneous transmission of dedicated radar sequences and communication signals. Nevertheless, communication users fail to cancel the interference caused by radar sequences at the receiver, which differs from the previous scheme.
% pros/cons
Although this scheme also achieves full radar DoFs, ensuring satisfactory sensing performance, it suffers from communication performance degradation due to the unavoidable inter-functionality interference triggered by the dedicated radar sequences.
%%%%%%%%%%%%%%%%%%
\par
With these joint dual-functional waveform designs established, we proceed to explore their impact on interference across different ISAC architectures, which is detailed as follows:
\begin{itemize}
    %%%%%%%%%%%%%%%%%%%%%%%
    \item \textit{Downlink communication and downlink mono-static sensing \cite{lf21_CRBoptimization, mxd23_NOMAISACmagazine,hhc23_ISAC_radarsequence,lx20_ISAC_radarsequence, ylf22_RSMApart2,lhc24_downc_downlinkmono}:} 
    As illustrated in Fig. \ref{fig:ISACcases}(c), the BS transmits downlink signals to not only communication users but also sensing targets, and simultaneously performs target detection or parameter estimation based on the received echo signals. 
    There exist inter-functionality, inter-user, inter-target, clutter interference as well as self-interference.
    It is worth noting that, there is generally no \textit{communication-to-sensing interference} when considering inter-functionality interference, since the radar sensing and information bit decoding are carried out at the BS and the users, respectively. This also attributes to the fact that the communication signals can be leveraged for the sensing functionality. 
    However, \textit{sensing-to-communication interference} arises when employing a radar-sequence-enabled waveform design without SIC, as the additional radar sequence becomes a source of interference. This interference, however, vanishes when using a radar-sequence-disabled design or a radar-sequence-enabled design with SIC \cite{lx20_ISAC_radarsequence,hhc23_ISAC_radarsequence,lf18_SDISAC}.
    \textit{Self-interference} also occurs in this architecture due to mono-static sensing nature.
    %%%%%%%%%%%%%%%%%%%%%%%
    \item \textit{Downlink communication and downlink bi-static sensing \cite{Pucci22_downlinkbistatic, Zabini23_downlinkbistatic, zqm23_downlinkbistatic,mth25_downlinkbistatic}:} 
    As illustrated in Fig. \ref{fig:ISACcases}(d), there exists one BS that sends dual-functional signals for both downlink communication and radar sensing. 
    The probing echo signals are received at a different BS, i.e., the transmit and receive antennas are well separated.
    In this architecture, there occur inter-functionality, inter-user, inter-target and clutter interference.
    Similar to Fig. \ref{fig:ISACcases}(c), there is generally no \textit{communication-to-sensing interference}, thanks to the use of communication signals for the sensing functionality. Meanwhile, the incorporation of extra radar sequence may lead to \textit{sensing-to-communication interference}.
    In such architecture, the nature of bi-static sensing generally helps alleviate \textit{self-interference}.
    %%%%%%%%%%%%%%%%%%%%%%%
    \item \textit{Uplink communication and uplink bi-static sensing \cite{pjh23_uplinkc_uplinkbistatics,Tapio24_uplinkbistaticISAC,hjm25_uplinkbistaticISAC}:}
    As illustrated in Fig. \ref{fig:ISACcases}(e), the uplink signals from communication users are exploited for both communication and sensing. 
    And the BS receives not only uplink communication signals, but also probing echo signals for target detection or parameter estimation. 
    Since communication users are typically not allowed to transmit extra radar sequences, we only consider the radar-sequence-disabled scheme in this architecture, where inter-functionality, inter-user, inter-target, and clutter interference arise.
    Specifically, there exist both \textit{communication-to-sensing interference} and \textit{sensing-to-communication interference} similar to the architectures in Fig.\ref{fig:ISACcases}(a) and (b).
    Moreover, \textit{self-interference} is effectively mitigated due to separated transmit and receive antennas, which ensures that the received signals remain uncorrupted by the transmit signals.
\end{itemize}
% other three
% uplink communication and uplink mono-static sensing
% (downlink communication and uplink mono-static sensing)
% (downlink communication and uplink bi-static sensing)
\par
%% mention the FD ISAC (to communication functionality)
% HD
% for sensing: mono-static in FD; bi-static no related concept.
% for communication: HD; FD
It is worth noting that the aforementioned distinct ISAC architectures can be typically referred to as half-duplexing (HD) ISAC, which are commonly considered in existing ISAC studies \cite{mxd23_NOMAISACmagazine, oycj22_uplinkISAC, oycj22_uplinkdownlinkISAC, qq24_uplinkISAC, wxy22_FDsensing,Kim22_upc_bis,lf21_CRBoptimization,hhc23_ISAC_radarsequence,lx20_ISAC_radarsequence, ylf22_RSMApart2,lhc24_downc_downlinkmono,Pucci22_downlinkbistatic, Zabini23_downlinkbistatic, zqm23_downlinkbistatic,mth25_downlinkbistatic,pjh23_uplinkc_uplinkbistatics,Tapio24_uplinkbistaticISAC,hjm25_uplinkbistaticISAC}. 
In these HD-ISAC systems, communication is confined to HD operation, functioning exclusively in either uplink or downlink mode at any given time \cite{tam24_FDISACinterference}. 
% FD
%In fact, the FD capabilities can be extended to the communication functionality. 
In comparison, there also exists full-duplex (FD) ISAC, i.e., integrated radar sensing and FD communication, where the system senses the targets, and enables both uplink and downlink communications \cite{tam24_FDISACinterference,hzy23_FDISAC,lrg24_FDISAC}. 
The distinction between HD/FD ISAC is primarily based on whether the communication subsystem can simultaneously support uplink and downlink transmissions \cite{tam24_FDISACinterference, hzy23_FDISAC,lrg24_FDISAC}. 
This classification is independent of the sensing functionality, since many existing ISAC studies typically consider simultaneous sensing signal transmission and echo reception, which is inherently full-duplex \cite{barneto21_FDsensing,Islam22_FDsensing,wxy22_FDsensing}.
It is obvious that FD ISAC naturally suffers from more complex interference.
Moreover, self-interference is inherent in FD ISAC with co-located transmit and receive antennas, as both desired sensing echo signals and uplink communication signals are likely to reach the receive antennas before transmission is complete, disrupting the received signal \cite{wl15_FDcommunication_SI}.
\par
In summary, though appealing with a high level of integration, NO-ISAC systems inevitably suffer from significant interference issues. 
Specifically, apart from inter-user, inter-target, clutter interference and self-interference, inter-functionality interference arises due to the overlapped resource sharing between communication and sensing.
In particular, this interference becomes more severe in uplink communication architectures, or downlink communication architectures with radar sequence enabled without SIC.
%Concretely, inter-functionality interference becomes severe in architectures involving uplink communication, where communication and sensing signals are simultaneously processed at the same BS. 
%This interference is also notable in downlink communication architectures with radar sequence enabled without SIC, since the radar sequence itself becomes one source of interference.
In fact, NO-ISAC systems fail to unlock its full potential without addressing these interference, highlighting the need for effective interference management strategies.

%% O-ISAC
\subsection{Orthogonal ISAC}\label{section2B}
% definition
In contrast to NO-ISAC, O-ISAC schedules the communication and sensing functionalities within a unified system leveraging orthogonal resource allocation. 
%This thus alleviates the interference between the two functionalities. 
% category: 4 domain
Specifically, 
%as per Fig. \ref{fig:O-ISAC}, 
O-ISAC can be realized in various resource domains such as the time, frequency, spatial and code domains, which are referred to as the time-division (TD), frequency-division (FD), spatial-division (SD) and code-division (CD) ISAC, respectively \cite{lf22_ISACsurvey,zd25_ISAC,cyh21_ISACsurvey}. In the following, we provide a detailed explanation of O-ISAC in each domain and discuss the potential interference within each system. 
\subsubsection{Time-division ISAC}
% definition
Leveraging the established commercial communication/sensing, TD-ISAC can be easily deployed with the communication and sensing tasks being fulfilled via orthogonal time resources \cite{zqx22_802.11}. 
%\cite{Kumari18_TDISAC802.11,zqx22_802.11}
Such low level of integration in time domain enables TD-ISAC to be a straightforward implementation of ISAC.
% category
To be specific, 
%as shown in Fig. \ref{fig:O-ISAC}(a), 
time division in O-ISAC can be achieved by different levels of time resource granularity—division on symbols \cite{Kumari17_TDISACsymbol}, time slots \cite{hl13_TDISACslot}, and subframes each containing an integer number of time slots \cite{zqx21_TDISACsubframe}.   
% some literature
%For instance, the authors of \cite{Kumari17_TDISACsymbol} introduced an adaptive preamble design with time division on symbols. There exist several symbols in a frame, with a portion of symbols being leveraged for data transmission. Such adaptive design achieves a balance between the communication data rate and sensing estimation accuracy.
%
%Additionally, in \cite{hl13_TDISACslot}, a dual-functional modulation waveform was designed, where each operation duration contains two time slots, i.e., the radar cycle and communication cycle. 
%
%More recently, the authors of \cite{zqx21_TDISACsubframe} proposed a dynamic TD architecture design, where the subframe containing an integer number of time slots is selected as the minimum unit for resource allocation. The communication subframes and sensing subframes are adjusted flexibly so as to fulfill different dual-functional requirements.
% pros/cons: minimized interference,low cost; no simultaneous operation, time synchronisation.
In summary, TD-ISAC can be implemented with low construction cost thanks to its straightforward extension from existing commercial systems \cite{zqx22_802.11}. %\cite{Kumari18_TDISAC802.11,zqx22_802.11}. 
Nevertheless, the interference between the two functionalities is alleviated at the expense of resource utilization efficiency, i.e., no simultaneous communication and sensing is allowed. Furthermore, extra time synchronization is required to be taken into account \cite{hl13_TDISACslot}.
%%%%%%%%%%%%%%%%%%    
%\begin{figure}[tb]
%    \centering
%    \includegraphics[width=0.95\linewidth]{Fig_pdf/Fig2_OISAC.pdf}
%    \caption{O-ISAC in various domains: (a) Time domain; (b) Frequency domain; (c) Spatial/Code domain \cite{cyh21_ISACsurvey}.}
%    \label{fig:O-ISAC}
%\end{figure}
%%%%%%%%%%%%%%%%%% 
\subsubsection{Frequency-division ISAC}
% definition
FD serves as an alternative option to achieve O-ISAC, which is typically established on the prevalent orthogonal frequency division multiplexing (OFDM) systems \cite{cyh21_ISACsurvey,lyh24_FDISAC}.
%\cite{cyh21_ISACsurvey,zj23_FDISAC,lyh24_FDISAC}.
%
This differs from its TD counterpart which demonstrates universality with any sensing and communication waveforms probably being allowed.
% category
In FD-ISAC, the communication and sensing functionalities are scheduled in different frequency bands or subcarriers according to the specific channel status, the required key performance indicators (KPIs), and the transmit power budget at the BS \cite{lf22_ISACsurvey}.
% some literature
For instance, the authors of \cite{scg18_FDISAC} proposed a subcarrier selection strategy for transmit power minimization, with constraints on mutual information for sensing and data rate for communication.
% pros/cons: low cost, simultaneous; reduced SE, intermodulation issue 
Overall, the primary advantages of FD-ISAC stem from its minor modifications to existing OFDM-based systems, as well as its capability to support simultaneous communication and sensing services \cite{cyh21_ISACsurvey}.
However, with the aim of alleviating the interference between communication and sensing, blank frequency bands or subcarriers are required between subchannels, leading to reduced SE. 
In addition, the frequency characteristics (such as signal strength, phase, and frequency response) are likely to be affected owing to the nonlinear distortion in channels. 
This therefore poses a risk of unpromising intermodulation issues in FD-ISAC \cite{xjj22_FDISAC}.
%%%%%%%%%%%%%%%%%%  
\subsubsection{Spatial-division ISAC}
% definition
The progressive development of the MIMO techniques opens the door for another strategy called SD-ISAC \cite{lf22_ISACsurvey}.
% category (literature)
To be specific, in order to alleviate inter-functionality interference, the communication and sensing tasks are realized via orthogonal spatial resources. 
On the one hand, different antenna groups are leveraged to form several spatial beams for either communication or sensing. For example, the authors of \cite{lf18_SDISAC} considered a separation architecture, where antennas are split into two groups: one for downlink communication and one for radar sensing. 
On the other hand, the radar waveform can be projected to fall into the null signal space of communication channels \cite{Sodagari12_SDISAC}.
% pros/cons: simultaneous, no interference; filter design, energy decrease
In summary, such SD strategy displays the capability to implement simultaneous data communication and radar sensing without interference between each other.
Nevertheless, alongside antenna separation, the energy utilized for either communication or sensing will inevitably decrease, prone to result with performance loss in SD-ISAC systems \cite{zd25_ISAC}.
%Moreover, when considering the waveform projection, additional spatial filter design is required for interference management \cite{cyh21_ISACsurvey}.
%%%%%%%%%%%%%%%%%%
\subsubsection{Code-division ISAC}
% definition/category
The main idea of CD-ISAC is to implement the communication and sensing functionality via orthogonal or quasi-orthogonal code sequences, such that there exists no inter-functionality interference. 
% literature
For example, the authors of \cite{cx21_CDISAC} designed a novel CD-ISAC system, where the CD-OFDM waveform together with a SIC-based processing method was considered for simultaneous communication and sensing with CD multiplexing gain.
% pros/cons: enhanced SE; high complexity
Overall, in CD-ISAC, the SE can be enhanced with the same time-frequency band being shared by the two functionalities. 
However, CD-ISAC struggles with a high design complexity since appropriate spreading code sequence is a significant concern which still requires further investigation. 
\par
It is worth noting that the five distinct O-ISAC architectures share similarities with their NO-ISAC counterparts mentioned in Section \ref{section2A}, which are actually simpler thanks to their orthogonal resource allocation.
Overall, there typically exists no inter-functionality interference in all O-ISAC architectures. However, other interference types such as inter-user, inter-target, clutter interference and self-interference remain due to the broadcast nature of wireless communications, co-existence of multiple targets, reflections from unrelated scatterers, and co-located transmit and receive antennas, respectively.
These unresolved interference, along with low SE and EE of O-ISAC, substantially limits its potential for future applications.
% advantages
% Despite its advantages of low construction cost, marginal interference between the two functionalities as well as adaptable individual waveform design, O-ISAC inevitably suffers from poor SE and EE, which substantially hinders its future applications.
%%%%%%%%%%%%%%%%%%%%%%%%%%%%%%%%%%%%
%% Semi-ISAC
\subsection{Semi-ISAC}\label{section2C}
% introduce to NO-ISAC
In fact, the deficiency of NO-ISAC (i.e., high interference level) and O-ISAC (i.e., low SE and EE) can be relieved with a medium integration level between communication and sensing. 
Accordingly, the concept of S-ISAC has been introduced, where the time, frequency, spatial, and hardware resources are partially shared between the two functionalities \cite{zc22_SemiISAC}.
% example: frequency domain
\par
For instance, in the frequency domain, some certain frequency bands have already been scheduled to specific purposes, as exemplified by the 4-8 GHz, i.e., the C band, which mainly focuses on the weather observation, weapon location and long range tracking \cite{Griffiths15_Cband}. 
This makes it impractical to exploit the whole spectrum for integrating the communication and sensing functionalities. 
In other words, a more feasible scheme is to allow a portion of the spectrum being shared by both communication and sensing, while the remaining being scheduled for either bandwidth-demanding wireless communication or sensing \cite{zc22_SemiISAC, zc23_SemiISAC}.
% explanation
%Such S-ISAC in the frequency domain is illustrated in Fig. \ref{fig:S-ISAC}. 
% 
%To be specific, the given bandwidth is split into three non-overlapping portions: 
To be specific, in such frequency domain S-ISAC, the given bandwidth is split into three non-overlapping portions: the sensing-only bandwidth $B_{s}$, the communication-only bandwidth $B_{c}$ and the integration bandwidth $B_{isac}$ \cite{zc23_SemiISAC}. 
Note that the spectrum is shared between communication and sensing solely in the integration bandwidth $B_{isac}$.
Interestingly, when $B_{s}=0$ and $B_{c}=0$, the frequency domain S-ISAC boils down to NO-ISAC. Conversely, when $B_{isac}=0$, S-ISAC reduces to FD-ISAC, i.e., O-ISAC.
%%%%%%%%%%%
%\begin{figure}[tb]
%    \centering
%    \includegraphics[width=0.55\linewidth]{Fig_pdf/Fig32_SISAC.pdf}
%    \caption{Typical Semi-ISAC model in the frequency domain.}
%    \label{fig:S-ISAC}
%\end{figure} 
% \usepackage{multirow}
\begin{table*}[t]
\centering
\caption{A Brief Comparison among Different ISAC Systems}
\label{Table:ISAC types}
\begin{tabular}{|c|c|c|c|c|c|}
\hline
%% the title
\textbf{ISAC System} & \textbf{Resource Allocation}& \textbf{Waveform Design} & \textbf{Ref.} & \textbf{Advantages} & \textbf{Disadvantages} \\ \hline
%% O-ISAC
\multirow{4}{*}{\begin{tabular}[c]{@{}c@{}}Orthogonal\\ ISAC\\ (O-ISAC)\end{tabular}} 
 & \multicolumn{1}{l|}{\scalebox{0.6}{$\bullet$} Time division (TD)} & \multirow{4}{*}{\begin{tabular}[c]{@{}c@{}}Separate communication and\\ sensing waveform design\end{tabular}} & \multicolumn{1}{c|}{\cite{Kumari17_TDISACsymbol,hl13_TDISACslot,zqx21_TDISACsubframe}} 
% advantages 
 & \multicolumn{1}{c|}{\multirow{4}{*}{\begin{tabular}[l]{@{}l@{}}\scalebox{0.6}{$\bullet$} Low construction cost\\ \scalebox{0.6}{$\bullet$} No mutual interference\\ \scalebox{0.6}{$\bullet$} Adaptable individual\\ \hspace{0.7em}waveform design \end{tabular}}} 
% disadvantages
 & \multicolumn{1}{c|}{\multirow{4}{*}{\scalebox{0.6}{$\bullet$} Poor SE and EE}} \\ 
 & \multicolumn{1}{l|}{\scalebox{0.6}{$\bullet$} Frequency division (FD)}  & & \cite{scg18_FDISAC} & & \\
 & \multicolumn{1}{l|}{\scalebox{0.6}{$\bullet$} Spatial division (SD)}    & & \cite{lf18_SDISAC,Sodagari12_SDISAC}& & \\
 & \multicolumn{1}{l|}{\scalebox{0.6}{$\bullet$} Code division (CD)}       & & \cite{cx21_CDISAC} & & \\ \hline
%% NO-ISAC
\multicolumn{1}{|c|}{\multirow{6}{*}{\begin{tabular}[c]{@{}c@{}}Non-orthogonal\\ ISAC\\ (NO-ISAC)\end{tabular}}} 
 & \multicolumn{1}{c|}{\multirow{6}{*}{\begin{tabular}[c]{@{}c@{}}Non-orthogonal\\ resource allocation\end{tabular}}} 
 & \multicolumn{1}{c|}{\begin{tabular}[c]{@{}c@{}}\scalebox{0.6}{$\bullet$} Joint design: radar sequence\\ disabled \end{tabular}}  & \cite{ylf22_RSMApart2}
% advantages
 & \multicolumn{1}{c|}{\multirow{5}{*}{\begin{tabular}[l]{@{}l@{}} \scalebox{0.6}{$\bullet$} Enhanced SE and EE\\\scalebox{0.6}{$\bullet$} Relieved dual-functional\\ \hspace{0.7em}resource rivalry \\\scalebox{0.6}{$\bullet$} Flexible hardware design\\ \hspace{0.7em}and signal processing \end{tabular}}} 
% disadvantages
 & \multicolumn{1}{c|}{\multirow{5}{*}{\begin{tabular}[l]{@{}l@{}}\scalebox{0.6}{$\bullet$} Unavoidable\\ \hspace{0.7em}inter-functionality\\ \hspace{0.7em}interference \\ \scalebox{0.6}{$\bullet$} High construction\\ \hspace{0.7em}complexity \end{tabular}}} \\
\multicolumn{1}{|c|}{} & \multicolumn{1}{c|}{} & \multicolumn{1}{c|}{\begin{tabular}[c]{@{}c@{}}\scalebox{0.6}{$\bullet$} Joint design: radar sequence\\ enabled with SIC  \end{tabular}} & \cite{hhc23_ISAC_radarsequence} & \multicolumn{1}{c|}{} & \\
\multicolumn{1}{|c|}{} & \multicolumn{1}{c|}{} & \begin{tabular}[c]{@{}c@{}}\scalebox{0.6}{$\bullet$} Joint design: radar sequence\\ enabled without SIC \end{tabular} & \cite{lx20_ISAC_radarsequence} & \multicolumn{1}{c|}{} & \\ \hline
%% S-ISAC
\begin{tabular}[c]{@{}c@{}}Semi-ISAC\\ (S-ISAC)\end{tabular}
& \multicolumn{1}{c|}{\begin{tabular}[c]{@{}c@{}}Partially orthogonal\\ or non-orthogonal\\ resource allocation \end{tabular}}  & \multicolumn{1}{c|}{\begin{tabular}[c]{@{}c@{}}Separate design in orthogonal\\ resource part; joint design\\ in non-orthogonal part \end{tabular}}         & \multicolumn{1}{c|}{\cite{zc23_SemiISAC,lf20_jointCandS}}
% advantages
& \multicolumn{1}{c|}{\begin{tabular}[l]{@{}l@{}}\scalebox{0.6}{$\bullet$} Satisfactory real-time\\ \hspace{0.7em}implementation\\\scalebox{0.6}{$\bullet$} Enhanced SE and EE\end{tabular} }   
% disadvantages
& \multicolumn{1}{c|}{\begin{tabular}[l]{@{}l@{}}\scalebox{0.6}{$\bullet$} Unavoidable\\ \hspace{0.7em}inter-functionality\\ \hspace{0.7em}interference \end{tabular}}\\ \hline
\end{tabular}
\end{table*}
%%%%%%%%%%%%%%%%%%%%%%%%%%%%%%%%%%%%
% \usepackage{multirow}
\begin{table*}[t]
\centering
\caption{Interference Types in Different ISAC Systems}
\label{Table:interference types}
\begin{tabular}{|cc|cc|c|c|c|c|}
\hline
\multicolumn{2}{|c|}{\multirow{4}{*}{\textbf{ISAC System}}} 
& \multicolumn{2}{c|}{\begin{tabular}[c]{@{}c@{}}\textbf{Inter-functionality}\\ \textbf{Interference}\end{tabular}} 
& \multirow{4}{*}{\begin{tabular}[c]{@{}c@{}}\textbf{Inter-user}\\ \textbf{Interference}\end{tabular}} 
& \multirow{4}{*}{\begin{tabular}[c]{@{}c@{}}\textbf{Inter-target}\\ \textbf{Interference}\end{tabular}} 
& \multirow{4}{*}{\begin{tabular}[c]{@{}c@{}}\textbf{Clutter}\\ \textbf{Interference}\end{tabular}} 
& \multirow{4}{*}{\begin{tabular}[c]{@{}c@{}}\textbf{Self-Interference}\end{tabular}} \\ \cline{3-4}
\multicolumn{2}{|c|}{} & \multicolumn{1}{c|}{\begin{tabular}[c]{@{}c@{}}\textbf{Communication}\\ \textbf{to Sensing}\\ \textbf{Interference}\end{tabular}} & {\begin{tabular}[c]{@{}c@{}}\textbf{Sensing to}\\ \textbf{Communication}\\ \textbf{Interference}\end{tabular}} &  &  &  & \\ \hline
%% O-ISAC
\multicolumn{2}{|c|}{\begin{tabular}[c]{@{}c@{}}O-ISAC\end{tabular}} & \multicolumn{1}{c|}{\scalebox{0.8}{$\bigcirc$}} & \scalebox{0.8}{$\bigcirc$} & $\odot$ & $\odot$ & $\odot$ & $\odot$\rule{0pt}{2ex} \\\hline
%% NO-ISAC
\multicolumn{1}{|c|}{\multirow{5}{*}{\begin{tabular}[c]{@{}c@{}}NO-ISAC\\ or S-ISAC\end{tabular}}} & \begin{tabular}[c]{@{}c@{}}Radar sequence\\ disabled\end{tabular} & \multicolumn{1}{c|}{\multirow{5}{*}{\scalebox{0.8}{$\bigcirc$}}} & \scalebox{0.8}{$\bigcirc$} & \multirow{5}{*}{\scalebox{0.9}{\ding{108}}} & \multirow{5}{*}{\scalebox{0.9}{\ding{108}}} & \multirow{5}{*}{\scalebox{0.9}{\ding{108}}} & \multirow{5}{*}{\scalebox{0.9}{\ding{108}}}\\ \cline{2-2} \cline{4-4} 
\multicolumn{1}{|c|}{} & \begin{tabular}[c]{@{}c@{}}Radar sequence\\ enabled with SIC \end{tabular} & \multicolumn{1}{c|}{} & \scalebox{0.8}{$\bigcirc$} & & & &\\ \cline{2-2} \cline{4-4} 
\multicolumn{1}{|c|}{} & \begin{tabular}[c]{@{}c@{}}Radar sequence\\ enabled without SIC \end{tabular} & \multicolumn{1}{c|}{} &$\odot$& & & &\\ \hline
\multicolumn{8}{l}{\begin{tabular}[l]{@{}l@{}}\textbf{\textit{Note:}} \scalebox{0.8}{$\bigcirc$} indicates no interference, while $\odot$ denotes the presence of this interference.\\ \scalebox{0.9}{\ding{108}} in NO-ISAC or S-ISAC signifies significantly stronger interference compared to the same type observed in O-ISAC.
\end{tabular}} 
%% S-ISAC
%\multicolumn{2}{|c|}{\begin{tabular}[c]{@{}c@{}}S-ISAC\end{tabular}} & \multicolumn{1}{c|}{$\surd$} & $\surd$ & $\surd$ & $\surd$ & $\surd$ & \begin{tabular}[c]{@{}c@{}}$\surd$\end{tabular}\\\hline
\end{tabular}
\end{table*}
% another example
\par
Apart from the frequency domain, S-ISAC has also been investigated in other domains such as the time domain. 
Recently, the authors of \cite{lf20_jointCandS} proposed a novel TD duplex (TDD) structure involving three stages in order to better unify the communication and sensing functionalities. 
In the stage of communication channel estimation and target searching, there exists a guard period between downlink and uplink pilots for the BS to merely receive radar echo signals. This thereby avoids the collision between communication and sensing tasks utilizing orthogonal time resources. 
While in other stages, simultaneous communication and sensing is enabled with joint transmit beamforming design and receiver design. 
\par
It is worth noting that in the five distinct S-ISAC architectures, the interference in dedicated radio resources is similar to that in O-ISAC architectures, while the interference in fully shared radio resources resembles that observed in NO-ISAC architectures.
For the sake of brevity, no further discussion will be included here.
% advantages
In summary, building upon the aforementioned O-ISAC and NO-ISAC, S-ISAC offers the advantages of practical feasibility as well as enhanced SE and EE. 
% disadvantages
However, the inter-functionality interference triggered by non-orthogonal resource sharing remains a challenge, which calls for efficient interference management approaches.
%%%%%%%%%%%%%%%%%%
%%%%%%%%%%%%%%%%%%%%%%%%%%%%%%%%%%%%
\subsection{Comparison among different ISAC systems}\label{section2D}
%% brief summary: table (type)
In this subsection, we compare the aforementioned three ISAC systems with different integration levels.
Table \ref{Table:ISAC types} highlights the key resource allocation and waveform design approaches for each ISAC system, along with their respective advantages and disadvantages, while Table \ref{Table:interference types} presents the primary interference types encountered in each ISAC system with the well-studied architecture downlink mono-static sensing and downlink communication \cite{lf21_CRBoptimization, mxd23_NOMAISACmagazine,hhc23_ISAC_radarsequence,lx20_ISAC_radarsequence, ylf22_RSMApart2, lhc24_downc_downlinkmono} as an example.
\par
%% interference
It is obvious that, by leveraging different resource allocation strategies and waveform design schemes, ISAC systems with varying integration levels experience different interference types and strengths. With a higher integration level, interference becomes an increasingly inevitable issue encountered in ISAC. 
% O-ISAC
Specifically, O-ISAC experiences the lowest interference level due to its orthogonal radio resource allocation between communication and sensing. 
Such dedicated resource utilization eliminates inter-functionality interference, albeit at the cost of low SE and EE. 
However, other interference types involving inter-user, inter-target, clutter interference and self-interference still necessitate effective interference management strategies.
% NO-ISAC
In contrast to O-ISAC, NO-ISAC systems encounter the highest interference level, stemming from the shared use of time, frequency, spatial, and hardware resources between communication and sensing.
This inevitably results with more complicated interference compared to that observed in O-ISAC.
Moreover, inter-functionality interference arises, which varies across different architectures and is influenced by different waveform designs. 
Specifically, this interference intensifies in uplink communication architectures, and downlink communication architectures with radar sequence enabled without SIC.
%% S-ISAC
Alternatively, by partially sharing radio and hardware resources between the two functionalities, S-ISAC bridges the aforementioned two systems and exhibits a medium level of interference.
%In fact, the deficiency of O-ISAC and NO-ISAC can be alleviated by S-ISAC, with radio and hardware resources being partially shared between the two functionalities. S-ISAC thereby bridges the aforementioned two ISAC systems and exhibits a medium level of interference.
%This is attributed to the fact that the orthogonal part of S-ISAC operates similarly to O-ISAC, resulting in low interference, whereas the non-orthogonal part resembles NO-ISAC, leading to high interference.
%
\par
Overall, interference management is a critical issue for all ISAC systems in addressing various types of interference, calling for efficient solutions.
%%%%%%%%%%%%%%%%%%%%%%%%%%%%%%%%%%%%
\section{Different MA-assisted O-ISAC Systems} \label{section3}
This section provides an overview of basic OMA-assisted O-ISAC systems, with a particular focus on TDMA/OFDMA-assisted O-ISAC, and evaluates their advantages and disadvantages.
\par
As mentioned, while O-ISAC systems eliminate inter-functionality interference, other types of interference such as inter-user interference, remain to be addressed. Fortunately, to tackle inter-user interference, the incorporation of MA into O-ISAC is indeed a straightforward extension from MA-assisted communication-only networks. 
This implies that all existing MA techniques such as OMA, SDMA, NOMA, and RSMA can be directly applied in O-ISAC systems. For clarity, this section only focuses on OMA-assisted O-ISAC, with other MA techniques discussed in the following sections.
%% interference management approach
\par
In general, with orthogonal radio resources being allocated to users, OMA adheres to interference management strategy of avoiding inter-user interference. 
% the principles of OMA
There are four OMA-assisted O-ISAC systems built upon well-established OMA schemes, which schedule communication users in orthogonal domains, i.e., the frequency domain—frequency-division MA (FDMA), the time domain—TDMA, the code domain—code-division MA (CDMA), the time and frequency domain—OFDMA \cite{bruno24_MA,m22_RSMAtutorial}.
% FDMA
Specifically, FDMA partitions the available spectrum into orthogonal frequency bands, each allocated to one communication user.
% TDMA
TDMA divides time resources into orthogonal time units, each designated for individual communication users.
% CDMA
CDMA utilizes orthogonal code sequences so as to accommodate communication users within the same time-frequency resource.
% OFDMA
OFDMA splits the time and frequency resources into non-overlapping time units and narrow subcarriers, where multiple communication users can be served in grouped resource blocks without inter-user interference.
All these OMA schemes can be readily applied to various O-ISAC systems, leading to four general types of OMA-assisted O-ISAC systems, i.e., FDMA, TDMA, CDMA, and OFDMA-assisted O-ISAC, which are detailed as follows:
%%%%%%%%%%%%%%%%%%%%%%
\begin{figure}[tb]
    \centering
    \includegraphics[width=1\linewidth]{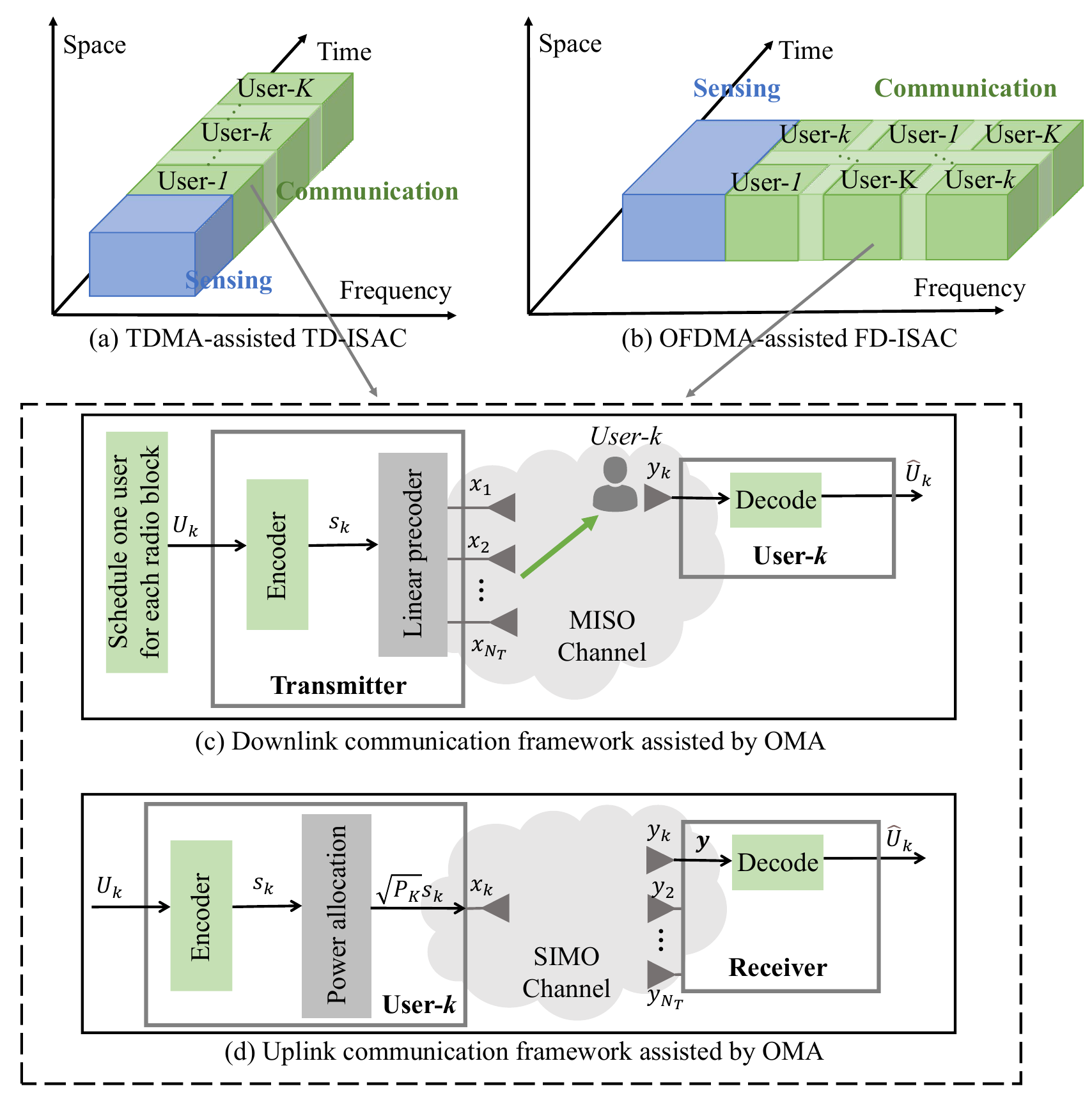}
    \caption{Typical O-ISAC systems assisted by OMA in the downlink or uplink.}
    \label{fig:MA OISAC}
\end{figure}
%%%%%%%%%%%%%%%%%%%%%%
%% model illustration
%\par
%Fig. \ref{fig:MA OISAC} illustrates TDMA-assisted TD-ISAC and OFDMA-assisted FD-ISAC in the downlink and uplink as examples. 
%The BS communicates with $K$ single-antenna downlink or uplink users and detects $Q$ moving sensing targets, which are indexed by $\mathcal{K}=\{1,\cdots, K\}$ and $\mathcal{Q}=\{1,\cdots,Q\}$, respectively.
%% FDMA-assisted O-ISAC
%\subsection{FDMA-assisted FD-ISAC}
%By leveraging different frequency bands or subcarriers, FD-ISAC systems simultaneously perform communication and sensing functionalities.
%The total transmit power budget is split between the two functionalities, since both multi-user communication and multi-target sensing operate over the entire time period.
%Fig. \ref{fig:MA OISAC}(a) illustrates the use of FDMA in the frequency bands designated for communication, with detailed downlink communication framework shown in Fig. \ref{fig:MA OISAC}(c) and uplink framework in Fig. \ref{fig:MA OISAC}(d).
%The available spectrum is split into orthogonal sub-bands, each assigned to a single communication user \cite{lx25_FDMA}.
%Scheduling communication users with such separate frequency sub-bands minimizes inter-user interference, albeit at the cost of constrained SE.
\subsection{TDMA-assisted O-ISAC}
As a typical case of O-ISAC, we take TD-ISAC systems assisted by TDMA as an example,
%\cite{Demir23_FDMA_O-ISAC,zqx21_TDISACsubframe},
where communication and sensing functionalities share the same frequency bands, eliminating the necessity to modify the current waveform design \cite{zqx21_TDISACsubframe}.
A portion of time is allocated to communication, while the remaining portion is dedicated to radar sensing. In other words, at different time units, the total transmit power budget is for either communication or sensing functionality.
Fig. \ref{fig:MA OISAC}(a) illustrates the use of TDMA for multi-user communication, with detailed downlink communication framework shown in Fig. \ref{fig:MA OISAC}(c) and uplink framework in Fig. \ref{fig:MA OISAC}(d).
Specifically, the time resources are partitioned into orthogonal time units, each designated for individual communication users.
This indicates that when specific time units are occupied by user-$k$, other users are required to wait for access. Inter-user interference is thereby tackled at the cost of low SE.
%% OFDMA-assisted FD-ISAC
\subsection{OFDMA-assisted O-ISAC}
In this subsection, OFDMA-assisted FD-ISAC is considered as a representative example to highlight the key features of OFDMA-assisted O-ISAC.
Specifically, by leveraging different frequency bands or subcarriers, FD-ISAC systems simultaneously perform communication and sensing functionalities.
The total transmit power budget is split between the two functionalities, since both multi-user communication and multi-target sensing operate over the entire time period.
Fig. \ref{fig:MA OISAC}(b) illustrates the use of OFDMA within the radio resources designated for communication, with detailed downlink and uplink communication framework in Fig. \ref{fig:MA OISAC}(c) and (d).
The available time and frequency resources are split into orthogonal subcarriers and time slots, where each grouped resource block is assigned to a single communication user, leading to minimized inter-user interference \cite{lyh24_FDISAC,scg18_FDISAC}.
%\cite{Demir23_FDMA_O-ISAC,lyh24_FDISAC,scg18_FDISAC}. 
\subsection{Other OMA-assisted O-ISAC}
It is worth noting that the other two OMA schemes, i.e., FDMA and CDMA, can also be incorporated into O-ISAC systems for managing inter-user interference. Specifically, in FDMA-assisted O-ISAC, the available spectrum for the communication functionality is split into orthogonal sub-bands, each assigned to a single communication user. In CDMA-assisted O-ISAC, multiple communication users are scheduled within the same time-frequency band available for the communication functionality via orthogonal code sequences, such that there exists no inter-user interference.
However, these schemes have garnered limited attention in existing O-ISAC studies, and are therefore not illustrated in detail in this work. This can be attributed to the fact that FDMA-assisted O-ISAC systems typically suffer from lower SE and reduced resource allocation flexibility compared to OFDMA-assisted ones, while CDMA-enabled O-ISAC systems still face challenges in maintaining orthogonality among code sequences. 
%% pros and cons
\par
The major advantages and disadvantages of incorporating OMA into O-ISAC can be summarized as follows:
\subsubsection{Advantages}
\begin{itemize}
    % advantages 
    \item[-] \textit{Low transceiver complexity:} OMA-assisted O-ISAC enables per-user signal processing without the need for computationally intensive multi-user detection (MUD) or SIC, and eliminates the reliance on advanced antenna arrays or adaptive beamforming for interference management. As a result, it features a simple transceiver architecture and can be implemented with minimal modifications to existing systems.
    \item[-] \textit{Interference-free and reliable communication:} OMA-assisted O-ISAC is capable to achieve transmissions free of inter-user interference, which thereby contributes to enhanced communication reliability in ISAC systems.
\end{itemize}
\subsubsection{Disadvantages}
\begin{itemize}
    \item[-] \textit{Low capacity:} With orthogonal radio resources being exclusively designated for individual communication users, O-ISAC systems assisted by OMA inevitably suffer from low capacity, where the amount of supported users is constrained by the limited availability of orthogonal resources. This deficiency further hinders the application of OMA to future ISAC networks with surging user demand. 
    \item[-] \textit{Low spectral efficiency:} In addition to the low SE inherent in O-ISAC systems, OMA is prone to result with further inefficient utilization of the spectrum, since low-rate communication users occupy the whole resource blocks far beyond their requirements.
    \item[-] \textit{Extra signaling overhead:} Another issue inherent in OMA-assisted O-ISAC is that achieving promising system performance typically requires well-designed user scheduling, which thereby yields a high signaling overhead.
\end{itemize}
%%%%%%%%%%%%%%%%%%%%%%%%%%%%%%%%%%%%
\section{Downlink MA-assisted NO-ISAC Systems}\label{section4}
In this section, we provide a detailed illustration of NO-ISAC systems\footnote{For clarity, this section takes the well-established architecture—downlink communication and downlink mono-static sensing \cite{lf21_CRBoptimization, mxd23_NOMAISACmagazine,hhc23_ISAC_radarsequence,lx20_ISAC_radarsequence, ylf22_RSMApart2, lhc24_downc_downlinkmono} as an example. It is worth noting that other downlink MA-assisted NO-ISAC architectures can be similarly achieved, which will not be redundantly discussed.}aided by SDMA, NOMA, and RSMA in the downlink, highlighting their respective advantages and disadvantages. 
\par
As mentioned, with non-orthogonal radio resources shared between communication and sensing, NO-ISAC systems inevitably experience inter-functionality interference, along with other interference types such as inter-user interference.
%% system model
To facilitate a better understanding of how different MA techniques address these interference in NO-ISAC, we begin with a general NO-ISAC system model.
%% in general
Specifically, the BS equipped with $N_{T}$ transmit antennas and $N_{R}$ receive antennas simultaneously communicates with $K$ single-antenna downlink users and detects $Q$ sensing targets, which are indexed by $\mathcal{K}=\{1,\cdots,K\}$ and $\mathcal{Q}=\{1,\cdots,Q\}$, respectively.
We consider $I$ transmissions within each coherent processing interval (CPI).
At time index $i$, the dual-functional transmit signal is expressed as $\mathbf{x}[i] \in \mathbb{C}^{N_{T}\times 1}$, where the transmit power budget is constrained by $\mathbb{E}\{\left\|\mathbf{x}[i]\right\|_2^{2}\}\le P_{T}$.
%% the communication user
\par
The received signal at user-$k$ of the $i$th time index is given as
\begin{equation} \label{Yc}
    y_{k}[i]=\mathbf{h}_{k}^{H}\mathbf{x}[i]+z_{k}[i],
\end{equation}
where the channel between the BS and user-$k$ is denoted by $\mathbf{h}_{k}\in \mathbb{C}^{N_{T}\times 1}$.
$z_{k}[i]$ is the additive white Gaussian noise (AWGN) received at user-$k$, following $\mathcal{CN}(0,\sigma _{c}^{2})$. 
%% the sensing target
\par
Leveraging the dual-functional transmit signal, the sensing echo signal received at the BS of time index $i$ is given as
\begin{equation} \label{Ys}
		\mathbf{y}_{r}[i]=\sum_{q\in \mathcal{Q}}\alpha _{q}e^{j2\pi \mathcal{F}_{D_{q}}iT}\mathbf{b}(\theta _{q})\mathbf{a}^{T}(\theta _{q})\mathbf{x}[i]+\mathbf{z}_{r}[i],
\end{equation}
where $\alpha _{q}, \forall q\in\mathcal{Q}$ represents the complex reflection coefficient of target-$q$. $\mathcal{F}_{D_{q}}=\frac{2v_{q}f_{c}}{c}, \forall q\in\mathcal{Q}$ denotes the Doppler frequency of each target, where $v_{q}$ is the velocity of target-$q$, and $f_{c}$, $c$ is the carrier frequency and speed of light, respectively. $T$ represents the symbol period. 
$\theta_{q}, \forall q\in\mathcal{Q}$ denotes both the direction of departure (DoD) and direction of arrival (DoA) of target-$q$, which coincide in mono-static sensing mode. 
With adjacent array elements spaced half wavelength apart, the transmit steering vector is $\mathbf{a}(\theta_{q})=[1,e^{j\pi\sin (\theta_{q})},\dots,e^{j\pi(N_{T}-1)\sin (\theta_{q})} ]^{T}\in \mathbb{C}^{N_{T}\times 1}$, $\forall q \in \mathcal{Q}$. And the receive steering vector $\mathbf{b}(\theta_{q})\in \mathbb{C}^{N_{R}\times 1}$ is calculated in the same way.
$\mathbf{z}_{r}[i] \in \mathbb{C}^{N_{R}\times 1}$ follows $\mathcal{CN}(\mathbf{0}_{N_{R}\times 1},\sigma _{r}^{2}\mathbf{I}_{N_{R}})$. 
%%
% this section
\par
The following subsections further detail NO-ISAC systems assisted by SDMA, NOMA, and RSMA, respectively, assuming perfect CSI is available at both the BS and the communication users.
\subsection{SDMA-assisted NO-ISAC}
%% the development of SDMA
Confronting with the unprecedented demands for enhanced wireless capacity and efficiency, MIMO ISAC, where multiple antennas are deployed at most access points, has become increasingly popular \cite{lf22_ISACsurvey, hhc23_ISAC_radarsequence, lx20_ISAC_radarsequence, zqm23_downlinkbistatic, lf18_SDISAC}. 
Capturing the additional spatial dimension introduced by MIMO enables the incorporation of a new MA scheme called SDMA into ISAC.
%% interference management approach
In general, SDMA displays the capability to accommodate streams for multiple communication users and sensing targets with the same time-frequency resource, adhering to the interference management strategy with interference being precanceled at the transmitter and treated as noise at the receiver.
%% the principles of SDMA
% general: non-linear/ linear. focus: MU-LP
\par
In SDMA-assisted NO-ISAC systems, precoding techniques are utilized at the transmitter for interference management, which can be typically divided into the non-linear and linear precoding \cite{bruno24_MA, m22_RSMAtutorial}.
Appealing from the perspective of practical application, the linear technique, e.g., MU-LP assisted SDMA becomes crucial in ISAC systems \cite{hhc23_ISAC_radarsequence, lx20_ISAC_radarsequence, lf18_SDISAC}, where linear precoding is leveraged at the transmitter, and communication users are allowed to directly decode their intended streams by regarding both inter-user and inter-functionality interference as noise. 
As mentioned, this is abbreviated as ``SDMA-assisted ISAC'' for simplicity.
Despite its suboptimal performance, SDMA-assisted ISAC showcases substantial efficacy particularly when perfect CSI at the transmitter (CSIT) is available and user channels are semi-orthogonal with comparable strengths or long-term signal-to-noise ratios (SNR). 
%% model illustration
\par
\begin{figure}[tb]
    \centering
    \includegraphics[width=1\linewidth]{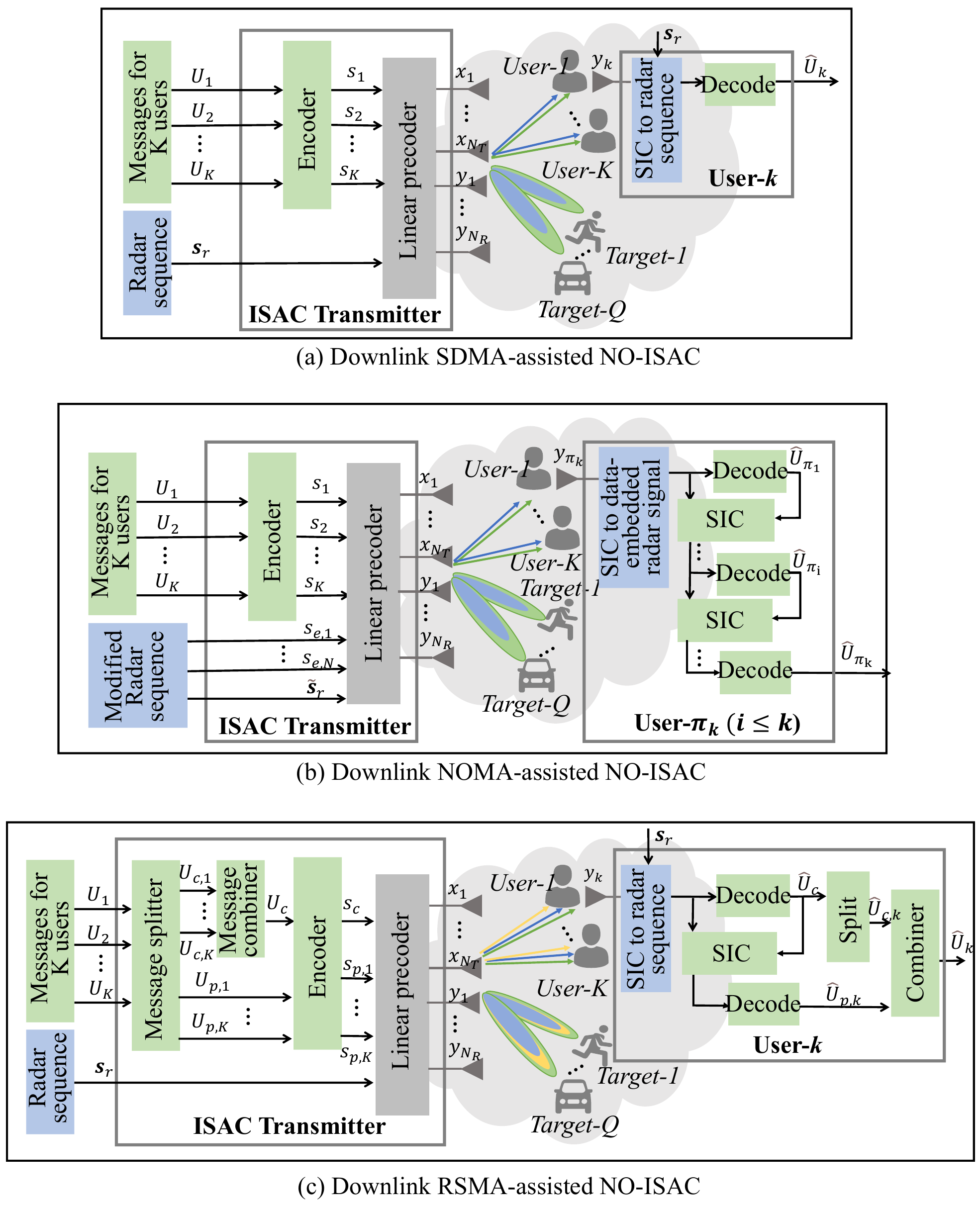}
    \caption{Non-orthogonal integrated downlink communication and downlink mono-static sensing systems assisted by different MA techniques—SDMA, NOMA, and RSMA in the downlink.}
    \label{fig:downlink MA NOISAC}
\end{figure}

 
As illustrated in Fig. \ref{fig:downlink MA NOISAC}(a), we consider a general SDMA-assisted NO-ISAC with multiple downlink users and sensing targets.
% transmitter side
At the transmitter, extra radar sequences $\mathbf{s}_r[i] \in \mathbb{C}^{N_T\times 1}$ are incorporated alongside the information streams \cite{lx20_ISAC_radarsequence,hhc23_ISAC_radarsequence}. 
Specifically, the transmit radar sequences $\mathbf{s}_{r}[i]$ containing no useful information for communication are pre-designed before transmission.
And the information streams $s_{k}[i], \forall k \in \mathcal{K}$ are obtained by encoding communication user messages $U_{k}[i], \forall k \in \mathcal{K}$. 
The transmit vector at time index $i$ is then expressed as $\mathbf{s}[i]=[\mathbf{s}_{r}[i],s_{1}[i],\dots,s_{K}[i]]^{T}\in \mathbb{C}^{(N_T+K)\times 1}$.
Subsequently, the streams are linearly precoded via the precoder $ \mathbf{W}=[\mathbf{W}_{r}, \mathbf{w}_{1},\dots, \mathbf{w}_{K}]\in \mathbb{C}^{N_{T}\times (N_T+K)}$, which remains unchanged throughout one CPI. We denote the power of extra radar sequences as $P_{r}=\left\|\mathbf{W}_{r}\right\|_F^{2}$.
The dual-functional signal transmitted at time index $i$ is modeled as
\begin{equation}
	\mathbf{x}[i]=\mathbf{W}\mathbf{s}[i]=\mathbf{W}_{r}\mathbf{s}_{r}[i]+\sum_{k\in \mathcal{K}}\mathbf{w}_{k}s_{k}[i],
\end{equation}
where the streams are assumed to be independent from each other, i.e., $\mathbb{E}\{\mathbf{s}[i]\mathbf{s}[i]^{H}\}=\mathbf{I}_{N_T+K}$. The covariance matrix of the dual-functional signal is then calculated as $\mathbf{R}_{X}\approx\frac{1}{I}\sum_{i\in \mathcal{I}}\mathbf{x}[i]\mathbf{x}[i]^{H}=\mathbf{W}\mathbf{W}^{H}$ with $I$ sufficiently long. The total transmit power budget constraint is thereby $\mathrm{tr}(\mathbf{W}\mathbf{W}^{H})\le P_{T}$. 
% receiver side: 3 schemes
\par
At the receiver, each user is allowed to decode its intended message while treating inter-user interference as noise. 
The inter-functionality interference, specifically the sensing-to-communication interference, can either be treated as noise or eliminated at each user through SIC.
An auxiliary variable $\eta_{r}$ is then incorporated with the aim of capturing both SIC-enabled and disabled cases. 
The general signal-to-interference-plus-noise ratio (SINR) for decoding $s_{k}$ at communication user-$k$ of each CPI is given as
\begin{equation}\label{R_sdma}
	\gamma _{k}=\frac{|\mathbf{h}_{k}^{H}\mathbf{w}_{k}|^{2}}{\sum_{j\in \mathcal{K},j\neq k}|\mathbf{h}_{k}^{H}\mathbf{w}_{j}|^{2}+\eta_{r}\|\mathbf{h}_{k}^{H}\mathbf{W}_{r}\|_2^{2}+\sigma_{c}^{2}}, \forall k\in\mathcal{K}.
\end{equation}
The achievable rate of user-$k$ is then expressed as $R_{k}=\log_{2}(1+\gamma _{k})$.
By varing $\eta_r$ and the power $P_r$ allocated to the radar transmit precoder $\mathbf{W}_{r}$, we obtain the following three different cases:
\begin{itemize}
    % radar sequence disabled
    \item[\textit{i)}] When $P_r=0$, this refers to the radar-sequence-disabled waveform design mentioned in Subsection \ref{section2A}. In this case, the extra radar sequence is turned off, leaving the information streams exploited for both communication and sensing functionalities. 
    % radar sequence enabled with SIC
    \item[\textit{ii)}] When $P_r>0$ and $\eta_r=0$, this refers to the rader-sequence-enabled waveform design with SIC.  
    In other words, confronting with the sensing-to-communication interference imposed by extra radar sequences, communication users in this case showcase the capability to cancel it via SIC before the intended messages are decoded. This attributes to the fact that the radar sequences containing no useful information can be pre-designed and shared between the transmitter and users. 
    % radar sequence enabled without SIC
    \item[\textit{iii)}] When $P_r>0$ and $\eta_r=1$, this corresponds to the radar-sequence-enabled waveform design without SIC. The radar sequences $\mathbf{s}_r$ are then considered as interference in addition to inter-user interference when communication users decode their intended messages.   
\end{itemize}
%%%%%%%%%%%%%%%%%%%%%%
%%%%%%%%%%%%%%%%%%
%% how to handle interference in ISAC
\par
As mentioned, in the former two cases—whether the radar sequences are disabled or enabled with SIC, there eventually occurs no inter-functionality interference. 
In the third case with radar sequences but without SIC, there exists sensing-to-communication interference caused by radar sequence $\mathbf{s}_{r}$. 
This therefore requires communication users to decode its intended streams by treating both inter-user and inter-functionality interference as noise. 
The second case with radar sequences and SIC naturally surpasses the other two cases, thanks to the extra DoF introduced by the dedicated sensing beam as well as the effective interference cancellation enabled by SIC.
%% pros and cons
\par
The major advantages and disadvantages of incorporating SDMA into NO-ISAC are summarized as follows:
\subsubsection{Advantages}
\begin{itemize}
% advantages
    % enhanced SE
    \item[-] \textit{Enhanced spectral efficiency:} The incorporation of the spatial domain from MIMO networks enables SDMA to serve communication and sensing with the same time-frequency resource. In particular, with perfect CSIT and an underloaded network (i.e., the number of transmit antennas is larger than the total number of receive antennas at users), SDMA is capable of eliminating inter-user interference, achieving maximum DoFs. This thus contributes to promising communication and sensing performance, as well as enhanced SE compared with OMA-assisted O-ISAC/NO-ISAC. 
    % low complexity
    \item[-] \textit{Low transceiver complexity:} Although SDMA-assisted NO-ISAC systems typically exhibit higher computational and hardware complexity compared to OMA-assisted ones, the resulting complexity is still acceptable, thereby facilitating their practical implementation.
    This attributes to the linear precoding leveraged at the transmitter and the direct decoding of the intended message with interference being treated as noise at the receiver \cite{bruno16_RSMA}.
    It is worth noting that when SIC is incorporated for eliminating the sensing-to-communication interference, this system inevitably entails increased hardware complexity at the receiver. 
\end{itemize}
\subsubsection{Disadvantages}
\begin{itemize}
%% disadvantages 
    % network load: only suitable for underloaded
    \item[-] \textit{Restricted flexibility in network load:} In contrast to the promising performance with underloaded network, SDMA becomes unsuitable when the network becomes overloaded, since sufficient transmit antennas are critical for SDMA to fulfill successful interference management \cite{ylf22_RSMApart2,ylf22_SAGIN}. Such inefficient alleviation of inter-user interference not only results with poor communication performance but also degrades the sensing performance.
    % user deployment: semi-orthogonal
    \item[-] \textit{Restricted flexibility in user deployment:} The performance of SDMA-assisted NO-ISAC drops significantly when user channels are aligned or with various channel strengths. 
    Such limitation in user deployment quests for precise scheduling in SDMA-assisted NO-ISAC, which results with extra signaling overhead and computational complexity.
    % CSIT
    \item[-] \textit{High sensitivity towards CSIT imperfections:} SDMA-assisted NO-ISAC suffers from severe performance degradation when CSIT becomes imperfect \cite{Rafael21_RSMAISAC}.
    Such sensitivity to CSIT inaccuracy comes from the fact that SDMA is designed based on perfect CSIT. 
    With further application to imperfect CSIT, there inevitably exists residual inter-user interference alongside inter-functionality interference, which is caused by the imprecise linear precoding at the transmitter.
\end{itemize}
%%%%%%%%%%%%%%%%%%
\subsection{NOMA-assisted NO-ISAC}
%% the development of NOMA
To enhance SE, power-domain NOMA \cite{islam17_NOMAtutorial, lyw22_NOMA} has been incorporated into ISAC systems, enabling non-orthogonal resource allocation. %\cite{Amhaz24_NOMAISAC}.
%% interference management approach
At the transmitter, users\footnote{Note that with extra radar sequences being allowed to convey information bits, sensing targets can be treated as virtual users \cite{mxd23_NOMAISACmagazine, wzl22_NOMAISAC2,Ahmed24_NOMAISACmagazine,xn25_NOMAISACvirtual}.}are allocated with varying power levels, and the signals are superposed within the same time-frequency resources via the power domain. 
At the receiver, NOMA handles inter-user and inter-functionality interference by mandating specific users to decode the messages of virtual users and other real communication users.
Leveraging such superposition coding (SC) at the transmitter and SIC at the receiver, NOMA (a.k.a. SC-SIC) adheres to the interference management scheme of fully decoding interference.
%% the principles of SDMA
\par
In fact, with the aim of capturing the spatial domain introduced by multi-antenna networks, communication users and sensing targets can also be partitioned into distinct groups \cite{Ahmed24_NOMAISACmagazine}. 
This thereby efficiently synergizes SDMA and NOMA into ISAC systems, where interference between different groups is handled by SDMA, while the interference within the same group is managed via SC-SIC. 
% single-group
It is worth noting that when only one group is considered, such approach boils down to the aforementioned SC-SIC.
%% model illustration: downlink. 
% transmitter side; 
\par
Fig. \ref{fig:downlink MA NOISAC}(b) illustrates a multi-user multi-target NO-ISAC system assisted by NOMA.
The primary difference between this transmitter and that in SDMA-assisted NO-ISAC lies in the design of embedding information bits within sensing waveforms.
For instance, by embedding $N, 0 \leq N \leq \mathrm{rank}(\mathbf{W}_r)$ information symbols, the modified sensing waveform can be expressed as $\mathbf{x}_r[i]=\sum_{j=1}^{N}\mathbf{w}_{e,j}s_{e,j}[i]+\tilde{\mathbf{W}}_r\tilde{\mathbf{s}}_r[i]$, where $\{s_{e,j}[i]\}_{j=1}^N$ denotes the information symbols, while $\tilde{\mathbf{s}}_{r}[i]\in\mathbb{C}^{(N_T-N)\times1}$ represents the remaining radar sequence \cite{wzl22_NOMAISAC2,xn25_NOMAISACvirtual}. 
$\{s_{e,j}[i]\}_{j=1}^N$ is assumed to be statistically mutually independent with zero-mean and unit-power, and is also independent of $\tilde{\mathbf{s}}_r[i]$. 
Note that in the case of $N=0$, i.e., without information embedding, the transmitter architecture coincides with that in SDMA-assisted NO-ISAC.
% receiver side: 3 schemes
\par
%At the receiver, users with lower power levels are typically mandated to decode the messages of that with higher power levels. 
At each receiver, the information-embedded sensing signal is treated as virtual communication signals, which can be first eliminated via SIC \cite{wzl22_NOMAISAC2,xn25_NOMAISACvirtual}. 
Furthermore, we consider all possible real user decoding orders with $\pi$ representing a specific one, due to their vital impact on performance enhancement. It is assumed that we decode the message of real user $\pi_{t}$ after decoding that of user $\pi_{k}, \forall t \geq k$.
Similar to \eqref{R_sdma}, an auxiliary variable $\eta_r$ is introduced in the SINR expression, which involves cases with or without extra SIC for radar sequence cancellation.
It is important to note that in schemes with radar sequence disabled or enabled with SIC, $N$ is set to 0, i.e., without information embedding. 
In other words, information embedding is considered only when the radar sequence is enabled without SIC, which facilitates more effective management of sensing-to-communication interference.
The general SINR for decoding the message of real user-$\pi_{k}$ at real user-$\pi_{t}$, $t \geq k$ is given by
\begin{equation} \label{R_noma}
	\gamma _{\pi_{t}\rightarrow \pi_{k}}=\frac{|\mathbf{h}_{\pi_{t}}^{H}\mathbf{w}_{\pi_{k}}|^{2}}{\sum_{j\in \mathcal{K},j>k}|\mathbf{h}_{\pi_{t}}^{H}\mathbf{w}_{\pi_{j}}|^{2}+\eta_{r}\|\mathbf{h}_{\pi_{t}}^{H}\tilde{\mathbf{W}}_{r}\|_2^{2}+\sigma_{c}^{2}}.
\end{equation}
The rate at real user-$\pi_t$ to decode the message of real user-$\pi_k$ is then calculated as $R_{\pi_{t}\rightarrow\pi_{k}}=\log_{2}(1+\gamma _{\pi_{t}\rightarrow\pi_{k}})$. The overall achievable rate of $s_{\pi_{k}}, \forall k\in\mathcal{K}$ is thus expressed as $R_{\pi_{k}}=\min_{t \ge k, t\in \mathcal{K}}\{R_{\pi_{t}\rightarrow\pi_{k}}\}$.
\par
In \eqref{R_noma}, the setting of $\eta_{r}$ and $\tilde{\mathbf{W}}_{r}$ follows the same way as $\eta_{r}$ and $\mathbf{W}_r$ in SDMA-assisted NO-ISAC. 
%% how to handle interference in ISAC
Notably, in the radar-sequence-enabled scheme without SIC, contrary to SDMA-assisted NO-ISAC systems where inter-functionality interference is typically treated as noise, one novel approach in NOMA-assisted NO-ISAC is to embed information bits into sensing waveforms, enabling sensing targets to be treated as virtual users \cite{mxd23_NOMAISACmagazine, wzl22_NOMAISAC2,Ahmed24_NOMAISACmagazine,xn25_NOMAISACvirtual}.
The messages of virtual users can be first detected and removed via SIC at the receiver, thereby effectively alleviating sensing-to-communication interference. 
%% pros and cons
\par
The major advantages and disadvantages of incorporating NOMA into NO-ISAC can be concluded as follows:
% advantages
\subsubsection{Advantages}
\begin{itemize}
    \item[-] \textit{Enhanced spectral efficiency in overloaded networks:} NOMA-assisted NO-ISAC facilitates the coexistence of multiple communication users and sensing targets via non-orthogonal resource allocation. 
    Particularly, in overloaded scenarios with high channel correlation, the additional DoFs provided by SIC at the receiver open the door to effective management of inter-user interference \cite{wzl22_NOMAISAC}.
    Moreover, by allowing extra radar sequences to convey information bits, NOMA-assisted NO-ISAC introduces a novel mechanism to alleviate sensing-to-communication interference, where information-embedded sensing signals can be first eliminated via SIC at the receiver \cite{wzl22_NOMAISAC2}. 
    This thereby contributes to an enlarged dual-functional trade-off region, as well as a promising SE compared to SDMA-assisted NO-ISAC under extremely overloaded network conditions.
\end{itemize}
\subsubsection{Disadvantages}
\begin{itemize}
% disadvantages
    \item[-] \textit{High transceiver complexity:} Although appealing for achieving notably enhanced performance in extremely overloaded network, NOMA-assisted NO-ISAC is impractical owing to high computational burdens and hardware complexities. On the one hand, the transmitter quests for a joint optimization of the user grouping, decoding orders and the precoders, which therefore calls for promising complexity reduction method. On the other hand, at each user, the number of SIC layers, i.e, the receiver complexity, increases proportionally with the number of users, leading to intensified SIC error propagation. 
    \item[-] \textit{Inefficient utilization of SIC:} NOMA-assisted NO-ISAC may experience a loss in DoF owing to the inefficient utilization of SIC \cite{bruno21_MA}. 
    To be specific, such system suffers from performance degradation when the network is underloaded with weak channel correlation or low requirement on sensing performance. This attributes to the fact that the spatial DoF available in this scenario is indeed sufficient, rendering the SIC operation redundant.
    \item[-] \textit{Restricted flexibility in network load and user deployment:} NOMA-assisted ISAC systems are sensitive to not only notwork load but also user deployment, which imposes a stricter demand on user scheduling. It is more suitable for severely overloaded systems with user channels being aligned. The trade-off performance between communication and sensing drops significantly when other scenarios are considered.
    \item[-] \textit{High sensitivity towards imperfect CSI:} Imperfect CSI complicates the user decoding order in SIC, which in turn disrupts beamforming design and leads to poor communication and sensing performance \cite{lrg24_FDISAC}.%\cite{Faezeh18_NOMA,lyh24_NOMAISAC,lrg24_FDISAC,Nassar24_NOMAISAC_CSI}.
\end{itemize}
%%%%%%%%%%%%%%%%%%
\subsection{RSMA-assisted NO-ISAC}
%% the development of RSMA
More recently, RSMA has been incorporated into NO-ISAC systems for flexible management of inter-user and inter-functionality interference \cite{xcc21_RSMAISAC}. The amounts of interference to be decoded and treated as pure noise can be dynamically adjusted with interference levels.
% RS
Such improvement in flexibility is facilitated by the concept of rate-splitting (RS), where user messages are split into common and private parts before transmission. 
Note that the common messages are decoded by multiple users, while the private messages are exclusively decoded by corresponding users.
%% interference management approach
Leveraging such RS at the transmitter and SIC at the receiver, RSMA fulfills interference management of partially decoding and partially treating interference as noise \cite{m22_RSMAtutorial}. 
%% the principles of RSMA
\par
In fact, leveraging different methods for message splitting and combining, RSMA displays various transmission schemes such as the linearly precoded one-layer RS, two-layer hierarchical RS (HRS), as well as generalized RS (GRS) \cite{m22_RSMAtutorial, m17_RSMAdownlink}.
%\cite{m22_RSMAtutorial, m17_RSMAdownlink, flores21_THPRS}.
% one-layer
Appealing for simple transceiver architecture, ISAC systems assisted by one-layer RS has been well studied \cite{xcc21_RSMAISAC,Rafael21_RSMAISAC,ylf22_RSMAISACsatellite,ylf22_RSMApart2,ckx24_RSMAISAC,Juha24_RSMAISAC}. It is the fundamental building block of all other schemes, where each user message is split into one common and one private part.
%% model illustration: downlink.
\par
% transmitter side; 
Fig. \ref{fig:downlink MA NOISAC}(c) illustrates a multi-user multi-target NO-ISAC system assisted by one-layer RS as an example. 
At the transmitter, the message $U_{k}$ of user-$k$ is split into a common sub-message $U_{c,k}$ and a private sub-message $U_{p,k}$.
We jointly encode all common sub-messages into a single common stream $s_{c}$, and separately encode each private sub-message into its respective private stream $s_{p,k}, \forall k \in \mathcal{K}$. 
Note that the model of the extra radar sequences $\mathrm{s}_r[i]$ is consistent with that in aforementioned SDMA-assisted NO-ISAC system. The overall transmit vector at time index $i$ is then given by $\mathbf{s}[i]=[\mathbf{s}_{r}[i],s_{c}[i],s_{p,1}[i],\dots,s_{p,K}[i]]^{T}\in \mathbb{C}^{(N_T+1+K)\times 1}$. 
Subsequently, the streams are linearly precoded via precoder $\mathbf{W} =[\mathbf{W}_{r},\mathbf{w}_{c},\mathbf{w}_{p,1},\dots, \mathbf{w}_{p,K}]\in \mathbb{C}^{N_{T}\times (N_T+1+K)}$, which remains consistent in one CPI. 
The dual-functional signal transmitted at time index $i$ is expressed as
\begin{equation}
	\mathbf{x}[i]=\mathbf{W}\mathbf{s}[i]=\mathbf{W}_{r}\mathbf{s}_{r}[i]+\mathbf{w}_{c}s_{c}[i]+\sum_{k\in \mathcal{K}}\mathbf{w}_{p,k}s_{p,k}[i],
\end{equation}
where the streams satisfy $ \mathbb{E}\{\mathbf{s}[i]\mathbf{s}[i]^{H}\}=\mathbf{I}_{N_T+1+K}$, indicating that the entries are independent from each other. The covariance matrix of the transmit signal is similarly calculated as $\mathbf{R}_{x}\approx\mathbf{W}\mathbf{W}^{H}$, where the total power budget is constraint by $\mathrm{tr}(\mathbf{W}\mathbf{W}^{H})\le P_{T}$.
% receiver side: 3 schemes
\par
At each communication receiver, the common stream is first decoded with all private streams being regarded as noise. 
After the successful decoding of the common stream via SIC, we subtract it from the received signal by re-encoding and precoding the common message.
Subsequently, the private streams are exclusively decoded by corresponding users, i.e., the private stream $s_k$ is decoded by user-$k$ with all other private streams being treated as noise.
Similarly, the general SINR expression for decoding $s_{c}$ and $s_{p,k}$ at user-$k$ of each CPI is given as
\begin{equation}\label{R_rsma}
	\begin{aligned}
    &\gamma_{c,k}=\frac{|\mathbf{h}_{k}^{H}\mathbf{w}_{c}|^{2}}{\sum_{j\in \mathcal{K}}|\mathbf{h}_{k}^{H}\mathbf{w}_{p,j}|^{2}+\eta_{r}\|\mathbf{h}_{k}^{H}\mathbf{W}_{r}\|_2^{2}+\sigma_{c}^{2}}, \forall k\in\mathcal{K},\\
	&\gamma _{p,k}=\frac{|\mathbf{h}_{k}^{H}\mathbf{w}_{p,k}|^{2}}{\sum_{j\in \mathcal{K},j\neq k}|\mathbf{h}_{k}^{H}\mathbf{w}_{p,j}|^{2}+\eta_{r}\|\mathbf{h}_{k}^{H}\mathbf{W}_{r}\|_2^{2}+\sigma_{c}^{2}}, \forall k\in\mathcal{K}.
    \end{aligned}
\end{equation}
The corresponding rates are calculated as $R_{c,k}=\log_{2}(1+\gamma_{c,k})$ and $R_{p,k}=\log_{2}(1+\gamma _{p,k})$. With the aim of ensuring that all users can successfully decode the common stream, the achievable common rate is expressed as $R_{c}=\min_{k\in \mathcal{K}}\{R_{c,k}\}=\sum_{k\in \mathcal{K}}C_{k}$, with $C_{k}$ denoting the allocated rate to transmit the common message of user-$k$.
The achievable rate of user-$k$ is thereby given as $R_{k}=C_{k}+R_{p,k}$, $\forall k\in\mathcal{K}$.
\par
In \eqref{R_rsma}, $\eta_{r}$ and $\mathbf{W}_{r}$ are set similar to aforementioned NO-ISAC systems assisted by SDMA and NOMA. 
%Similarly, in the cases where extra radar sequences are disabled or enabled with SIC, no inter-functionality interference arises. 
%Confronting with more intricate inter-user interference in NO-ISAC systems, RSMA establishes the common stream to enable a smart interference management of partially decoding and partially treating interference as noise.
%Moreover, considering the case with extra radar sequences but without SIC, sensing-to-communication interference arises together with inter-user interference. 
%Indeed, the common stream is capable to realize the beampattern approximation for sensing functionality, thereby eliminating the need for dedicated sensing signals \cite{xcc21_RSMAISAC}. The initial sensing-to-communication interference caused by extra radar sequences is thereby equivalently handled with the common stream being decoded at each communication receiver.
%% how to handle interference in ISAC
It is worth noting that the interference management strategy utilized in RSMA-assisted NO-ISAC systems contrasts with those in SDMA and NOMA-assisted ones, enabling a new paradigm suitable for all interference levels \cite{xcc21_RSMAISAC,Rafael21_RSMAISAC,ylf22_RSMAISACsatellite,ylf22_RSMApart2,ckx24_RSMAISAC,Juha24_RSMAISAC}.
This aligns with the established findings in RSMA-assisted communication-only systems, where the common stream is leveraged to enable a smart inter-user interference management of partially decoding and partially treating interference as noise \cite{m22_RSMAtutorial,bruno24_MA,park24_RSMAtutorial}.
Concretely, when the level of inter-user interference is low or high, RSMA-assisted ISAC automatically boils down to SDMA or NOMA-assisted ISAC by turning off the common or private streams. 
When the level of inter-user interference is medium, i.e., neither low or high, RSMA-assisted ISAC is still capable to realize promising trade-off performance between communication and sensing, potential to gain widespread use in the future.
Furthermore, contrary to SDMA-assisted NO-ISAC systems where inter-functionality interference is treated as noise, and NOMA-assisted systems where interference decoding is facilitated via information embedding, RSMA-assisted NO-ISAC enables a new solution thanks to the capability of the common stream in approximating the desired sensing beampattern \cite{xcc21_RSMAISAC,Juha24_RSMAISAC}. 
A portion of the initial sensing-to-communication
interference caused by extra radar sequences is thereby equivalently handled with the common stream being decoded at each communication receiver.
%% pros and cons
\par
The major advantages and disadvantages of incorporating RSMA into NO-ISAC can be concluded as follows:
\subsubsection{Advantages}
\begin{itemize}
%% advantages
    \item[-] \textit{Efficient use of the common stream:} 
    Leveraging the common stream, RSMA-assisted ISAC not only simplifies ISAC systems by omitting extra radar sequences, but also contributes to performance enhancement with effective interference management.
    To be elaborated, this improvement is facilitated by the triple functions of common streams, which showcase the capability to manage inter-user interference, tackle inter-functionality interference, and act as radar sequences to fulfill the beampattern requirements for sensing functionality \cite{xcc21_RSMAISAC,Juha24_RSMAISAC}. 
    \item[-] \textit{Enhanced system performance:} The incorporation of RSMA opens the door to a smart management of both inter-user and inter-functionality interference in NO-ISAC systems.
    Thanks to its flexibility in deciding the amount of interference to be decoded and treated as noise, RSMA-assisted NO-ISAC shows obvious trade-off gain over SDMA and NOMA-assisted ISAC in both single and multi-target scenario \cite{ckx24_RSMAISAC}.
    %\cite{GP23_RSMAISAC, ckx24_RSMAISAC,lzw24_RSMAISAC}. 
    % detection capability
    Moreover, it is capable of detecting more sensing targets than systems assisted by SDMA and NOMA while maintaining QoS of communication users \cite{ckx24_RSMAISAC}.
     \item[-] \textit{High flexibility in network load and user deployment:} Leveraging the flexible interference management strategy suitable for all interference levels, RSMA-assisted NO-ISAC system exhibits its capability to not only handle different network loads, i.e., both underloaded and overloaded networks, but also adapts to different user deployments featured by varying channel strengths and directions \cite{ylf22_RSMApart2}.
     %\cite{ylf22_RSMApart2,zjs25_RSMAISAC}. 
    \item[-] \textit{Low transceiver complexity:} 
    The one-layer RSMA-assisted NO-ISAC has become increasingly popular thanks to its simplicity in transmitter and receiver architecture \cite{jsb24_RSMAISAC}. 
    Specifically, at the transmitter, the design of common stream alleviates the necessity for extra radar sequences. Moreover, considering its tolerance for various network loads and user deployments, the optimization of user grouping as well as user scheduling becomes unnecessary, which further simplifies the transmitter design.
    At the receiver, only one layer of SIC is required to decode the common stream. 
    This is contrast to aforementioned NOMA-assisted ISAC, which quests for the user grouping and scheduling at the transmitter, and requires SIC layers proportional with user numbers at the receiver.
    \item[-] \textit{Robustness towards CSIT imperfections and user mobility:}
    RSMA-assisted ISAC shows strong resilience to CSIT imperfections caused by different sources of impairment such as user mobility \cite{Rafael21_RSMAISAC}.  
    It demonstrates more flexibility as well as robustness compared to aforementioned systems assisted by SDMA and NOMA, which are tailored for perfect CSIT and are susceptible to performance degradation when CSIT is imperfect.
\end{itemize}
\subsubsection{Disadvantages}
%% disadvantages
\begin{itemize}
    \item[-] \textit{High encoding and optimization complexity:} In previously mentioned SDMA and NOMA-assisted ISAC systems, the transmitter only needs to encode a single information stream for each communication user. In contrast, RSMA-assisted NO-ISAC requires to encode additional information streams owing to the message splitting and recombination \cite{jsb24_RSMAISAC}. This thereby results with increased encoding complexity.
    Furthermore, RSMA-assisted ISAC requires joint optimization of the common rate allocation and the precoders, which increase the computational complexity. 
    \item[-] \textit{Extra signaling burden:} RSMA-assisted NO-ISAC systems demand extra signaling overhead to align the transmitter and receivers, so as to
    guarantee a mutual understanding of the message splitting and combining methodology \cite{m22_RSMAtutorial,bruno24_MA}.
\end{itemize}
%%%%%%%%%%%%%%%%%%%%%%%%%%%%%%%%%%%%
\section{Uplink MA-assisted NO-ISAC Systems}\label{section5}
%% interference in NO-ISAC with uplink communication
To provide a holistic view, this section incorporates uplink MA techniques—uplink NOMA and RSMA—for managing interference in NO-ISAC systems, where uplink communication and downlink mono-static sensing architecture is considered. 
\par
\begin{figure}[t]
    \centering
    \includegraphics[width=1\linewidth]{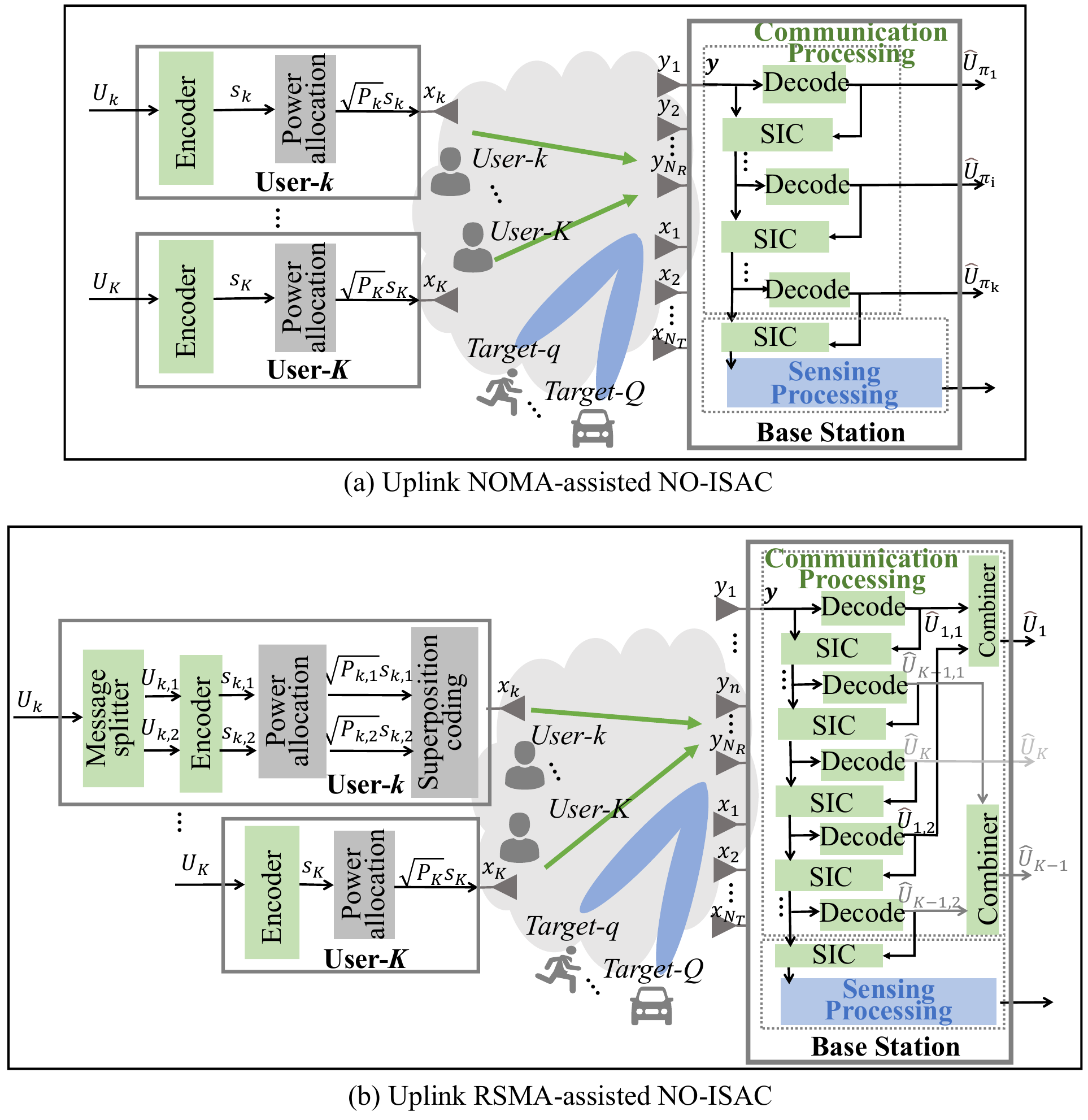}
    \caption{Non-orthogonal integrated uplink communication and downlink mono-static sensing systems assisted by different MA techniques—NOMA and RSMA in the uplink.}
    \label{fig:uplink MA NOISAC}
\end{figure}

As mentioned earlier, in addition to inter-user interference, such NO-ISAC architectures with uplink communication experience both sensing-to-communication and communication-to-sensing interference. 
%% this subsubsection
Uplink NOMA and uplink RSMA can be utilized to manage these interference. 
Fig. \ref{fig:uplink MA NOISAC} illustrates two examples of uplink NOMA and uplink RSMA-assisted NO-ISAC architectures, where the BS simultaneously serves $K$ single-antenna uplink communication users and detects $Q$ sensing targets indexed by $\mathcal{K}=\{1,\cdots,K\}$ and $\mathcal{Q}=\{1,\cdots,Q\}$, respectively.
%%%%%%%%%%%%%%%%%%%%%%
%%%%%%%%%%%%%%%%%%%%%%
\subsection{Uplink NOMA-assisted NO-ISAC}
%% model illustration 
Fig. \ref{fig:uplink MA NOISAC}(a) illustrates the use of uplink NOMA for managing interference in such NO-ISAC architecture. 
% transmitter
To be elaborated, at the $k$th communication user, message $U_k$, $\forall k \in \mathcal{K}$ is directly encoded into the information stream $s_k$ with unit variance, i.e., $\mathbb{E}\{|s_k|^2\}=1$. Subsequently, the stream $s_k$ is allocated with certain power $P_k$ for transmission.
% receiver
\par
Confronting with the mixed signals received at the BS, NOMA-assisted NO-ISAC is required to follow a fixed communication-to-sensing decoding order.
This arises from the fact that the NO-ISAC BS lacks prior knowledge of sensing echo signals, and only communication signal contains information bits \cite{mxd23_NOMAISACmagazine}.
In other words, communication messages are first decoded in sequence based on SIC with the existence of sensing-to-communication interference. 
Once all communication messages are decoded, the sensing echo signals are processed in a interference-free manner \cite{mxd23_NOMAISACmagazine,zc23_SemiISAC}. 
The major advantages and disadvantages of incorporating uplink NOMA into NO-ISAC are:
\subsubsection{Advantages} It is worth noting that in practice, the varying propagation characteristics of uplink communication and sensing typically give rise to a disparity among their power levels received at the BS. 
This substantial power differences enable NOMA to realize its potential in tackling inter-user and inter-functionality interference \cite{zc23_SemiISAC}.
\subsubsection{Disadvantages}  
%% suitable for sensing-prior design 
The main disadvantage is that the uplink information signals are always decoded by treating sensing-to-communication interference as noise due to the restriction on communication-to-sensing decoding order. This inevitably degrades communication performance and restricts the system to sensing-prior designs \cite{mxd23_NOMAISACmagazine}.
\subsection{Uplink RSMA-assisted NO-ISAC}
Leveraging the message splitting at users and SIC at the ISAC BS, uplink RSMA opens the door to a smart management of interference in NO-ISAC systems.
%% model illustration: transmitter, receiver 
Fig. \ref{fig:uplink MA NOISAC}(b) illustrates the system model of uplink RSMA-assisted NO-ISAC.
% transmitter
In this $K$-user scenario, performing message splitting for $K-1$ users is sufficient \cite{m22_RSMAtutorial}. Without loss of generality, it is assumed that the messages of all users except the $K$th user are split.
To be specific, at communication user-$k$, $\forall k \in \{1,\cdots K-1\}$, its message $U_k$, $\forall k \in \mathcal{K}$ is split into two sub-messages $U_{k,1}$ and $U_{k,2}$, equivalent to two virtual users.
The two sub-messages are then separately encoded into the information streams $s_{k,1}$ and $s_{k,2}$ with unit variance, i.e., $\mathbb{E}\{|s_{k,n}|^2\}=1, n=1,2$. Subsequently, the two streams are allocated with certain powers $P_{k,1}$ and $P_{k,2}$, and superposed at the $k$th user before transmission.
While at communication user-$K$, its message $U_K$ is directly encoded into the information stream $s_k$ and allocated with certain power $P_K$ for transmission.
% receiver
\par
Similar to the aforementioned NOMA-assisted NO-ISAC, the ISAC BS consecutively decodes uplink communication signals based on a certain decoding order. After that, the sensing signal is processed free from communication-to-sensing interference. The major advantages and disadvantages of incorporating uplink RSMA into NO-ISAC are:
\subsubsection{Advantages} Contrary to uplink NOMA-assisted NO-ISAC where each user involves a single message, RSMA-aided systems achieve more flexible inter-user interference management by splitting user messages into sub-messages. %\cite{Katwe23_uplinkRSMA}. 
With dynamic power allocation among sub-messages and optimized decoding order at the receiver, RSMA enables enhanced communication performance and improved SE, demonstrating significant potential for future applications.
\subsubsection{Disadvantages} Due to the fixed communication-to-sensing decoding order, NO-ISAC assisted by RSMA shares a similar disadvantage of that aided by NOMA, i.e., more suitable for sensing-prior scenario where sensing is prioritized over communication.
%% suitable for sensing-prior design 
Another disadvantage lies in the high computational complexity of optimizing decoding order and power allocation, which increases exponentially as the number of users grows. 
%\cite{hjw24_uplinkRSMA}.
Moreover, multiple layers of SIC are required to decode almost twice the number of messages in uplink NOMA-assisted NO-ISAC, leading to intensified hardware complexity at the receiver \cite{yzh22_uplinkRSMA}.
\par
Fortunately, the strength of sensing echo signal is typically weaker than communication signals due to long distance of round-trip transmission \cite{zc23_SemiISAC, mxd23_NOMAISACmagazine}. 
This characteristic facilitates the effective implementation of SIC in uplink NOMA or uplink RSMA-assisted NO-ISAC systems and may result in acceptable communication performance. 
Moreover, when target tracking is involved, prior knowledge of sensing parameters from previous observations enables the alleviation of sensing-to-communication interference during communication bit decoding, as the predicted sensing echoes can be subtracted from the received signal \cite{Mishra24_uplinkRSMA}. 
This further opens up opportunities for integrating uplink NOMA or RSMA with ISAC.
%Future research can explore potential techniques such as selective interference cancellation based on the correct frequency offset estimation \cite{wzq23_ISACsurvey_SI} to address sensing-to-communication interference, further enabling the integration of uplink NOMA or RSMA with ISAC.
%%%%%%%%%%%%%%%%%%%%%%%%%
%\begin{figure*}[tb]
%    \centering    \includegraphics[width=0.85\linewidth]{Fig_pdf/Fig9_SCMA.pdf}
%    \caption{One example of SCMA encoding and multiplexing \cite{bruno24_MA}.}
%    \label{fig:SCMA}
%\end{figure*}
%%%%%%%%%%%%%%%%%%%%%%%%%
%%%%%%%%%%%%%%%%%%%%%%%%%%%%%%%%%%%%
\section{Other Advanced MA-assisted ISAC Systems} \label{section7}
In addition to prominent SDMA, power-domain NOMA, and RSMA, a few MA schemes in other domains such as the code domain and the delay-Doppler (DD) domain have been proposed for interference management in ISAC. In the following, we review code-domain NOMA (CD-NOMA) and DD-domain MA, followed by an outlook on unified MA (UMA) for future ISAC systems.
%%%%%%%%%%%%%%%%%%%%%%%%%
%\begin{figure*}[tb]
%    \centering    \includegraphics[width=0.8\linewidth]{Fig_pdf/Fig8_OTFS.pdf}
%    \caption{Diagram of OTFS and ODDM \cite{dqw25_OTFSODDM,lh22_ODDM}. DFT: discrete Fourier transform; DTFT: discrete-time Fourier transform; CP: cyclic prefix.}
%    \label{fig:OTFSODDM}
%\end{figure*}
%%%%%%%%%%%%%%%%%%%%%%%%%
\subsection{Code-domain NOMA-assisted ISAC}
% introduction
Inspired by conventional CDMA, CD-NOMA has also unveiled its value in ISAC, which utilizes sparse or non-orthogonal sequences/codebooks with low correlation coefficients to distinguish different users \cite{lyx24_MAISAC}. 
This contrasts to the dense and orthogonal spreading sequences used in CDMA, making CD-NOMA more suitable for future ISAC demanding high user/target capacity and efficient spectrum utilization.
In general, existing CD-NOMA techniques can be typically classified into two types: dense and sparse CD-NOMA \cite{lzk21_CDNOMA}, which differs from each other based on whether their sequences/codebooks exhibit sparsity. 
Sparse CD-NOMA is one important research direction in CD-NOMA \cite{lzk21_CDNOMA}. There are in general two typical sparse CD-NOMA—low-density signatures based CD-NOMA (LDS-CDMA) and sparse code MA (SCMA)—in ISAC, which are discussed as follows.
% LDS-CDMA
\par
In LDS-CDMA-assisted ISAC systems, the information symbols utilized for both communication and sensing are spread by a unique LDS with a limited portion of non-zero entries, and then superimposed before transmission.
By leveraging the sparsity of LDS, ISAC systems aided by LDS-CDMA induce a sparse inter-user interference pattern that can be modeled as a low-density graph, thereby alleviating inter-user interference on each chip.
Symbol detection is then fulfilled utilizing the message passing algorithm (MPA) \cite{Hoshyar08_LDSCDMA}, while the environment sensing (such as imaging) can be modeled as signal recovery problems based on compressive sensing (CS) theory \cite{Donoho06_CS}.
% SCMA
This LDS-CDMA-assisted ISAC can be further extended to that aided by SCMA, where bit streams are directly mapped to sparse codewords chosen from the pre-defined sparse codebooks of users \cite{lzk21_CDNOMA}. After that, all codewords are multiplexed over shared orthogonal resources, e.g., OFDM subcarriers \cite{dll15_CDNOMA}. 
This contrasts to the spreading of modulated symbols via LDS in LDS-CDMA-assisted ISAC. 
Actually, SCMA can be considered as an enhanced version of LDS-CDMA, which offers additional shaping gain \cite{Taherzadeh14_SCMA} thanks to a multi-dimensional constellation in comparison to simple repetition of modulated symbols in LDS-CDMA. 
%One example of SCMA encoding and multiplexing is illustrated in Fig. \ref{fig:SCMA}, where six communication users, each transmitting two bits, are multiplexed across four orthogonal resources. 
The major advantages and disadvantages of CD-NOMA assisted ISAC are:
\subsubsection{Advantages} 
CD-NOMA-aided ISAC allocates multiple users with the same time-frequency resources, therefore improving the multi-user access efficiency and SE in ISAC. By exploiting the sparse sequences/codebooks, ISAC systems assisted by LDS-CDMA or SCMA face relieved inter-user interference, which contributes to enhanced dual-functional performance with the aid of effective detection methods.
Additionally, various spreading sequence designs enable the flexibility in CD-NOMA-assisted ISAC, making it a potential solution for future ISAC.
\subsubsection{Disadvantages} 
Despite its advantages in SE, CD-NOMA introduces substantial processing overhead at the ISAC receiver due to complex detection techniques. The design of spreading sequences also imposes high computational complexity. 
In fact, due to these challenges, the investigation of CD-NOMA for ISAC is still in its infancy. 
The authors of \cite{tx21_SCMAISAC} investigated a SCMA-aided ISAC system where multiple uplink users communicate with the access point (AP), and part of the transmit signals are utilized to sense the environment. 
This study, however, only showcases the capability of CD-NOMA to manage inter-user interference, without focus on inter-functionality interference.
\subsection{Delay-Doppler-domain MA-assisted ISAC}
Orthogonal time-frequency space (OTFS) modulation has emerged as an effective technique for high-mobility ISAC scenarios, which multiplexes data in the DD domain rather than the conventional time, frequency, spatial, and code domain \cite{ywj24_OTFSISAC}.
It exploits the DD domain to identify the time-varying channel characteristics and is therefore robust to significant delay and Doppler shifts \cite{Shtaiwi24_OTFSISACsurvey}.
% ODDM
Inspired by OTFS, orthogonal delay-Doppler division multiplexing (ODDM) has been newly proposed \cite{lh22_ODDM}, which achieves orthogonality in fine resolutions of the DD domain with realizable pulses. 
%As illustrated in Fig. \ref{fig:OTFSODDM}, 
Specifically, the transceiver aided by OTFS or ODDM involves two main tasks: at the transmitter, converting information symbols from the DD domain to a time-domain signal; and at the receiver, reconstructing the DD-domain signal from the received signal in the time domain \cite{ywj23_OTFSISAC}. 
The primarily difference between OTFS and ODDM lies in how they transform symbols between the DD domain and the time domain \cite{lh22_ODDM}. 
For instance, at the transmitter, OTFS first maps the DD domain symbols to the TF domain via the inverse symplectic finite Fourier transform (ISFFT), followed by a Heisenberg transform to generate the time-domain signal. 
In contrast, ODDM bypasses explicit TF-domain mapping and directly maps DD-domain symbols to the time domain by leveraging a stagger \cite{dqw25_OTFSODDM,lh22_ODDM}.
It is worth noting that the unique mapping strategy of the ODDM-aided system enables its orthogonality in the DD domain and enhances its roubtness in high-mobility scenarios \cite{dqw25_OTFSODDM}. 
The communication functionality is then fulfilled through demodulating DD-domain signals, while sensing is achieved by extracting parameters from the established DD matrix \cite{ywj24_OTFSISAC}.
\par
Recently, the surge of users and targets in high-mobility ISAC has stimulated the development of DD-domain MA for efficient multi-user access.
Specifically, ODDM can directly facilitate OMA in the DD domain, which avoids inter-user interference with orthogonal DD domain resource elements allocated to multiple users.
Moreover, OTFS/ODDM can be combined with other advanced MA techniques such as SDMA, NOMA and RSMA to efficiently manage interference and achieve high SE in ISAC systems \cite{bruno24_MA}. 
For instance, the authors of \cite{xlp24_OTFSNOMAISAC} combined OTFS with NOMA for managing interference in ISAC, where an unmanned aerial vehicles (UAV)-based BS extracts parameters of users (such as the velocity) from DD-domain echo signals, and employs non-orthogonal power allocation to enhance data rates. 
RSMA also holds the potential to be combined with ODDM by multiplexing common and private streams into real and imaginary parts of DD-domain components, enabling efficient interference management \cite{wxh25_ODDMRSMA}.
Indeed, current research on DD-domain MA for ISAC remains limited, which calls for further in-depth investigation.
The major advantages and disadvantages of DD-domain MA-assisted ISAC are summarized as follows: 
\subsubsection{Advantages}
By exploiting the DD domain, ISAC systems aided by DD-domain MA have great potential to enhance both communication and sensing functionalities, especially in high-mobility scenarios. On the one hand, this system ensures both reliable and efficient transmission, thanks to its resilience to significant delay and Doppler shifts. On the other hand, the direct mapping of DD-domain parameters to target range and velocity makes it well-suited for sensing \cite{Shtaiwi24_OTFSISACsurvey}. Another advantage lies in its unified transceiver design, eliminating the need for extra modules to support communication (i.e., demodulating DD domain signals), and sensing (i.e., computing the range-Doppler matrix), since both functionalities can be achieved with the same module such as the inverse discrete Zak transform (IDZT) \cite{ywj24_OTFSISAC}.
\subsubsection{Disadvantages}
DD-domain MA-assisted ISAC suffers from high receiver complexity due to its reliance on computationally intensive sensing algorithms. The preliminary approaches typically rely on random coding with maximum likelihood decoding, which exhibit exponential complexity with respect to signal length \cite{ywj24_OTFSISAC}. This thereby hinders their application in real-time systems with strict power and latency constraints.
\subsection{Unified MA in Future ISAC}
% introduction
Inspired by various MA-assisted ISAC, we provide an outlook on ISAC systems aided by future unified MA (UMA), which unifies RSMA with all other MA schemes and ultimately establishes a conceptually simple understanding of different MA \cite{bruno24_MA}. 
This subsection addresses two key questions: ``Why is UMA essential in ISAC?'' and ``how to design UMA-assisted ISAC?''.
% the first question
\par
The first question, why UMA-assisted ISAC is needed, can be answered from three perspectives: \textit{(1)} hardware and computational complexity, \textit{(2)} the performance demands of future ISAC systems, and \textit{(3)} resource allocation design.
% 1 common problem
Specifically, there exist a series of common issues in existing MA-assisted ISAC, such as high optimization complexity (involving precoders, and resource allocation between the two functionalities or among multiple users), high receiver complexity (involving complex operations or algorithms), as well as difficult channel estimation. 
Future ISAC systems, featuring diverse network deployments and a wide range of service requirements, may require optimization across multiple MA-assisted ISAC schemes, each designed for specific operational conditions. However, this multi-scheme switching approach inevitably incurs high computational complexity. 
A unified UMA-assisted ISAC framework offers an efficient alternative by inherently supporting diverse requirements, eliminating the need for such complex joint MA optimization.
% 2 MA performance comparison
Moreover, catering to the growing demand for fully exploiting radio resources, the past decade has witnessed an explosion of MA techniques, which aims to make better use of different resource domains. Apart from the conventional time, frequency, spatial, and code domain, there exist some other dimensions such as the power, message split and combiner \cite{bruno24_MA}.  
This makes it challenging to obtain a comprehensive performance comparison among ISAC systems aided by different MA in several domains.
Indeed, the studies on NOMA-assisted ISAC always compare with that aided by SDMA and OMA, but not with RSMA and other MA. RSMA-assisted ISAC systems are compared with OMA/SDMA/NOMA-aided systems, without consideration for code-domain MA. The exploration of CD-NOMA in ISAC typically includes the comparison of different types of sequences/codebooks.
This therefore necessitates the study of UMA-assisted ISAC, which is expected to give a holistic understanding of performance benefits triggered by MA techniques in different domains.
% 3 efficient solution
Additionally, facing critical issues such as massive connectivity and multi-functionality co-existence, future ISAC calls for more efficient and flexible utilization of radio resources, improved management of inter-user and inter-functionality interference, and the development of unified hardware architectures. 
By deeply understanding the essence of unified MA design—maximizing the utilization of all dimensions in the simplest way, UMA demonstrates its potential to efficiently enable intelligence and multi-functionality for future ISAC systems.
% the second question
\par
There are no definite answers to the second question on how to design UMA-assisted ISAC, since UMA doesn’t exist.
% 1 RSMA
However, this tendency for UMA-ISAC can be exemplified by the exploration of RSMA-assisted ISAC.
Specifically, RSMA serves as a superset that includes other MA techniques such as SDMA and NOMA, and can be easily reduced to these techniques by switching off certain streams. 
This unified capability inherent in RSMA-assisted ISAC is essential for the long-term advancement of both theoretical research and practical implementation.
On the one hand, by providing an in-depth understanding of SDMA and NOMA in the spatial and power domain, RSMA fully exploits these domains and unified these MA schemes in a simple way. This signifies the advent of future unified research direction.
On the other hand, the unification of MA techniques in ISAC contributes to shared design of hardware architecture and software framework, which makes practical implementation easier.
% combine CD-NOMA
Though RSMA-assisted ISAC exists as a good example, the development of generalized UMA-assisted ISAC framework remains an ongoing research frontier.
For example, CD-NOMA such as LDS-CDMA and SCMA itself has not been unified or bridged yet, let alone the unification of RSMA and CD-NOMA.
In other words, the major challenge in UMA-assisted ISAC lies in whether there exists a unified MA-ISAC in the code domain, and the relationship between ISAC systems aided by RSMA and MA in other domains such as the code and DD domain.
Moreover, inspired by the aforementioned multi-function of the common streams in RSMA-assisted ISAC, particularly its capability to act as radar sequences for sensing, whether UMA has potential to enhance dual-functional trade-offs while further simplifying the ISAC transceiver architecture (e.g., no need for extra radar sequences) deserves future exploration.  
%%%%%%%%%%%%%%%%%%%%%%%%%%%%%%%%%%%%
\section{Comparison among Various MA-assisted ISAC}\label{section6}
This section provides a performance comparison of different MA-assisted ISAC systems, measured by established communication and sensing metrics.
%%%%%%%%%%%%%%%%%%
\subsection{Metrics for communication}
In general, communication metrics can be typically classified into two types, i.e., efficiency and reliability. 
\subsubsection{Efficiency} The main idea of efficiency metrics is to quantify how much information is successfully delivered from transmitter to receiver within limited resources, e.g., the weighted sum rate (WSR) \cite{xcc21_RSMAISAC}, max-min fairness rate (MFR) \cite{ylf22_RSMApart2}, and EE \cite{zjq22_ISAC_EE}, which are detailed as follows:
\begin{itemize}
    \item \textit{Weighted sum rate:} The WSR is commonly adopted in multi-user scenarios, which is capable of indicating the user priority via the assigned rate weight. With a specific weight vector $\bm{\mu}=[\mu_1, \dots, \mu_K]$, the WSR is given by 
    \begin{equation}
        \mathrm{WSR}=\sum_{k \in \mathcal{K} } \mu_{k} R_k,
    \end{equation}
    where $R_k$ denotes the achievable rate of communication user-$k$, as defined below \eqref{R_sdma}, \eqref{R_noma}, and \eqref{R_rsma}.
    \item \textit{Max-min fairness rate:} The MFR ensures fairness among communication users, which is defined as 
    \begin{equation}
        \mathrm{MFR}=\mathop{\min}_{k\in \mathcal{K}} R_{k}.
    \end{equation}
    \item \textit{Energy efficiency:} The EE is defined as the ratio between the achievable data rate and the total consumed power, which represents bits per energy consumption and is calculated as \cite{zjq22_ISAC_EE}
\begin{equation}
    \mathrm{EE} = \frac{\sum_{k\in \mathcal{K}}R_{k}}{\frac{1}{\rho_e }P_T+P_{\mathrm{cir}}},
\end{equation}
where $\rho_e \in (0,1]$ denotes the amplifier efficiency of the BS, $P_T$ and $P_{\mathrm{cir}}$ represent the transmission power consumption and circuit power consumption, respectively.
\end{itemize}
%% reliability
\subsubsection{Reliability} Reliability metrics focus on evaluating how effectively a system can correct or minimize errors in information bits received at the receiver. These metrics are essential since systems inevitably suffer from interference, noise and fading effects. The primary reliability metrics include the bit error rate (BER), symbol error rate (SER), and frame error rate (FER) \cite{lf22_ISACsurvey}.
\par
Specifically, the BER is defined as the ratio of the number of incorrectly received bits to the total number of transmitted bits over a given time interval, while the SER quantifies the probability of incorrect reception at the symbol level, where each symbol may involve multiple bits depending on the modulation scheme. The FER measures the probability that a received data frame has errors in any of its bits or symbols. 
It is important to note that these error rates share a similar ratio-based definition, while their exact expressions vary with specific modulation schemes and are therefore omitted here for brevity.
%%%%%%%%%%%%%%%%%%
\subsection{Metrics for sensing}
% radar metric
In the realm of radar sensing tasks within the physical (PHY) layer, existing sensing metrics can be divided into three categories, i.e., target detection, parameter estimation, and other generalized requirements. 
\subsubsection{Target detection} This metric makes a binary or multiple decision regarding the target state (i.e., whether it is present or absent) under noisy or disturbed observations \cite{lhl24_ISACdetection}. 
For instance, the binary detection problem is typically formulated as
\begin{equation}
        \mathbf{Y}_{r}=\left\{\begin{matrix}
\mathcal{H}_{1}: &\mathbf{H}_{r}\mathbf{X}+\mathbf{Z}_{r},  \\
\mathcal{H}_{0}: &\mathbf{Z}_{r},
\end{matrix}\right.  
\end{equation}
where $\mathbf{X}\in\mathbb{C}^{N_T\times I}$ denotes the transmit signal, $\mathbf{H}_{r} \in \mathbb{C}^{N_{R} \times N_{T}}$ represents the target response matrix (TRM), and $\mathbf{Z}_{r} \in \mathbb{C}^{N_{R} \times I}$ is the AWGN with each column following $\mathcal{CN}(\mathbf{0}_{N_{R}\times 1},\sigma _{r}^{2}\mathbf{I}_{N_{R}})$. $\mathcal{H}_{1}$ corresponds to the hypothesis that the target is present, i.e., both echo signals and the noise are received at the radar receiver, whereas $\mathcal{H}_{0}$ denotes the null hypothesis, indicating that the target is absent and thus only noise is received.
To be specific, the performance of target detection can be measured via the detection probability $P_{d}$, false-alarm probability $P_{fa}$ \cite{Naghsh12_radarmetric}, and Kullback-Leibler divergence (KLD, a.k.a. the relative entropy) \cite{Cover99_radarmetric,tb15_KLD}, which are detailed as follows:
\begin{itemize}
    \item \textit{The detection probability and false-alarm probability:} The detection probability $P_d$ signifies that $\mathcal{H}_1$ is true and the detector correctly chooses $\mathcal{H}_1$, while the false-alarm probability $P_{fa}$ indicates that $\mathcal{H}_0$ is true but the detector incorrectly chooses $\mathcal{H}_1$ \cite{steven93_CRB}. The Neyman-Pearson detector defined as $\Lambda_d \mathop{\gtrless}_{\mathcal{H}_0}^{\mathcal{H}_1} \delta_d$ \cite{steven93_CRB}
    %\cite{steven93_CRB,bekkerman06_pd}
    is commonly utilized to decide whether the target is present, where $\delta_d$ denotes the decision threshold. The probabilities of detection $P_d$ and false-alarm $P_{fa}$ are calculated as
\begin{equation}
\begin{aligned}
    P_d &= \mathbb{P}(\Lambda_d>\delta_d| \mathcal{H}_1), \\
    P_{fa} &= \mathbb{P}(\Lambda_d>\delta_d| \mathcal{H}_0).
\end{aligned}
\end{equation}
Note that considering the highly complex closed-form expressions for $P_d$ and $P_{fa}$, the radar SNR given by $\mathrm{SNR}_r=\frac{\|\mathbf{H}_{r}\mathbf{X}\|_F^{2}}{I\sigma_{r}^{2}}$ has emerged as a more tractable alternative for evaluating detection performance, since $P_d$ typically increases with the radar SNR \cite{lr23_SNRCRBISAC}.
\item \textit{Kullback-Leibler divergence:} The KLD is used to quantify the difference between the probability distributions under $\mathcal{H}_{1}$ and $\mathcal{H}_{0}$, which is defined as \cite{tb15_KLD} 
\begin{equation}
\begin{aligned}
    D(P_{0}\|P_{1} )=& \int P_{0}(\mathbf{Y}_{r})\mathrm{log} \frac{P_{0}(\mathbf{Y}_{r})}{P_{1}(\mathbf{Y}_{r})}\mathrm{d}\mathbf{Y}_{r} \\
= &\mathbb{E}_{P_{0}}\left [ \mathrm{log} \frac{P_{0}(\mathbf{Y}_{r})}{P_{1}(\mathbf{Y}_{r})} \right ],
\end{aligned}
\end{equation}
where $P_{1}(\mathbf{Y}_{r})$ and $P_{0}(\mathbf{Y}_{r})$ represent the probability density functions (PDF) of distributions under $\mathcal{H}_1$ and $\mathcal{H}_0$, respectively. $\mathbb{E}_{P_{0}}$ denotes the expectation with respect to the distribution $P_{0}(\mathbf{Y}_{r})$. 
It is worth noting that based on Stein’s lemma \cite{Cover99_radarmetric}, the relationship between the KLD and $P_{d}$ under any fixed $P_{fa}$ is formulated as
\begin{equation}
    D(P_{0}\|P_{1} )=\lim_{N_d \to \infty} (-\frac{1}{N_d}\log (1-P_{d} ) ),
\end{equation}
where $N_d$ represents the observation numbers. This indicates that for sufficiently large $N_d$, maximizing the KLD becomes asymptotically equivalent to maximizing the detection probability $P_{d}$ for any fixed false-alarm probability $P_{fa}$ \cite{tb15_KLD}. 
\end{itemize}
\subsubsection{Parameter estimation} Parameter estimation involves determining the parameters of interested sensing targets such as angles, velocities and ranges. An effective estimator is required so as to map the received signal $\mathbf{Y}_r$ to the corresponding estimates, i.e., $\hat{\bm{\xi}}=f(\mathbf{Y}_{r})$ \cite{lf23_seventyISAC}. Specifically, the performance of parameter estimation can be measured by the mean squared error (MSE) \cite{xcc21_RSMAISAC}, Cram\'er-rao bound (CRB) \cite{ylf22_RSMApart2, lj07_CRB}, and radar estimation information rate (REIR) \cite{bliss14_radarmetric}, which are defined as follows:
\begin{itemize}
    \item \textit{Mean squared error:} The MSE quantifies the average squared error between the estimated value $\hat{\bm{\xi}}$ and the true value of parameters $\bm{\xi}$, which is given as
    \begin{equation}
        \varepsilon=\mathbb{E} \left \{ \|\bm{\xi} -\hat{\bm{\xi}} \|^{2}\right \},
    \end{equation} 
    where the expectation $\mathbb{E}$ is taken over the parameter distribution when the parameters are modeled as random variables, or solely over the noise when the parameters are deterministic \cite{lf23_seventyISAC}.
    \item \textit{Cram\'er-Rao bound:} The CRB is a a fundamental estimation metric that serves as a lower bound on the variance of any unbiased estimator (i.e., $ \mathbb{E} (\mathbf{\hat{\bm{\xi}}} )=\bm{\xi}$). It characterizes the theoretical limit of estimation accuracy and is defined as the inverse of the Fisher information matrix (FIM), i.e., $\mathbf{CRB}=\mathbf{F}^{-1}$ \cite{steven93_CRB,lj07_CRB}. 
    With the parameter set $\bm{\xi }= \left \{ \bm{\theta}, \bm{\alpha}_{\Re}, \bm{\alpha}_{\Im}, \bm{\mathcal{F}}_{D}\right \}^{T}\in \mathbb{R}^{4Q \times 1}$, where $\bm{\theta}=[\theta_1,\dots,\theta_Q]$, $\bm{\alpha}=[\alpha_1,\cdots,\alpha_Q]$, and $\bm{\mathcal{F}}_{D}=[\mathcal{F}_{D_{1}},\dots,\mathcal{F}_{D_{Q}}]$ denotes the angular direction, the complex reflection coefficient, and the Doppler frequency, respectively, the FIM is expressed as \cite{lj07_CRB}
    \begin{equation}\label{eq:classicCRB}
    \begin{aligned}
        \mathbf{F}&=-\mathbb{E}\bigg[\frac{\partial^2 \mathrm{log} P(\mathbf{Y}_{r}|\bm{\xi}  )}{\partial \bm{\xi}^2} \bigg]\\
        &=\frac{2}{\sigma_r^2}\left(\mathrm{tr}\left[ \sum_{i\in\mathcal{I}} \frac{\partial\mathbf{v}^{H}[i] }{\partial\bm{\xi} }\frac{\partial\mathbf{v}[i]}{\partial\bm{\xi}}\right]\right)_{\Re},
    \end{aligned}
    \end{equation}
 where $\mathbf{Y}_{r}=[\mathbf{y}_{r}[1],\cdots, \mathbf{y}_{r}[I]]$, $P(\mathbf{Y}_{r}|\bm{\xi})$ refers to the likelihood function, $\mathbf{v}[i]=\mathbf{y}_{r}[i]-\mathbf{z}_{r}[i]$ with $\mathbf{y}_{r}[i]$ and $\mathbf{z}_{r}[i]$ defined in \eqref{Ys}.
 It is also worth noting that when $\bm{\xi}$ is modeled as random variables with a priori distribution $P(\bm{\xi})$, we select the Bayesian CRB (BCRB) to evaluate the estimation performance, and the corresponding FIM is defined as \cite{lsh24_ISACtenchallenges}
\begin{equation}
        \mathbf{F}_{B}=-\mathbb{E}\bigg[\frac{\partial^2 \mathrm{log} P(\mathbf{Y}_{r}|\bm{\xi})}{\partial \bm{\xi}^{2}} \bigg]-\mathbb{E}\bigg[\frac{\partial^2 \mathrm{log} P(\bm{\xi})}{\partial \bm{\xi}^{2}} \bigg],
\end{equation}
where with no available prior distribution, the BCRB, i.e., ${\mathbf{F}_B}^{-1}$, reduces to the classical CRB in \eqref{eq:classicCRB} \cite{lsh24_ISACtenchallenges}. 
\item \textit{Radar estimation information rate:} Inspired by the communication capacity, the REIR refers to the cancellation of the uncertainty in estimating sensing parameters per second \cite{lf22_ISACsurvey}, which is upper-bounded by \cite{bliss14_radarmetric}
\begin{equation}
R_{\mathrm{est}}\leq \frac{H_\mathrm{rec}-H_\mathrm{est}}{T_\mathrm{pri}},
\end{equation}
where $T_\mathrm{pri}$ is the pulse repetition interval, $H_\mathrm{rec}$ denotes the entropy of the received signal accounting for both process and estimation uncertainties, and $H_\mathrm{est}$ represents the entropy of estimation associated solely with the estimation uncertainty \cite{bliss14_radarmetric}. This bound reflects the maximum amount of information about the target parameters that can be extracted per unit time.
\end{itemize}
\subsubsection{Other generalized requirements} In addition to the aforementioned metrics, there exist some other metrics related to other generalized requirements, such as radar mutual information (MI) \cite{ylf22_RSMApart2}, beampattern MSE \cite{xcc21_RSMAISAC}.
\begin{itemize}
    \item \textit{Radar mutual information:} The radar MI represents the mutual information between the TRM $\mathbf{H}_{r}$ and sensing echo signal $\tilde{\mathbf{y}}_{r}=\mathrm{vec}(\mathbf{Y}_{r})$, which is calculated as \cite{ylf22_RSMApart2}
\begin{equation}
\begin{aligned}
    \mathrm{MI}=&I(\tilde{\mathbf{y}}_{r};\mathbf{{H}_{r}|\mathbf{X}})=\mathrm{log}_{2}(|\mathbf{I}_{N_{R}}+\frac{\mathbf{H}_{r}\mathbf{R}_{X}\mathbf{H}_{r}^{H}}{\sigma_{r}^{2}}|)\\
\overset{(a)}=&\mathrm{log}_{2}(|\mathbf{I}_{N_{R}}+\frac{|\alpha |^{2}\mathbf{b}(\theta)\mathbf{a}^T(\theta)\mathbf{R}_{X}\mathbf{a}^\ast(\theta)\mathbf{b}^{H}(\theta)}{\sigma_{r}^{2}}  |),
\end{aligned}
\end{equation}
where $\mathbf{R}_X$ denotes the covariance matrix of the transmit signal, and step (a) is derived with \eqref{Ys} in a single-target scenario. In this case, maximizing the radar MI is equivalent to maximizing the transmit beampattern gain defined as $\mathbf{P}_{d}(\theta)=\mathbf{a}^{T}(\theta)\mathbf{R}_{X} \mathbf{a}^{\ast}(\theta)$.
\item \textit{Beampattern MSE:} The transmit beampattern can be designed to approximate a desired highly directional pattern $\mathbf{P}_{d}$, with the beampattern MSE defined as \cite{xcc21_RSMAISAC}
\begin{equation} \label{beampatternMSE}
    \mathrm{MSE}_B=\sum_{j=1}^{J}|\mathbf{P}_{b}(\theta_{j})-\mathbf{a}^{T}(\theta_{j})\mathbf{R}_{X}\mathbf{a}^{\ast}(\theta_{j})|^{2},
\end{equation}
where $\theta_{j}$ denotes the jth angle grid among all $J$ grids, and $\mathbf{P}_{b}(\theta_{j})$ is the desired beampattern at $\theta_{j}$. 
Note that a lower beampattern MSE typically results with an increased radar SNR at the receiver, thereby enhancing the detection probability and the estimation accuracy.
\end{itemize}
\subsection{Performance comparison}\label{section3C3}
%% this subsubsection
In the following, we provide a performance comparison of various MA-assisted ISAC—TDMA-assisted TD-ISAC, and SDMA/NOMA/RSMA-assisted NO-ISAC, where WSR/MFR and CRB are applied to evaluate the communication and sensing performance, respectively. The well established downlink communication and downlink momo-static sensing is considered in this subsection. 
%% problem formulation
\par
To clarify, we begin with the problem formulation of the ISAC transmission design. Specifically, we focus on jointly maximizing the WSR/MFR for communication and minimizing the largest eigenvalue of the CRB matrix for sensing. Note that the latter corresponds to maximizing the smallest eigenvalue of $\mathbf{F}$ \cite{lj07_CRB}. 
The regularization parameter $\phi$ is incorporated to synergize WSR/MFR and CRB in the objective function, facilitating a trade-off between communication and sensing. The optimization problem is formulated as \cite{ylf22_RSMApart2}
\begin{subequations}\label{p1}
	\begin{align}
&\mathop{\max}_{\mathbf{W},t}\,\,\, \mathrm{WSR/MFR}+\phi t \\
		s.t.\,\,\,\,&\mathbf{F}\succeq  t\mathbf{I}_{4Q}, \label{constraint 1}\\
		&\mathrm{diag}(\mathbf{W}\mathbf{W}^{H})=\frac{P_T\mathbf{1}^{N_{T}\times 1}}{N_{T}}, \label{constraint 2}
	\end{align}
\end{subequations}
where $\mathbf{I}_{4Q}$ denotes the identity matrix with the same dimension of $\mathbf{F}$.
Constraint \eqref{constraint 1} ensures that the matrix $\mathbf{F}-t\mathbf{I}_{4Q}$ is positive semi-definite, where the auxiliary value $t$ is equivalent to the smallest eigenvalue of $\mathbf{F}$. 
The per-antenna transmit power budget is constraint by \eqref{constraint 2}.
The classical successive convex approximation (SCA)-based algorithm introduced in \cite{ckx24_RSMAISAC} is directly applied to address problem \eqref{p1}.
%% MA-assisted O-ISAC/NO-ISAC
\par
Note that with the common stream switched off, RSMA-assisted ISAC boils down to SDMA-assisted ISAC. Moreover, to reduce computational complexity, we directly adopt a specific decoding order according to the ascending channel strengths in NOMA-assisted ISAC.
The additional radar sequences are omitted in this point-target scenario, since the parameters to be estimated are typically no larger than the number of users, rendering $K$ DoFs sufficient to ensure promising sensing performance measured by CRB \cite{lf21_CRBoptimization,Attiah24_CRB}.
When considering varying numbers of targets, we select the average tr(CRB), i.e., $\mathrm{tr}(\mathbf{F}^{-1})/Q$, rather than tr(CRB) as the sensing metric to maintain fairness in comparison.
The key simulation parameters are presented in Table \ref{Table:parameter}.
The following results are obtained by averaging over 100 channel realizations.
%%%%%%%%%%%%
\begin{table}[t]
\centering
\caption{Simulation Parameters.}
\label{Table:parameter}
\begin{tabular}{cc}
\hline
\rowcolor{gray!20}
\textbf{Parameters} & \textbf{Values} \rule{0pt}{2ex}\\ \hline
Transmit power budget at the BS ($P_T$)& 20 dBm\\  Length of one CPI ($I$)& 1024 \\
Communication noise ($\sigma_c^2$) & 0 dBm \\
Sensing noise ($\sigma_r^2$) & 0 dBm \\
Complex reflection coefficient ($\{\alpha_q\}_{q=1}^Q$) & $\{|\alpha_q|\}_{q=1}^Q=1$ \\
%Radar SNR ($\mathrm{SNR}_{\mathrm{radar}}=\frac{|\alpha|P_T}{\sigma_r^2}$) & -20 dBm \\
Angles of targets 1-7 & \begin{tabular}[c]{@{}c@{}}45$^{\circ}$, 0$^{\circ}$, 34$^{\circ}$, 18$^{\circ}$,\\ 9$^{\circ}$, 30$^{\circ}$, 15$^{\circ}$\end{tabular}\\
Velocities of targets 1-5 & 10 $\mathrm{m/s}$\\
Velocities of targets 6 and 7 & 14 $\mathrm{m/s}$, 18 $\mathrm{m/s}$\\
Channels between the BS and users & \begin{tabular}[c]{@{}c@{}}Rayleigh fading \\with i.i.d. $\mathcal{CN}(0,1)$ entries\end{tabular}\\ \hline
\end{tabular}
\end{table}
%%%%%%%%%%%%
%% simulation
%% MFR 6741; 
%%%%%%%%%%%%%%%%%%%%%%%%%
\begin{figure}[tb]
    \centering
    \includegraphics[width=0.85\linewidth]{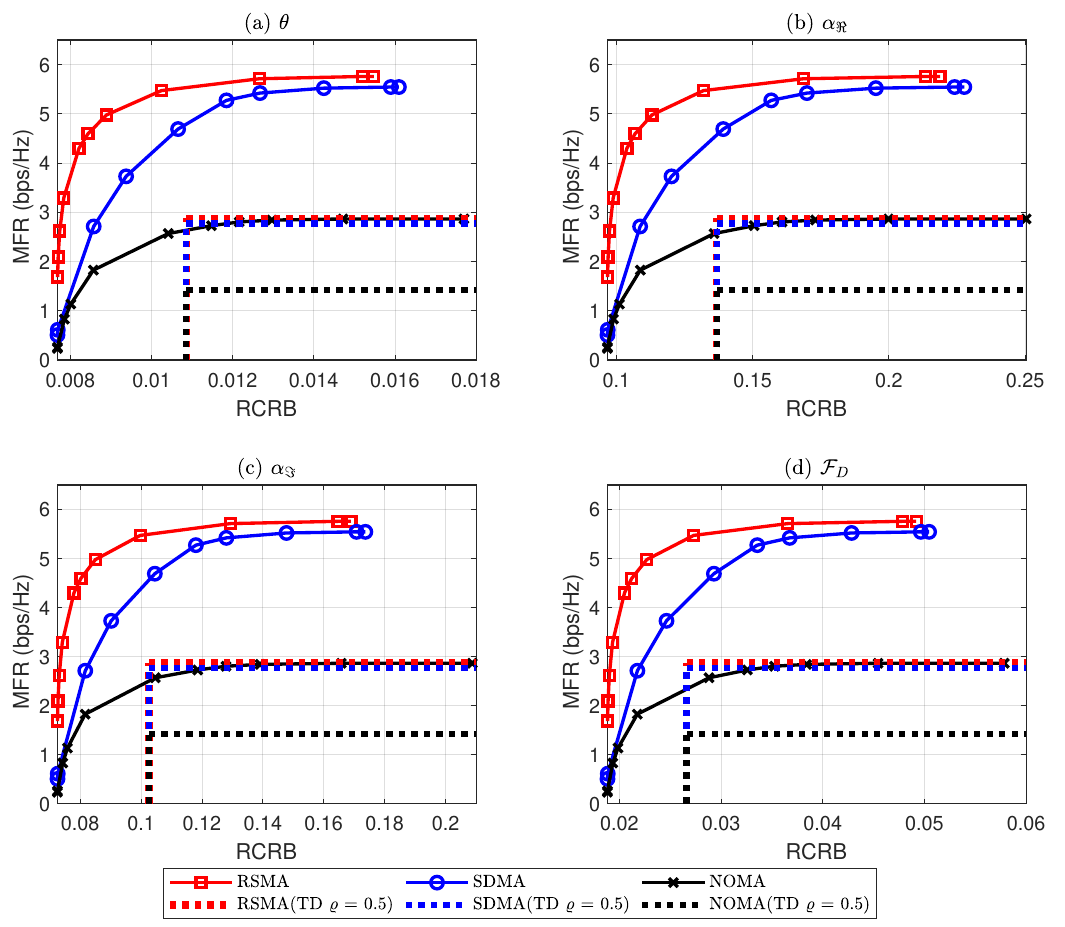}
    \caption{MFR versus RCRB. $N_{T}=6$, $N_{R}=7$, $K=4$, $Q=1$. (a) $\theta$, (b) $\alpha _{\Re}$, (c) $\alpha _{\Im}$, (d) $\mathcal{F}_{D}$.}
    \label{fig:simulation_MFR6741}
\end{figure}
%%%%%%%%%%%%%%%%%%%%%%%%%
\subsubsection{Single-target scenarios} Fig. \ref{fig:simulation_MFR6741} depicts trade-offs between communication and sensing in an underloaded scenario, where the root CRB (RCRB) of each estimation parameter, i.e., $\sqrt{\mathbf{F}_{i,i}^{-1}}, \forall i\in\{1,\cdots,4\}$, is utilized for better illustration.
In general, MA-assisted NO-ISAC systems exhibit a notably enhanced communication MFR versus sensing RCRB over their O-ISAC counterparts, thanks to the integration gain triggered by the seamless interplay between the two functionalities.
To be specific, in NO-ISAC systems, when we consider a communication-prior design (i.e., higher MFR), NO-ISAC aided by SDMA is capable of realizing a MFR comparable to that assisted by RSMA. 
While with priority given to sensing (i.e., lower RCRB), RSMA-assisted NO-ISAC outperforms SDMA-assisted NO-ISAC, as it retains the spatial DoF provided by multiple antennas via SIC at the receiver.
This NO-ISAC system aided by NOMA experiences the worst trade-offs due to its DoF loss.
Similarly, we observe that O-ISAC systems (with a portion $\rho$ of time for communication and $1-\rho$ for sensing) aided by SDMA can achieve a promising performance comparable to that aided by RSMA, while NOMA-assisted O-ISAC suffers from a poor communication performance.
%%%%%%%%%%%%%%%%%%%%%%%%%
\begin{figure}[tb]
    \centering
    \includegraphics[width=0.85\linewidth]{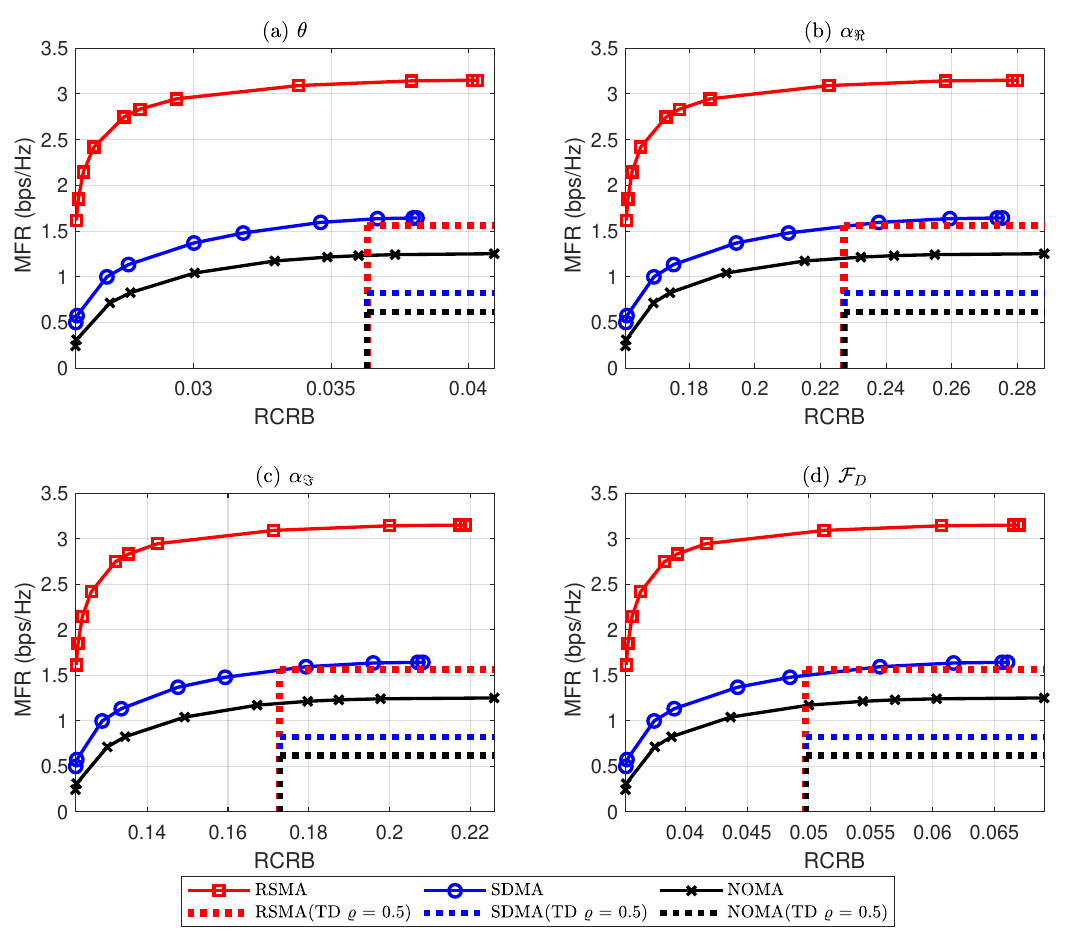}
    \caption{MFR versus RCRB. $N_{T}=3$, $N_{R}=4$, $K=4$, $Q=1$. (a) $\theta$, (b) $\alpha _{\Re}$, (c) $\alpha _{\Im}$, (d) $\mathcal{F}_{D}$.}
    \label{fig:simulation_MFR3441} 
\end{figure}
%%%%%%%%%%%%%%%%%%%%%%%%%
%% MFR 3441
\par
In Fig. \ref{fig:simulation_MFR3441}, we further illustrate communication MFR versus sensing RCRB in an overloaded scenario.
It is obvious that in both O-ISAC and NO-ISAC systems, the performance gap between RSMA and SDMA-aided one widens compared to that in Fig. \ref{fig:simulation_MFR6741}, thanks to the additional design of common stream in RSMA-assisted ISAC. By partially decoding interference and partially treating interference as noise, RSMA-assisted systems realize flexible and effective interference management, leading to promising system performance even in this overloaded scenario with severe interference. 
Moreover, both O-ISAC and NO-ISAC systems assisted by NOMA fail to exhibit performance gain in this overloaded case compared to SDMA-assisted ones, which may arise from the non-aligned user channels and the fixed decoding order at the receiver. 
%%%%%%%%%%%%%%%%%%%%%%%%%
\begin{figure}[tb]
    \centering
    \includegraphics[width=0.8\linewidth]{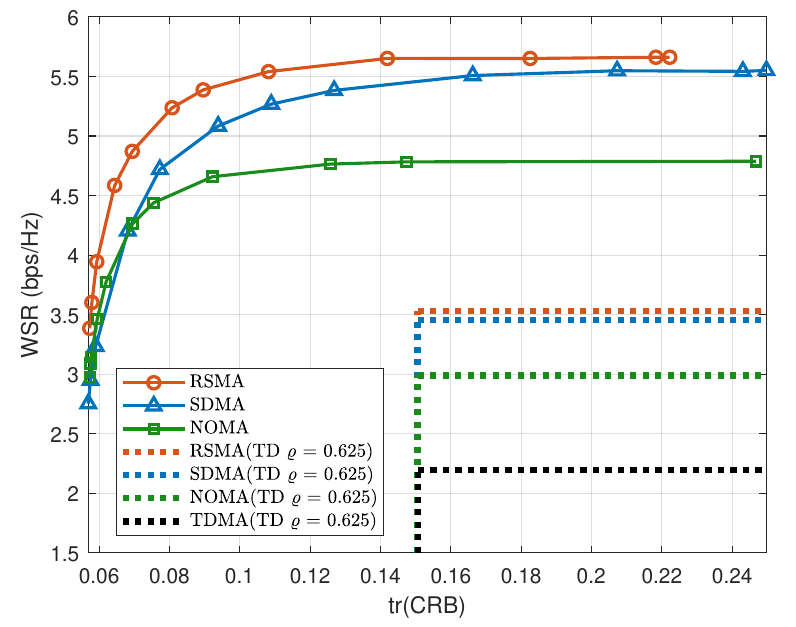}
    \caption{WSR versus tr(CRB). $N_{T}=4$, $N_{R}=5$, $K=3$, $Q=1$, $\mathbf{\mu}=[0.4, 0.3, 0.3]$.}
    \label{fig:simulation_WSR4531}
\end{figure}
%%%%%%%%%%%%%%%%%%%%%%%%%
%% WSR 4531
\par
Fig. \ref{fig:simulation_WSR4531} illustrates the trade-off region between the communication WSR and sensing tr(CRB) in an underloaded scenario. 
We observe that MA-assisted NO-ISAC systems consistently outperform conventional O-ISAC systems in both communication and sensing performance. 
This is primarily due to the non-orthogonal resource allocation between communication and sensing in NO-ISAC, which enables mutually beneficial integration of the two functionalities.
Specifically, in NO-ISAC systems, when sensing is prioritized, NO-ISAC assisted by NOMA slightly outperforms that assisted by SDMA, as SDMA has limited spatial DoFs for managing interference. While in communication-prior designs, NOMA-aided NO-ISAC exhibits the worst trade-offs among all NO-ISAC systems due to its DoF loss. RSMA-assisted ISAC systems consistently deliver the best performance compared to SDMA and NOMA-assisted ones.
Moreover, it is obvious that in O-ISAC systems, the case aided by TDMA suffers from the worst performance since it serves one single user each time.
\subsubsection{Multi-target scenarios}
%%%%%%%%%%%%%%%%%%%%%%%%%
\begin{figure}[t]
    \centering
    \includegraphics[width=0.8\linewidth]{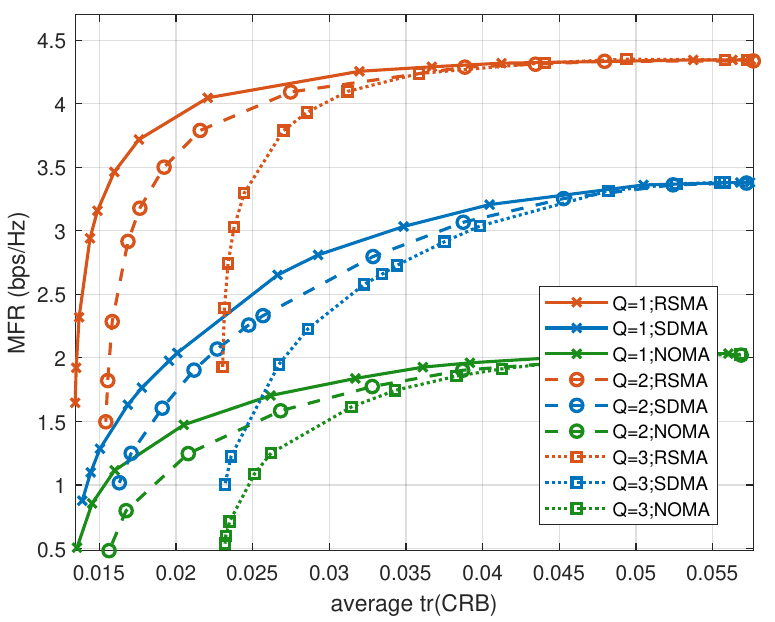}
    \caption{MFR versus average tr(CRB) under different numbers of targets. $Q=1, 2, 3$, $N_{T}=4$, $N_{R}=9$, $K=4$.}
    \label{fig:simulation_numbers}
\end{figure}
%%%%%%%%%%%%%%%%%%%%%%%%%
%% numbers
\par
The trade-offs in NO-ISAC systems for various numbers of targets are illustrated in Fig. \ref{fig:simulation_numbers}, where $K=4$ users and $Q=1$ (target 1), $Q=2$ (target 1, 6) and $Q=3$ (target 1, 6, 7) targets are considered.
As observed, RSMA-assisted NO-ISAC exhibits an explicit trade-off gain over SDMA and NOMA-assisted NO-ISAC. Thanks to the extra DoF provided by the common stream, RSMA-assisted ISAC demonstrates a superior MFR rate than systems assisted by SDMA and NOMA at the rightmost point.
As the number of targets increases, the trade-off regions under all three techniques deteriorate due to the reduced beamforming power allocated to each target. 
Interestingly, we observe that RSMA is capable to sense more targets without compromising the QoS of communication users. This therefore showcases the remarkable potential of RSMA to boost the sensing performance in ISAC.
%%%%%%%%%%%%%%%%%%%%%%%%%
\begin{figure}[tb]
    \centering
    \includegraphics[width=0.8\linewidth]{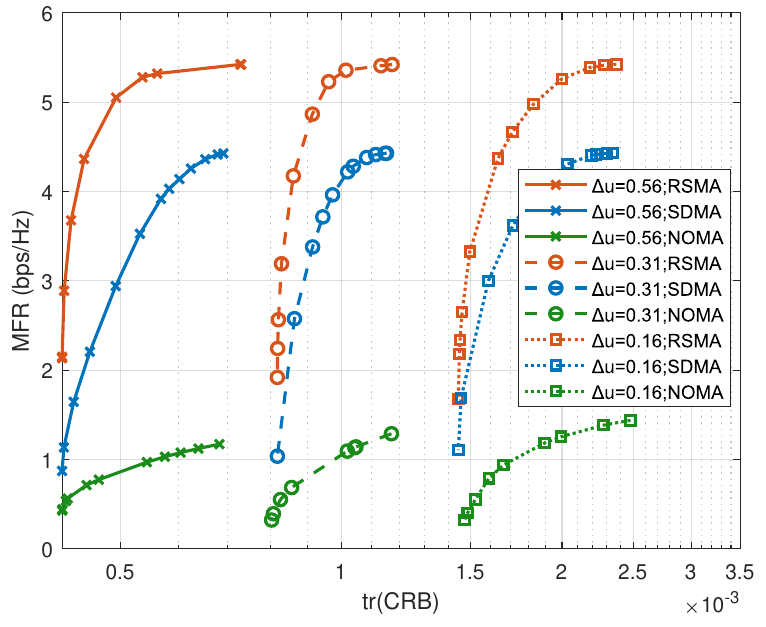}
    \caption{MFR versus tr(CRB) under various angle differences. $N_{T}=4$, $N_{R}=5$, $K=4$, $Q=2$, $I=256$, $\sigma_r^2=-30\ \mathrm{dBm}$.}
    \label{fig:simulation_angle}
\end{figure}
%%%%%%%%%%%%%%%%%%%%%%%%%
%% angles
\par 
In Fig. \ref{fig:simulation_angle}, we illustrate the trade-offs of NO-ISAC with variations in the angle difference between $Q=2$ targets, which is defined as $\Delta u=\sin(\theta_{1})-\sin(\theta_{2})$.
To better capture the impact of the angle difference, we consider the Doppler frequency difference $\Delta f=\mathcal{F}_{D_{1}}-\mathcal{F}_{D_{2}}=0$. 
%Moreover, the user channels are randomly generated following a complex Gaussian distribution \cite{xcc21_RSMAISAC}.
%
With an increasing angle difference from $ \Delta u=0.16$ (target 5, 2), $ \Delta u=0.31$ (target 4, 2) to $ \Delta u=0.56$ (target 3, 2), the tr(CRB) tends to decrease as a result of the reduced interference between received echo signals. However, regardless of the angle difference, the NO-ISAC system assisted by RSMA consistently achieves better performance than that assisted by SDMA and NOMA.
This thereby highlights the robust and efficient capability of RSMA-assisted NO-ISAC for alleviating interference under varying angular conditions.
%%%%%%%%%%%%%%%%%%%%%%%%%
\begin{figure*}[tb]
    \centering    \includegraphics[width=0.8\linewidth]{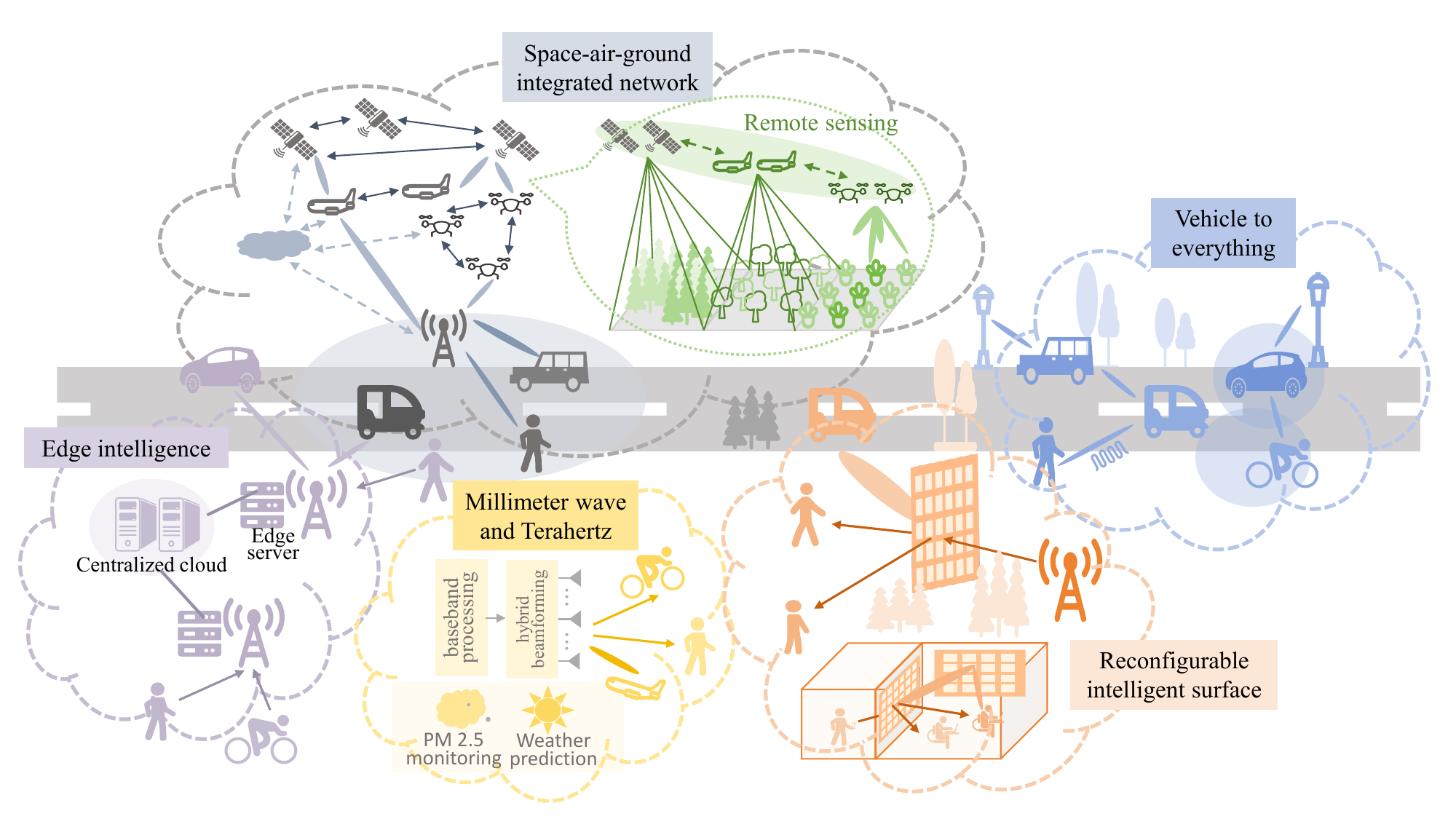}
    \caption{A few applications of MA-ISAC in next-generation wireless networks.}
    \label{fig:application}
\end{figure*}
%%%%%%%%%%%%%%%%%%%%%%%%%
%%%%%%%%%%%%%%%%%%%%%%%%%%%%%%%%%%%%
\section{Applications, Challenges, and Research Directions of MA-assisted ISAC in 6G}\label{section8}
This section discusses some open challenges in MA-ISAC such as precoding toward random ISAC signals, cross-layer optimization and so on. Subsequently, as shown in Fig. \ref{fig:application}, we summarize a few applications of MA-assisted ISAC for enabling potential technologies in next-generation wireless networks, followed by some future research directions.
\subsection{The Road Ahead in MA-ISAC}\label{section8A}
%\subsubsection{Fundamental limits of MA-ISAC} 
%Although substantial efforts have been dedicated to investigating different MA schemes in various ISAC architectures, the fundamental limits of MA-ISAC are far from being thoroughly studied.
%Specifically, existing works in MA-ISAC mainly focus on optimizing sensing performance under communication constraints or vice versa, or a weighted combination of communication and sensing performance using traditional metrics (e.g., the WSR for communication, and the CRB for sensing). 
%However, they typically overlook the investigation of the fundamental limits of MA-ISAC, where the core challenge lies in characterizing the trade-off with a unified performance metric from an information-theoretic perspective.
%In particular, the capacity-distortion trade-off \cite{la22_ISAClimitsurvey, lf22_ISACsurvey} based on the unified capacity-distortion performance metric captures the fundamental limit of simultaneously conveying information and extracting environmental parameters in ISAC, yet remains unexplored in the context of MA-ISAC.
%Importantly, such fundamental limit offers insights not only for theoretical understanding but also for practical system design, as they reveal the intrinsic advantages of various MA-assisted ISAC such as improved SE, and serve as benchmarks for evaluating practical strategies.
%Furthermore, future research can consider more practical conditions such as various mobility models and the imperfect CSI. 
\subsubsection{Precoding toward random ISAC signals}
Radar systems prefer well-designed deterministic signals with favorable ambiguity properties, while communication signals are required to be as random as possible.
Due to the shared use of dual-functional signals between the two functionalities, ISAC signals are typically random to convey useful information, which inevitably leads to sensing performance degradation.
Recent studies explored the impact of this random nature inherent in ISAC signals, which, however, are still in its infancy. Specifically, new sensing metrics such as the ergodic linear minimum mean square error (ELMMSE) \cite{lsh24_randomISAC}, and ergodic least-squares error (ELSE) \cite{lsh24_randomISAC2} were proposed to evaluate the average estimation error over random signals. 
However, existing MA-ISAC studies typically overlook the signal randomness, where the transmit data frames are assumed to be sufficiently long without practical consideration.
The dual-functional performance of MA-ISAC under random ISAC signals worth investigation, and other average sensing performance metrics such as the ergodic CRB, ergodic mutual information remains unexplored.
\subsubsection{Architecture extension toward multi-static sensing} Multi-static sensing mode can be regarded as an extension of the bi-static mode, which involves two or more spatially separated transmitters and receivers for the sensing functionality \cite{lxw25_ISAC}.
This architecture enables enhanced spatial diversity and broader coverage through multi-angle observations facilitated by the spatially distributed sensing transceivers \cite{lrg24_FDISAC}. 
Moreover, multi-static sensing offers additional performance gain by exploiting the joint transceiver beamforming across the network \cite{Behdad24_multistatic}.
Another advantage lies in the reduction of the overall energy consumption, since echo signals from a single transmitter can be reused by multiple receivers.
Multi-static sensing has therefore emerged as an enabler for future cooperative distributed ISAC systems, where multiple spatially distributed transceivers collaborate to support high-resolution sensing and reliable communication.
%\cite{hkw25_distributedISAC,lxl25_multistatic}. 
However, such systems inevitably encounter several challenges, including stringent time synchronization requirements, complex coordination among distributed transceivers, and more intricate interference such as inter-user and inter-functionality interference.
Different MA techniques show great promise for effective interference management and dynamic resource allocation in distributed ISAC architectures \cite{lzw24_RIS_ISACRSMA}, which, however, have not yet received widespread attention and require further investigation.
\subsubsection{Cross-layer optimization}
Due to the diversity of sensing services and dynamic system deployment, future ISAC is prone to experience an emerging burst in random and unexpected sensing quests \cite{lsh24_ISACtenchallenges}.
To fulfill dynamic resource management, the cross-layer optimization has drawn a great deal of attention, since the conventional layer-based optimization is no longer sufficient to meet the demands of ISAC systems or future multi-functional scenarios.
In addition to MA techniques for resource allocation at the PHY layer, another promising solution is to prioritize the dual-functional quests according to their specific requirements at the media access control (MAC) layer, where requests from high-mobility devices can be served more promptly and allocated with more resources.
By leveraging advanced queuing theory and clustering algorithms, network layer optimization holds potential to efficiently schedule multiple tasks, leading to reliable and real-time system performance.
Furthermore, efficient source data encoding and adaptive sensory data transmission can be considered at the application layer, which is crucial to optimize bandwidth utilization and computational efficiency \cite{lsh24_ISACtenchallenges}.
Future research can therefore focus on improving layer-specific strategies and investigating cross-layer design to enhance communication and sensing trade-offs.
\subsubsection{Artificial intelligence (AI)-aided design and optimization}
Existing optimization algorithms for resource allocation and waveform design may be infeasible for practical implementation of MA-aided ISAC.
This is primarily due to their inflexibility to dynamic conditions, high computational complexity, as well as reliance on unrealistic assumptions \cite{zxq25_ISACAI}.
%\cite{zxq25_ISACAI,Aldirmaz25_ISACAI}.
For instance, NOMA-assisted ISAC introduces extra computational complexity in waveform design induced by the joint optimization of user grouping, decoding orders and precoders, while RSMA-aided ISAC involves high-dimensional precoder design along with the common rate allocation arising from message splitting and recombination.
This inevitably results in complicated non-convex optimization problems with more variables to be optimized, which is therefore difficult to acquire low-complexity solutions.
Fortunately, the widespread development of artificial intelligence (AI) has motivated its application to MA-ISAC to address these issues, and it is widely expected to provide efficient and robust solutions, including but not limited to the prevailing reinforcement learning (RL) and deep learning (DL) techniques.
\subsubsection{Standardization of advanced MA and ISAC}
In addition to the aforementioned open challenges in MA-assisted ISAC systems, extensive efforts are needed for promoting the standardization of both advanced MA (such as RSMA) and ISAC in the future.
% MA
For the standardization of MA, 3GPP has evolved systems based on TDMA (2G), CDMA (3G), and (O)FDMA (4G and 5G) over the past several decades.
And from the early stage of 4G, SDMA was expected to enable multi-user MIMO systems. 
Subsequently, NOMA with non-orthogonal resource allocation was considered as a study item in 3GPP Release 15, but its standardization was decided not to continue and NOMA is left for possible use in the future \cite{makki20_NOMAtutorial}.
Recently, the brand new technique RSMA has showcased its potential for improving transmission performance thanks to its universal capability. This enables its application to future intelligent, multi-functional networks, and makes it attractive for 6G, though it remains uncertain whether 3GPP will adopt it as a unified MA scheme or continue with multiple tailored schemes for specific conditions in the future \cite{bruno24_MA}.
% ISAC
For ISAC standardization, key organizations such as 3GPP and European Telecommunication Standard Institution (ETSI) are focusing on integrating ISAC into future networks. 
Specifically, 3GPP is identifying key requirements for ISAC in the next phase of its study, with an evaluation of whether ISAC should be integrated into the PHY layer as part of the 5G-Advanced standard.
Moreover, the industry specification group (ISG) established by ETSI mainly focuses on pre-standard research for ISAC \cite{Ahmed24_NOMAISACmagazine}. Its scope includes important 6G-related ISAC aspects, such as defining prioritized use cases, establishing superior channel models, identifying performance indicators, addressing private and secure transmission, and so on.
%%%%%%%%%%%%%%%%%%%%%%%%
\subsection{MA-ISAC for Enabling Technologies in 6G}
\subsubsection{mmWave/THz MA-ISAC}
The surge in wireless data traffic, along with the scarcity of available spectrum resources, has motivated the implementation of ISAC in mmWave (i.e., 30 GHz to 300 GHz) and THz (i.e., 300 GHz to 10 THz) bands. 
The huge bandwidth contributes to the enhancement of both communication capacity and sensing resolution \cite{jw24_THz}.
However, mmWave/THz ISAC is susceptible to several issues, among which the predominant one is the severe path loss arising from its short wavelength characteristic 
\cite{Ahmed24_NOMAISACmagazine}.
One effective solution to overcome this attenuation is to employ large antenna arrays at the ISAC transmitter for achieving high gains, whereas this makes the conventional fully digital beamforming impractical.
Fortunately, hybrid analog/digital beamforming has been proposed as an alternative method, which is more feasible involving a high-dimensional analog beamformer and a low-dimensional digital beamformer \cite{wx18_mmWave}.
Such system, however, suffers from a high-complexity feedback procedure, which involves feedback of codewords indices for analog precoders and quantized effective channel.
Moreover, the promising performance demands perfect CSIT, which is a challenging task due to the channel variability in mmWave/THz \cite{wx18_mmWave}. 
It is also worth noting that the dual-functional performance is constrained by the severe inter-user and inter-functionality interference, which arises from the reduced spatial DoFs imposed by the hybrid beamforming architecture in ISAC systems.
Inspired by the interference management capability of MA schemes in mmWave/THz communication-only systems \cite{lz19_RSMAhybrid}, recent studies begin exploring their potential to mitigate increased interference in mmWave/THz ISAC, which is still in its early stage \cite{gj24_RSMA_mmWaveISAC} and deserves further investigation.  
%For instance, the authors of \cite{gj24_RSMA_mmWaveISAC} investigated a RSMA-assisted mmWave ISAC system, which tackled the joint optimization of common rate allocation and hybrid beamforming, and showed that mmWave ISAC aided by RSMA achieves higher EE and better trade-offs compared to NOMA and SDMA.
%%
\par
\textit{Future research directions:} To address the substantial feedback overhead in existing MA-assisted mmWave/THz ISAC systems, CS and DL techniques can be utilized to fully exploit the inherent spatial sparsity of mmWave/THz channels, thereby reducing the amount of CSI feedback while enhancing both communication and sensing performance. Additionally, MA techniques such as RSMA demonstrate robustness to quantization distortion, making MA a promising solution when combined with CS and DL methods to further improve dual-functional performance in the future.
%%%%%%%%%%%%%%%%%%%%%%%%
\subsubsection{MA-ISAC for V2X}
V2X networks involving vehicle-to-pedestrian (V2P), vehicle-to-vehicle (V2V), vehicle-to-infrastructure (V2I) and vehicle-to-network (V2N) plays a pivotal role in intelligent transportation, which has been included in 3GPP release 14 as a long time evolution (LTE)-supported technology \cite{zy22_V2Xsurvey}.
The extensive growth of V2X and rising demand of new services (such as automatic driving) motivates the development of the next-generation V2X reinforced by ISAC, which holds promise for simultaneously sensing surroundings, and communicating with other pedestrians, vehicles and roadside units (RSUs). 
This thereby facilitates V2X networks with both communication and sensing, leading to enhanced energy, spectral and cost efficiency. 
Despite the appealing prospect of V2X-ISAC, there exist several unique demands such as low latency, mobility support as well as exceptional reliability, which are difficult to be satisfied. MA techniques, which showcase the capability of reducing communication latency and improving reliability in V2X communication systems \cite{dby17_V2XNOMA},
%\cite{dby17_V2XNOMA, dby17_NOMAV2X2}, 
are therefore promising to address those challenges. 
\par
\textit{Future research directions:} 
One promising direction is to exploit the DD-domain MA for V2X-ISAC, thanks to its capability of alleviating traffic congestion and robustness in high-mobility V2X scenarios.
It is also worth noting that catering to extensive services, there typically exists private information transmission in V2X communications, which can be extracted easily by malicious targets due to the dual-functional ISAC signal design. The investigation of trade-offs between secrecy rate and sensing performance is therefore another potential direction for MA-assisted ISAC in V2X. 
%%%%%%%%%%%%%%%%%%%%%%%%
\subsubsection{MA-ISAC supported by reconfigurable intelligent surface (RIS)}
RIS, which features numerous passive elements, demonstrates its capability to independently alter the electromagnetic properties (such as phases and amplitudes) of incident signals by regulating the parameters of electronic circuits on each element \cite{wqq21_RIStutorial}. 
This therefore enables it to manipulate the wireless propagation environment for both communication and sensing functionalities in ISAC systems. 
Specifically, the DoFs introduced by RIS hold promise for enhancing signal strength and suppressing inter-user and inter-functionality interference.
The deployment of RIS also enables the ISAC transmitter to establish virtual links for users and targets in invisible areas, which therefore extends system coverage and improves overall dual-functional performance. 
Meanwhile, studies have begun incorporating MA techniques along with RIS into ISAC systems, which further enlarges its trade-off performance thanks to more DoFs \cite{czc24_RISISACRSMA}. 
Among them, secure transmission has emerged as an important research direction \cite{ph24_RIS_ISACRSMAsecure, Salem24_RIS_ISACRSMAsecure}, since there inherently involves security issue arising from the resource sharing between communication and sensing.
Additionally, growing efforts have been dedicated to investigating new RIS architectures for MA-ISAC, such as simultaneously transmitting and reflecting RIS (STAR-RIS)-assisted MA-ISAC \cite{wwj24_STARRIS_ISACNOMA}.
More recently, a revolutionary RIS architecture named beyond-diagonal RIS (BD-RIS) has been proposed \cite{lhy25_BDRIStutorial}.
%\cite{lhy22_BDRIS,lhy25_BDRIStutorial}. 
However, although substantial efforts have been dedicated to BD-RIS-assisted ISAC \cite{WBW23_BDRISISAC,ckx24_transmitterBDRIS}, the incorporation of both advanced MA and BD-RIS for effective interference management in ISAC systems remains unexplored.
\par
\textit{Future research directions:} 
Existing MA-RIS-assisted ISAC studies consider single RIS with certain location, where the signaling to control RIS is previously known and fixed. 
The scenarios with multiple RISs, vehicle-mounted RIS, new architectures of RIS (e.g., BD-RIS), as well as the dynamic design of RIS deployment, size and signaling merits investigation for better performance with enhanced efficiency.
It is also worth noting that the acquisition of CSI at the RIS is difficult. One common solution is to estimate the concatenated channel at the ISAC transmitter leveraging RIS reflection patterns, whereas this typically results in imperfect CSIT. 
The ergodic sensing and communication performance of MA-RIS-assisted ISAC systems with imperfect CSIT deserves further exploration.  

%%%%%%%%%%%%%%%%%%%%%%%%
\subsubsection{MA-ISAC with space-air-ground integrated network (SAGIN)}
With the extensive deployment of satellites (such as low Earth orbit (LEO) satellites), aerial platforms (such as UAVs), and terrestrial BSs, SAGIN has been envisioned as an appealing architecture in next-generation wireless systems. 
%\cite{xy24_SAGIN}.
Apart from the seamless global connectivity in communication, some emerging sensing services call for geographic location and status acquisition functions provided by the network. 
ISAC therefore becomes one effective solution to facilitate multi-functionality in SAGIN, ultimately leading to adaptable, robust, and high-performing integrated networks for diverse services \cite{zyb24_SAGIN}.
For instance, drone or satellite-based remote sensing has gained substantial attention as a promising component of SAGIN thanks to its high-resolution, day-and-night, and all-weather imaging capability, which supports applications such as atmospheric monitoring, disaster assessment and military surveillance \cite{lf22_ISACsurvey}.
However, the integrated network is susceptible to severe co-channel interference among the space, air and ground segments, and inter-functionality interference between communication and sensing.
This issue primarily arises from the radio resource sharing between communication and sensing, and among multiple sensing tasks, and is prone to result with performance degradation. 
Inspired by the demonstrated efficacy of MA in resource allocation and interference management in satellite, air, and terrestrial communication-only \cite{ylf22_SAGIN,seong25_LEOsatellite,Seong25_satellitecom} 
%\cite{ylf22_SAGIN, lx21_NOMAsatellite,zym22_NOMAsatellite, lz21_RSMAsatellite,ryu24_RSMAsatellite,Toka24_satellitecom,seong25_LEOsatellite,Seong25_satellitecom}
and ISAC \cite{ylf24_ISAC,Juha24_RSMAISAC}
%\cite{ylf24_ISAC, hmy24_NOMAsatelliteISAC,Juha24_RSMAISAC,lz24_LEOISAC}
networks, MA holds potential for addressing interference in ISAC with SAGIN, which, however, is still in its early stages \cite{zyx25_SAGINNOMAISAC} and requires further investigation.
\par
\textit{Future research directions:}
Due to the computational limitation in satellites and aerial platforms, there exists a growing demand for jointly optimizing computation in addition to communication and sensing functionalities within a unified framework \cite{cq25_SAGINcomputation}, whereas this leads to difficult resource allocation under dynamic network topology and diverse services. The incorporation of MA has great potential to tackle this issue and unlock the full potential of multi-functional SAGIN in the future.

%%%%%%%%%%%%%%%%%%%%%%%%
\subsubsection{MA-ISAC meets edge intelligence}
The next generation wireless networks are envisioned to drive forward intelligence of everything for extensive intelligent services. 
In contrast to conventional DL algorithms involving high computation loads at terminals, edge intelligence exists as one promising concept to support AI tasks at the network edge, which typically involves sensing for environment information acquisition, communication for data sharing, and computation for information processing 
\cite{wdz24_EIISAC}.
Benefiting from the concept of ISAC, edge intelligence tasks—communication, sensing and computation can be merged for enhanced performance rather than competing for resources to avoid signal suppression.
This emerging paradigm is also known as integrated sensing, computation, and communication (ISCC) \cite{wdz24_ISCCsurvey}.
%\cite{hyh24_ISCC,wdz24_ISCCsurvey}.
Though appealing as an integrated network with low-delay and reliable performance, the resource sharing among different tasks is one major challenge, which calls for effective solutions. 
Thanks to superior performance of MA in ISAC and edge intelligent computing systems \cite{cpx23_RSMAEI}, 
%\cite{ff20_NOMAEI, cpx23_RSMAEI}, 
it shows great potential to realize promising trade-offs among computation accuracy, communication efficiency and sensing accuracy.
\par
\textit{Future research directions:} 
To unlock the potential of edge intelligence, various edge servers are required to collaboratively work with each other, where local data with private information may be transmitted to unrelated devices. MA techniques-aided secure ISAC for edge intelligence is therefore one potential direction.
Additionally, despite progress in individual theoretical analysis  for communication, sensing, and computation, as well as their partial integration such as ISAC, 
a unified information-theoretic analysis that captures the joint performance of all three functionalities remains unexplored, leaving a gap in guiding the design of effective ISCC strategies.
MA-ISAC with edge intelligence also raises new challenges in ensuring robustness against unexpected device failures, as well as in protocol standardization, necessitating further efforts in this research area.
%%%%%%%%%%%%%%%%%%%%%%%%
\section{Conclusion}\label{section9}
The existing literature has shown that MA techniques play a crucial role in ISAC systems by enabling enhanced spectrum utilization, efficient ISAC waveform designs, as well as flexible interference management between communication and sensing, or among communication users. 
This synergy between MA and ISAC not only extends the interference management capability of MA techniques beyond communication-only networks, but also mitigates inter-user and inter-functionality interference in ISAC, thereby unlocking the potential of future multi-functional wireless networks.
In this paper, we provide the first holistic tutorial on the incorporation of MA into ISAC systems.
The scope of this tutorial spans the fundamental principles of ISAC, the diverse interference types encountered in different ISAC systems, as well as a comparison of various MA-assisted ISAC systems with respective advantages and disadvantages.  
In addition to presenting the foundation of the interplay between MA and ISAC, we also provide an outlook on the emerging applications and future research directions of MA-assisted ISAC. 
We hope this tutorial will serve as a valuable reference on the use of MA in ISAC, and ultimately expedite meaningful and interesting future research in this field.
%%%%%%%%%%%%%%%%%%%%%%%%
\bibliographystyle{IEEEtran}
\bibliography{reference}
\end{document}